\documentclass[aip,cha,reprint]{revtex4-1}

\newcommand{\arXiv}[1]{{\tt \href{http://arXiv.org/abs/#1}{arXiv:#1}}}

\usepackage{graphicx,amsmath,amssymb,comment,bbold,bbm,eufrak,yfonts,subfigure,color}
\usepackage[normalem]{ulem}

\definecolor{mygreen}{rgb}{0,0.7,0}

\begin{document}

\title{Memory effects, transient growth, and wave breakup in a model of paced atrium}
\author{Alejandro Garz\'{o}n}
\email{alejandro.garzon@correo.usa.edu.co}
\affiliation{Department of Mathematics, Universidad Sergio Arboleda, Bogot\'{a} 110221, Colombia}
\affiliation{School of Physics, Georgia Institute of Technology, Atlanta, Georgia 30332-0430, USA}
\author{Roman O. Grigoriev}
\affiliation{School of Physics, Georgia Institute of Technology, Atlanta, Georgia 30332-0430, USA}

\date{\today}

\begin{abstract}
The mechanisms underlying cardiac fibrillation have been investigated for over a century, but we are still finding surprising results that change our view of this phenomenon.
The present study focuses on the transition from normal rhythm to atrial fibrillation associated with a gradual increase in the pacing rate.
While some of our findings are consistent with existing experimental, numerical, and theoretical studies of this problem, one result appears to contradict the accepted picture.
Specifically we show that, in a two-dimensional model of paced homogeneous atrial tissue, transition from discordant alternans to conduction block, wave breakup, reentry, and spiral wave chaos is associated with transient growth of finite amplitude disturbances rather than a conventional instability.
It is mathematically very similar to subcritical, or bypass, transition from laminar fluid flow to turbulence, which allows many of the tools developed in the context of fluid turbulence to be used for improving our understanding of cardiac arrhythmias. 
\end{abstract}

\maketitle

\begin{quotation}
Atrial fibrillation is a common type of cardiac arrhythmia characterized by spatially uncoordinated high-frequency patterns of electrical excitation waves.
Several different explanations of how fibrillation arises from a highly coordinated and regular normal rhythm have been offered.
The most common is a dynamical scenario that involves several stages. 
First, a milder type of arrhythmia known as alternans, characterized by regular spatial and temporal modulation in the duration of the excitation, is created as a result of a period doubling or Hopf bifurcation.
In the second stage, the modulation grows until some portion of an excitation wave fails to propagate (which is known as conduction block), producing wave breakup.
In later stages, spiral waves segments, or wavelets, form and undergo subsequent breakups until a complicated dynamical equilibrium 
is established, with wavelets continually annihilating and reemerging.
The current paradigm is that each of these stages represents a separate bifurcation that leads to either a change in the stability, or the disappearance, of the solution describing a previous stage.
Our analysis of a simple model of paced cardiac tissue shows that transitions between different stages can happen without any bifurcations taking place.
In particular, conduction block can be caused by strong transient amplification of finite amplitude disturbances, which is a flip side of a very well-known memory effect that describes the sensitivity of the asymptotic dynamics of paced cardiac tissue to the entire pacing history. 
\end{quotation}

\section{Introduction}
Fibrillation, a state of uncoordinated contraction of the heart muscle, can occur in the ventricles, leading to sudden cardiac arrest, a major cause of death \cite{Mozaffarian2016}, or the atria, where it creates significant health complications \cite{Chugh2014}. 
Understanding of the transition from the normal cardiac rhythm, also known as the 1:1 response, to fibrillation is critical for the development of therapies for the prevention or termination of this dangerous arrhythmia.  
Abundant experimental evidence points to the role of alternans \cite{Nolasco1968}, a beat-to-beat alternation in the action potential duration (APD) at high excitation rates, also known as the 2:2 response, as a precursor of fibrillation both in the ventricles \cite{rosenbaum94,Bloomfield2006} and in the atria \cite{Hiromoto2005,Narayan2011,Narayan2016}.
Strong correlation between alternans and fibrillation was also found in the simulations employing high-fidelity models of two-dimensional ventricular \cite{Majumder2016} and atrial \cite{Chang2016} tissue.

In tissue, alternans manifests as spatial modulation in the width of the excitation waves \cite{Pastore1999}, which leads to dynamical variation in the tissue refractoriness. 
In particular, spatially discordant alternans, in which APD alternates out of phase in different regions of the heart, markedly enhances dispersion of refractoriness, increasing the likelihood of conduction block and, in two- or three-dimensional tissue, reentry \cite{Weiss2006}.
Theoretical analysis of propagating excitation waves in one dimension based on amplitude equations \cite{Echebarria2002,echebarria2007} show that voltage-driven alternans are described by stable periodic or quasiperiodic solutions, with the tissue size and dispersion of conduction velocity determining whether the alternans is concordant or discordant.
The same is true of calcium-driven alternans \cite{Skardal2014}, which qualitatively behave like their voltage-driven cousins.
Theoretical studies of alternans in higher dimensions \cite{echebarria2007} are too limited in scope to make any specific predictions regarding the transition from alternans to fibrillation, but based on experiments and numerical simulations \cite{Fenton2002} we know that conduction block that leads to wave breakup plays a key role.

As the pacing rate or the tissue size in increased, conduction block is predicted to arise when alternans becomes unstable leading to unbounded growth of the amplitude of modulation \cite{karma94,Echebarria2002,fox2002conduction}. 
Conduction block leads to a time-periodic response such as 2:1 (every other excitation wave is blocked) in one dimension, but typically generates  wave breakup and transition to spiral wave chaos (SWC) in two dimensions or scroll wave chaos in three dimensions (respectively, atrial and ventricular fibrillation) \cite{Alonso2016}.
As we show in the present study, there is another possibility: conduction block can be triggered by very small, but finite, perturbations of {\it stable} discordant alternans.

Although this result appears counterintuitive at first, it is not completely unexpected.
It is well-known that paced cardiac tissue displays multistability in zero \cite{Surovyatkina2010}, one \cite{skardal2014coexisting}, and two \cite{Comtois2005,Oliver2005} dimensions.
Transitions between different stable attractors in spatially extended dissipative nonequilibrium systems are in fact not uncommon.
Transition from stable laminar flow to turbulence in shear fluid flows is perhaps the oldest and best known example.
In the case of fluid turbulence, this is known as a bypass transition \cite{Henningson1993} 
and involves strong transient amplification of small disturbances \cite{Trefethen1993}.

Although no direct evidence of the role of transient amplification in the dynamics of cardiac tissue is available at present, indirect evidence is provided by the so-called memory effect \cite{Chialvo1990,rosenbaum1982,Hund2000}. 
This term refers to the dependence of asymptotic states of paced tissue preparations on the entire pacing history rather than just the final pacing interval. 
For instance, in canine ventricles, a pacing-down protocol (with pacing interval decreased in steps of 50 ms to 10 ms) produced a different stationary pattern of alternans than a protocol with constant pacing rate applied to quiescent tissue, for the same asymptotic rate \cite{Gizzi2013}. 
Similar dependence was found in normal rabbit hearts \cite{Ziv2009}: while decreasing the pacing interval by 10 ms steps failed to induce discordant alternans, this state was reached by using smaller steps of 5 ms to 2 ms. 

The objective of this paper is to establish a connection between transient amplification, memory effects, and the transition to atrial fibrillation in a simple two-dimensional model of cardiac tissue featuring discordant alternans. 
This paper is structured as follows.
Section \ref{sec:model} describes the model used in this study. 
The results and their discussion are presented in Sections \ref{sec:results} and \ref{sec:summary}, respectively.

\section{Model of paced atrial tissue}\label{sec:model}

Since our focus here is on fundamental mechanisms rather than detailed comparison with experiments, we chose a simple monodomain model of atrial tissue.  
It has the form of a reaction-diffusion partial differential equation (PDE)
\begin{equation}
\label{eq:p_tu}
\partial_t {\bf u} = D\nabla^2 {\bf u} + {\bf f}({\bf u}) + {\bf I}_p,
\end{equation}
where the field ${\bf u}=[u,v]({\bf r},t)$ describes the state of cardiac cells (cardiomyocytes), $D$ is a diagonal matrix of diffusion coefficients that represents the coupling between adjacent cells, ${\bf f}({\bf u})$ is the {\it ionic model} that describes the dynamics of individual cells, and ${\bf I}_p$ is the (scaled) density of the external pacing current.

Furthermore, we chose the ionic model introduced by Karma \cite{karma94}, which captures the essential alternans instability responsible for initiating and maintaining complex arrhythmic behaviors. 
To make it differentiable, the model has been modified \cite{ByMaGr14,Marcotte2015} such that
\begin{equation}
{\bf f}({\bf u}) =
\left[
\begin{array}{c}
(u^*-v^M) \{1-\tanh(u-3)\} u^2/2 -u \\
\epsilon \{ \beta \Theta_\alpha(u-1) + \Theta_\alpha(v-1)(v-1) - v \}
\end{array}
\right],
\end{equation}
where $\Theta_\alpha(u)=[1+\tanh(\alpha u)]/2$, $u$ is a scaled transmembrane voltage, and $v$ is a gating variable. Unless otherwise stated, 
the parameters values used were $u^*=1.5415$, $M=4$, $\epsilon=0.01$, $D_{11}=1.1 \times 10^{-3}$ ${\rm cm}^2/{\rm ms}$, $D_{22}=D_{11}/20$, $\alpha=32$, and $\beta=[1-\exp(-R)]^{-1}$, where $R=1.273$ is the restitution parameter. 
We also used no-flux boundary conditions $\nabla{\bf u}\cdot\hat{\bf n}=0$, where $\hat{\bf n}$ is a unit vector normal to the domain boundary, as a physiologically accurate representation of the boundary of excitable tissue.

At the chosen value of the restitution parameter, in two dimensions, this model reproduces the transition from normal rhythm to alternans to fibrillation observed in experiments as pacing frequency is increased. 
Yet, with just two variables, as opposed to several tens of variables in the most complex models \cite{FentonCherry08mcc}, the Karma model is simple enough to facilitate the computationally demanding analysis presented in this work.

All numerical simulations were carried out on a square domain of side length $L=2.489$ cm, which is close to the minimal size required for sustained SWC in Karma model with the present choice of parameters.
The PDE \eqref{eq:p_tu} was discretized in space using finite differences on a $96 \times 96$ computational grid, which corresponds to the size of one computational cell $\Delta x = \Delta y = 0.0262$ cm roughly equal to the physical length of one cardiac cell. 
The values of $u$ and $v$ on this grid define the state of the system as a point in a 18432-dimensional state space.
The discretized equations were solved using 4th order Runge-Kutta method with time step $\Delta t=0.01$ ms chosen to be much smaller than the shortest characteristic time scale of the model, $\tau_u=2.5$ ms.

The pacing was accomplished by applying a current to a $5 \times 5$ patch of grid points located near the upper-left corner of the domain. 
The patch was displaced by one grid point vertically in order to break the symmetry with respect to reflection about the diagonal of the computational domain. 
The pacing current was in the form of square pulses of 5 ms duration,
\begin{equation}
p(t)=\left\{
\begin{array}{ll}
I_0, &0 \le t \le 5 \mbox{ ms},\\
0,   &\mbox{otherwise}.
\end{array}
\right.
\end{equation}
Denoting $t_n$ the time at which the $n$-th pulse is applied, the pacing current can be written as
\begin{equation}
{\bf I}_p({\bf r},t) = \sum_{n=0}^\infty p(t-t_n)\,g({\bf r})
\left[\begin{array}{l}1\\ 0\end{array}\right],
\end{equation}
where $g({\bf r})=1$ for paced grid points and zero otherwise.

\section{Results}\label{sec:results}

\subsection{Wave breakup induced by fast pacing}

We investigated stability of the excitation waves produced by pacing with gradually decreasing pacing interval $T_n = t_n - t_{n-1}$, $n=1,2,...$, starting at $t=t_0=0$. 
For some tissue models, a decrease in the pacing interval does not produce wave breakup unless (i) the tissue is heterogeneous or (ii) an external current is applied at a location other than the pacing site (ectopic beat) \cite{Qu2000a}. 
However, for the Karma model considered here, neither of these conditions is necessary for wave breakup.

\begin{figure}
\centering
\includegraphics[width=2.8in]{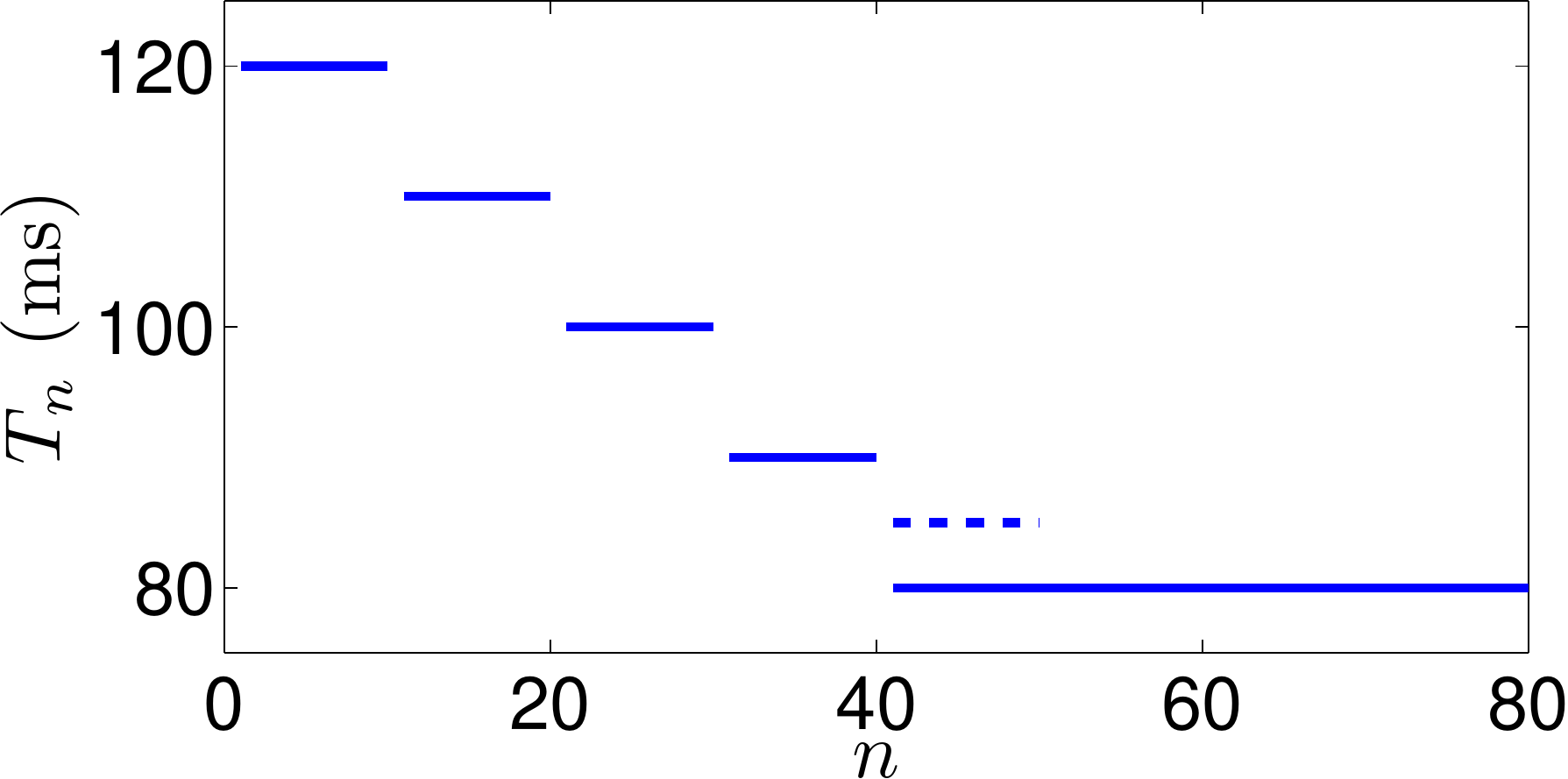}
\caption{\label{fig:Tn_n} Pacing protocol A (solid line) and values at which pacing protocol B differs from A (dashed line).}
\end{figure}

\begin{figure}
\centering
\subfigure[]{\includegraphics[trim=-10 0 -2 0,clip,height=1.38in]{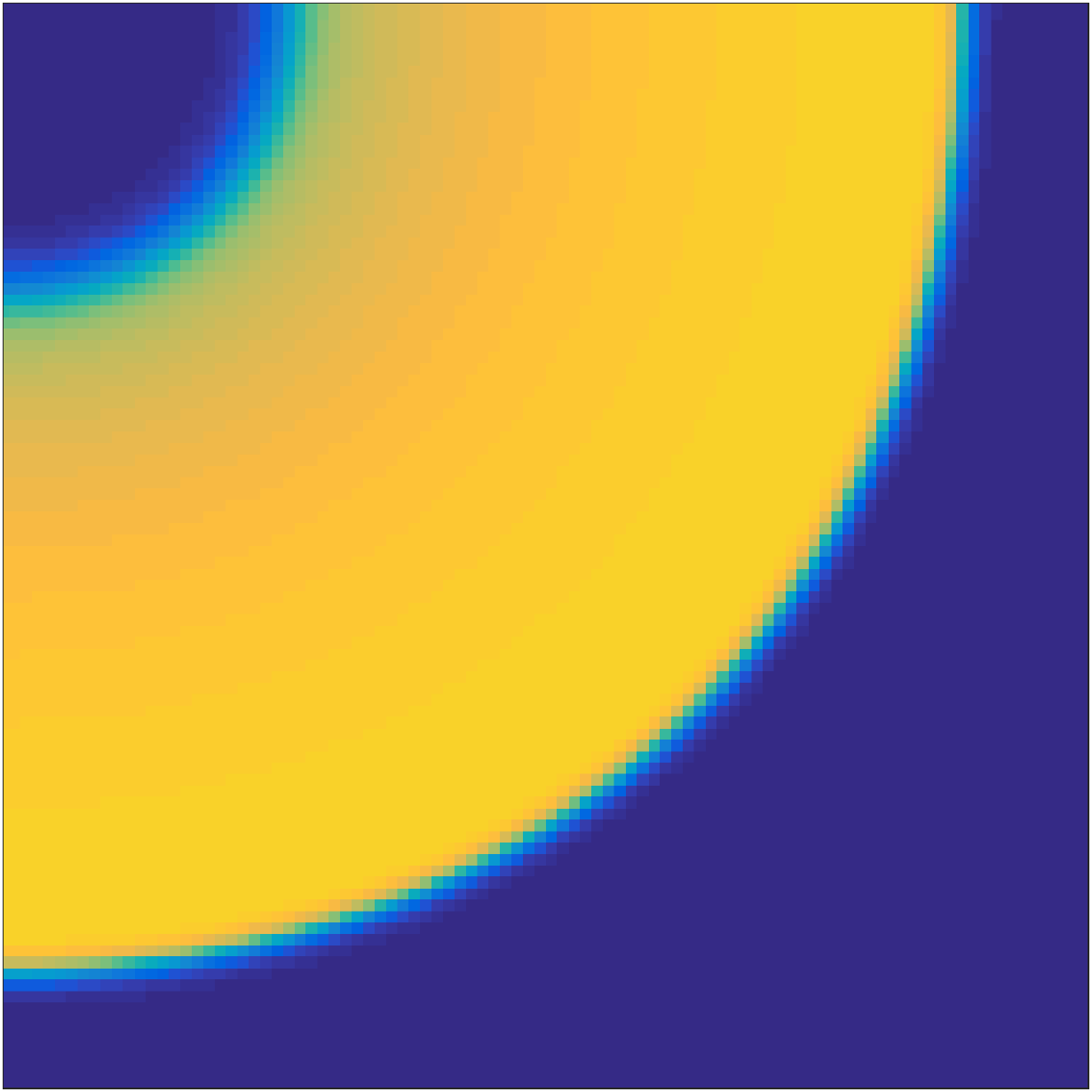}}\hspace{2mm}
\subfigure[]{\includegraphics[trim=0 0 -65 0,clip,height=1.38in]{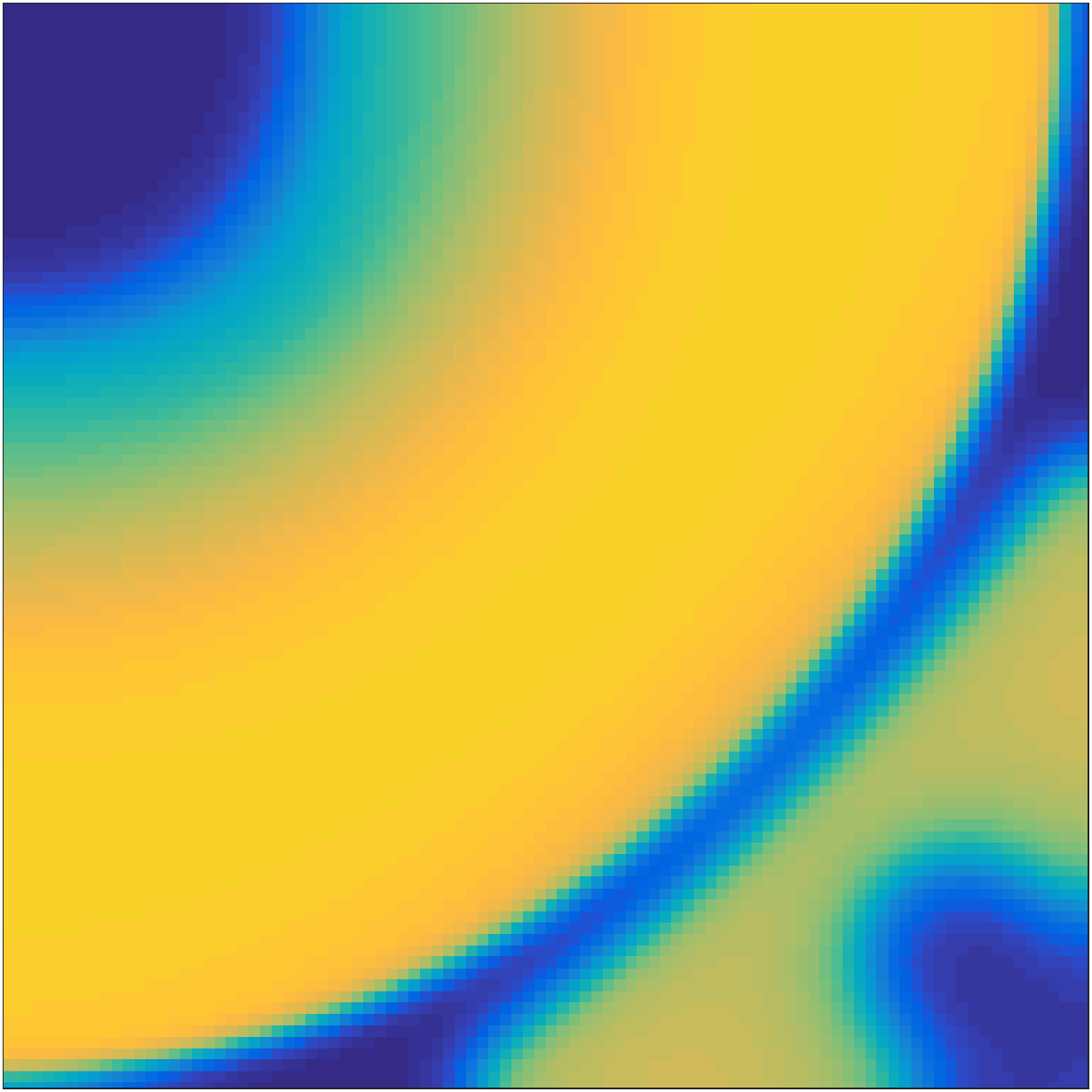}}\\
\subfigure[]{\includegraphics[trim=0 -11 0 0,clip,height=1.42in]{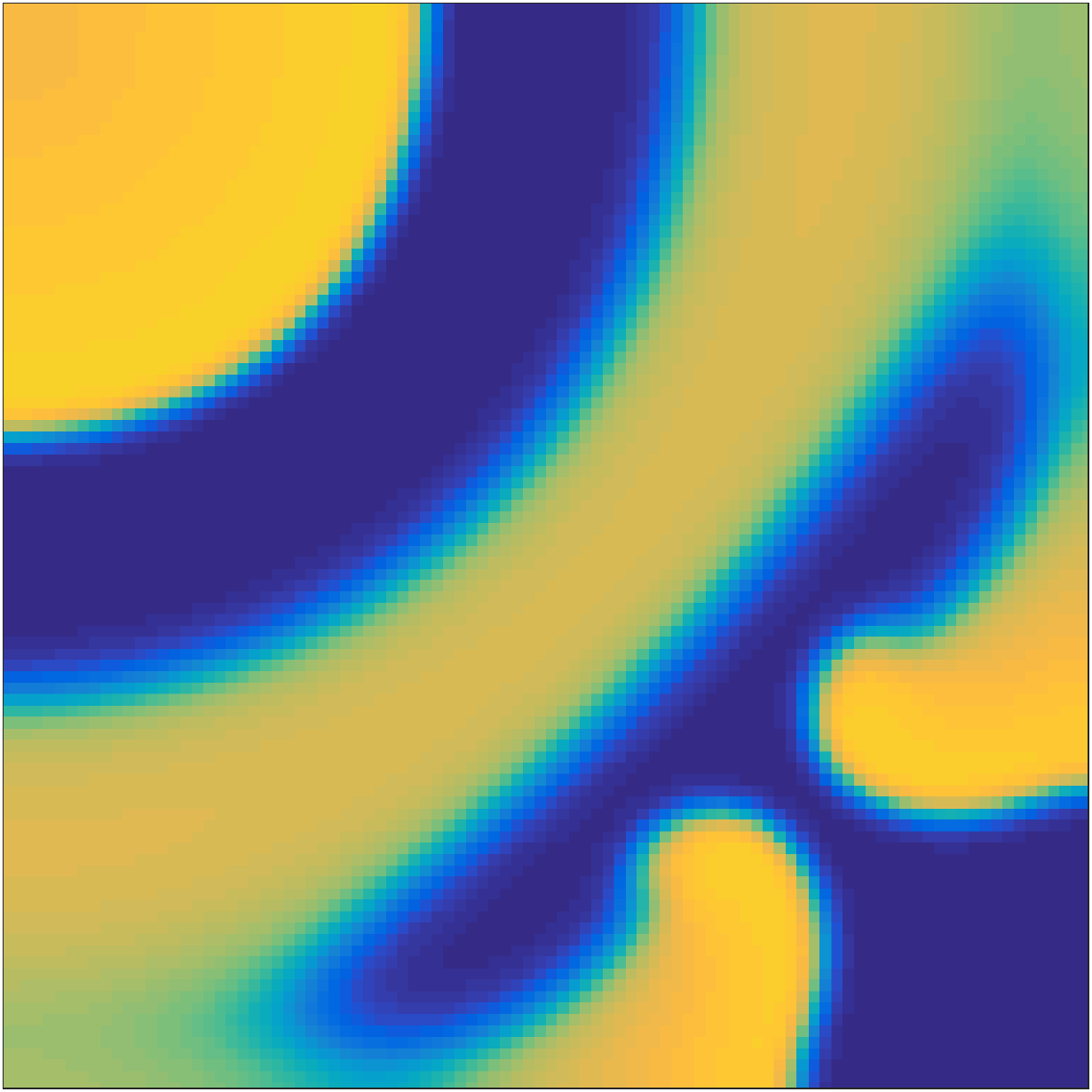}}\hspace{2mm}
\subfigure[]{\includegraphics[height=1.5in]{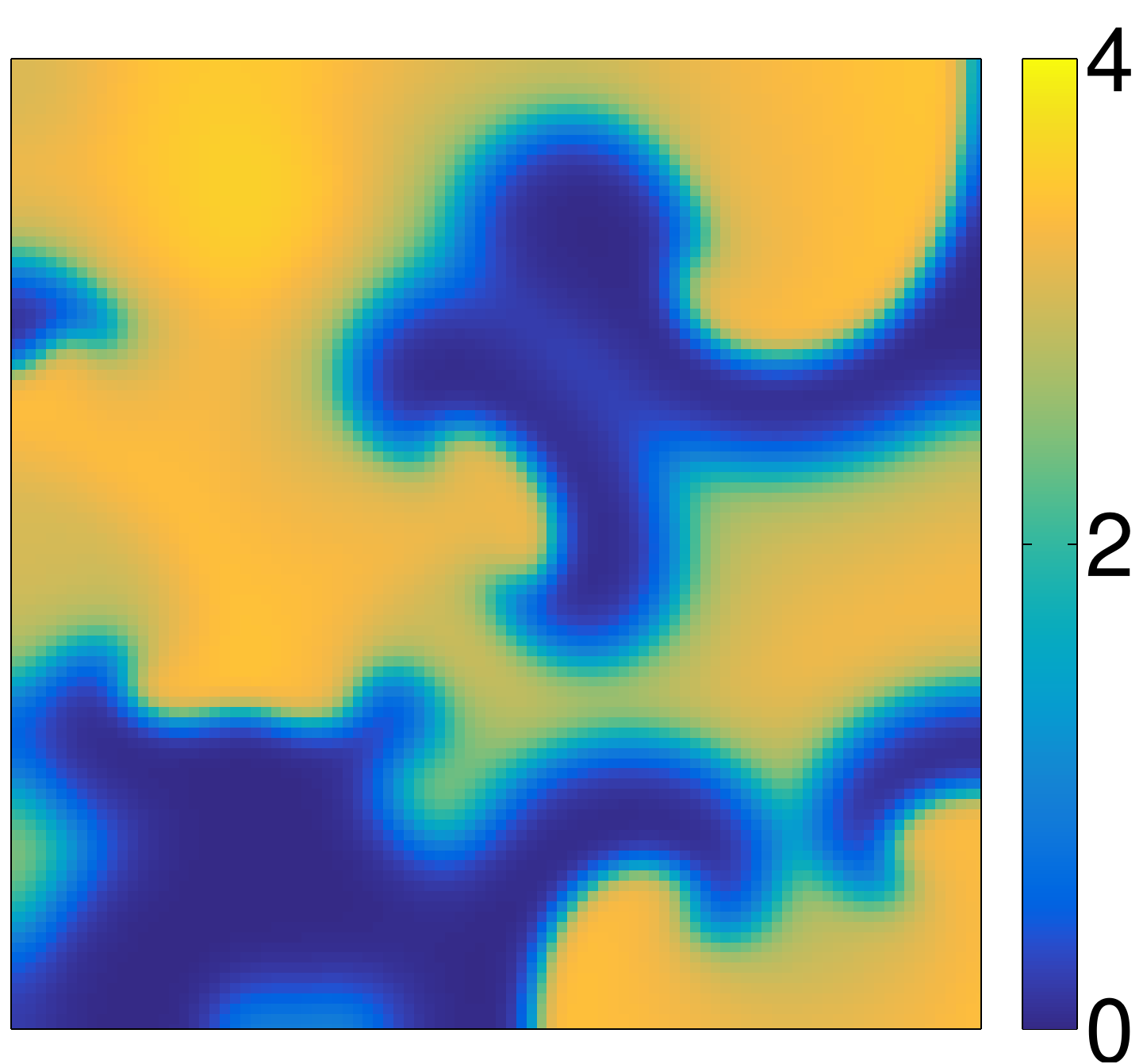}}
\caption{\label{fig:circ_wave} Voltage component, $u({\bf r},t)$, of excitation waves during protocol A at (a) $t=t_3 + 68$ ms, (b) $t=t_{55}+ 72$ ms, (c) $t=t_{56} + 28$ ms, and (d) $t=t_{72} + 32$ ms. The initial condition at $t_0=0$ is the rest state. The pacing site is located at the upper-right corner. The same color bar is used in all subsequent figures showing the voltage field.}
\end{figure}

Let us define $T_{[i,j]}$, $j>i$, to be a sequence of pacing stimuli with a constant interval: $T_i=T_{i+1}=...=T_j$.  
Consider the pacing protocol A: $T_{[1,10]}=120$ ms, followed by $T_{[11,20]}=110$ ms, $T_{[21,30]}=100$ ms, $T_{[31,40]}=90$ ms, and $T_{[41,80]}=80$ ms (cf. Fig. \ref{fig:Tn_n}) with initial condition corresponding to the stable uniform equilibrium ${\bf u}=[0,0]$.
The protocol B is identical to protocol A, except for $T_{[41,50]} = 85$ ms (dashed line in Fig. \ref{fig:Tn_n}).
For protocol B, each pacing stimulus generates a traveling wave with nearly circular wavefront and waveback (cf. Fig. \ref{fig:circ_wave}(a)) that is absorbed at the bottom and right domain boundaries. The asymptotic state is time-periodic, with period $2T=160$ ms, and corresponds to discordant alternans as we will see below.
During the first 55 pacing intervals, pacing protocol A produces qualitatively the same dynamics as protocol B.
On the 56th pacing interval, however, conduction block causes wave breakup (cf. Fig. \ref{fig:circ_wave}(b)), creating two spiral waves on the 57th interval (cf. Fig. \ref{fig:circ_wave}(c)), which undergo additional breakups and eventually transition to SWC (cf. Fig. \ref{fig:circ_wave}(d)). 
SWC corresponds to atrial fibrillation; both feature re-entrant waves and are sustained irrespective of the pacing.  

So how can we understand the difference in the outcomes of the two protocols?
Although this difference is a clear indication that the memory effect plays an important role in our system, this does not explain much.
The memory is actually quite long: indeed, 15(!) pacing stimuli preceding conduction block in protocol A were delivered at the same asymptotic rate.
Given that the asymptotic alternans state is {\it stable}, it would have been natural to expect that any deviation present at $t=t_{50}$ would all but disappear after such a long period.
In fact, as we show below, infinitesimal initial perturbations about this stable alternans state can {\it grow}, albeit transiently, by several orders of magnitude over long time intervals. 
Small, but finite, initial perturbations can grow enough to cause a transition to a completely different (here chaotic) attractor, as it happens for protocol $A$.
This is a signature of multistability often associated with subcritical instabilities; similar phenomenology is found during bypass transition to turbulence in fluid flows \cite{Orszag1980,Chate1987}.

\subsection{Periodic solutions}

\begin{figure}
\includegraphics[width=1.1in]{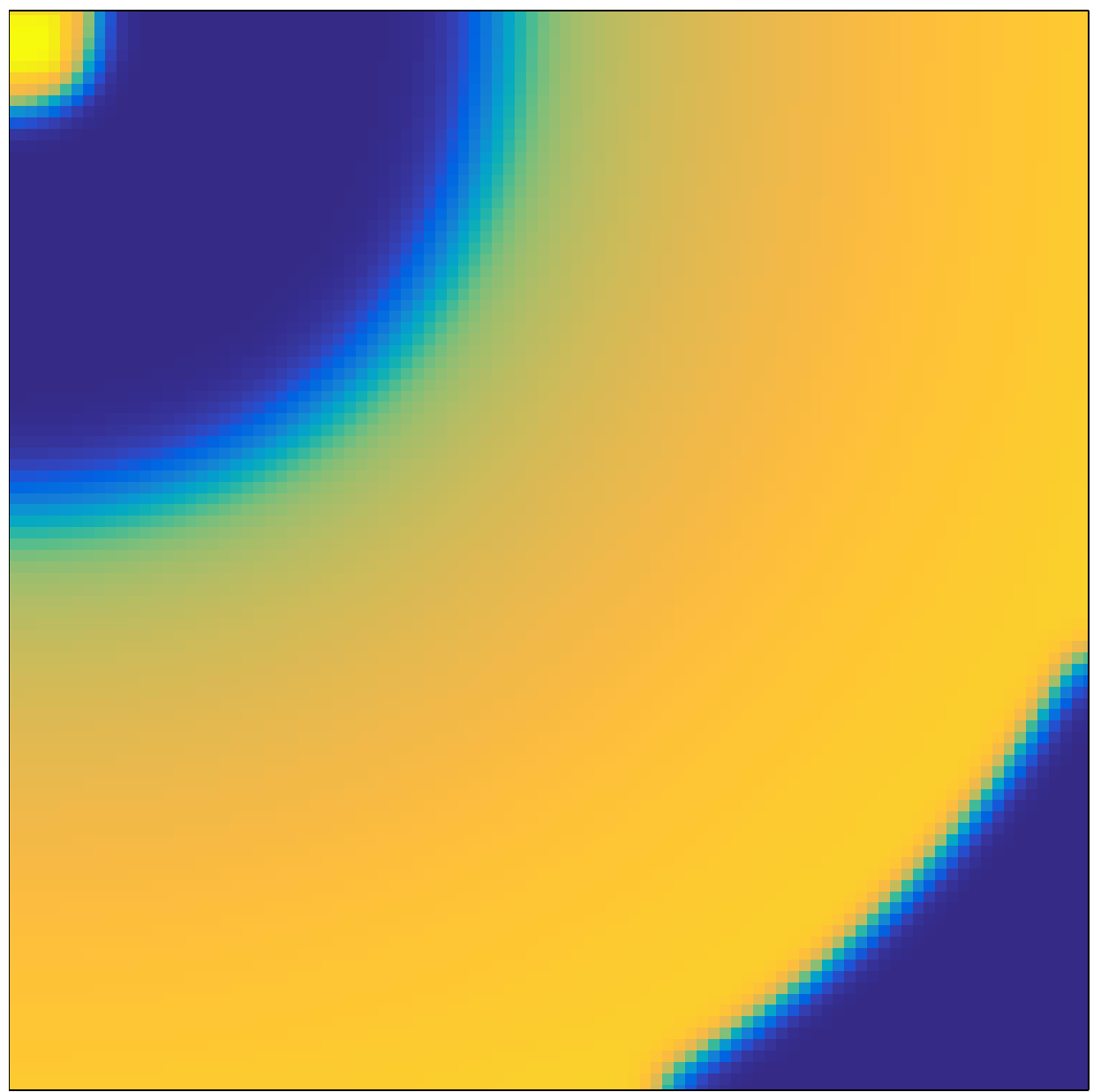}
\includegraphics[width=1.1in]{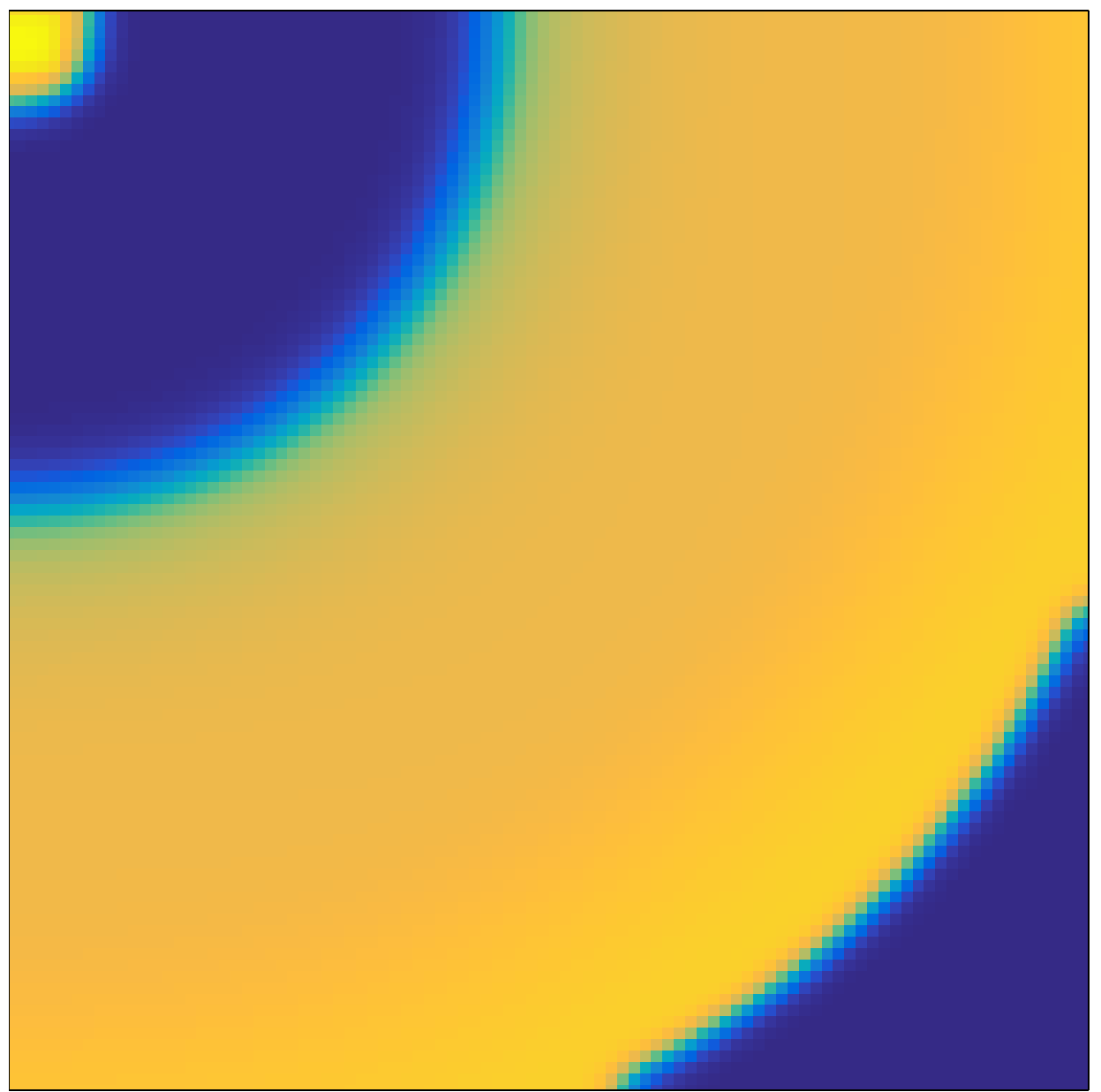}
\includegraphics[width=1.1in]{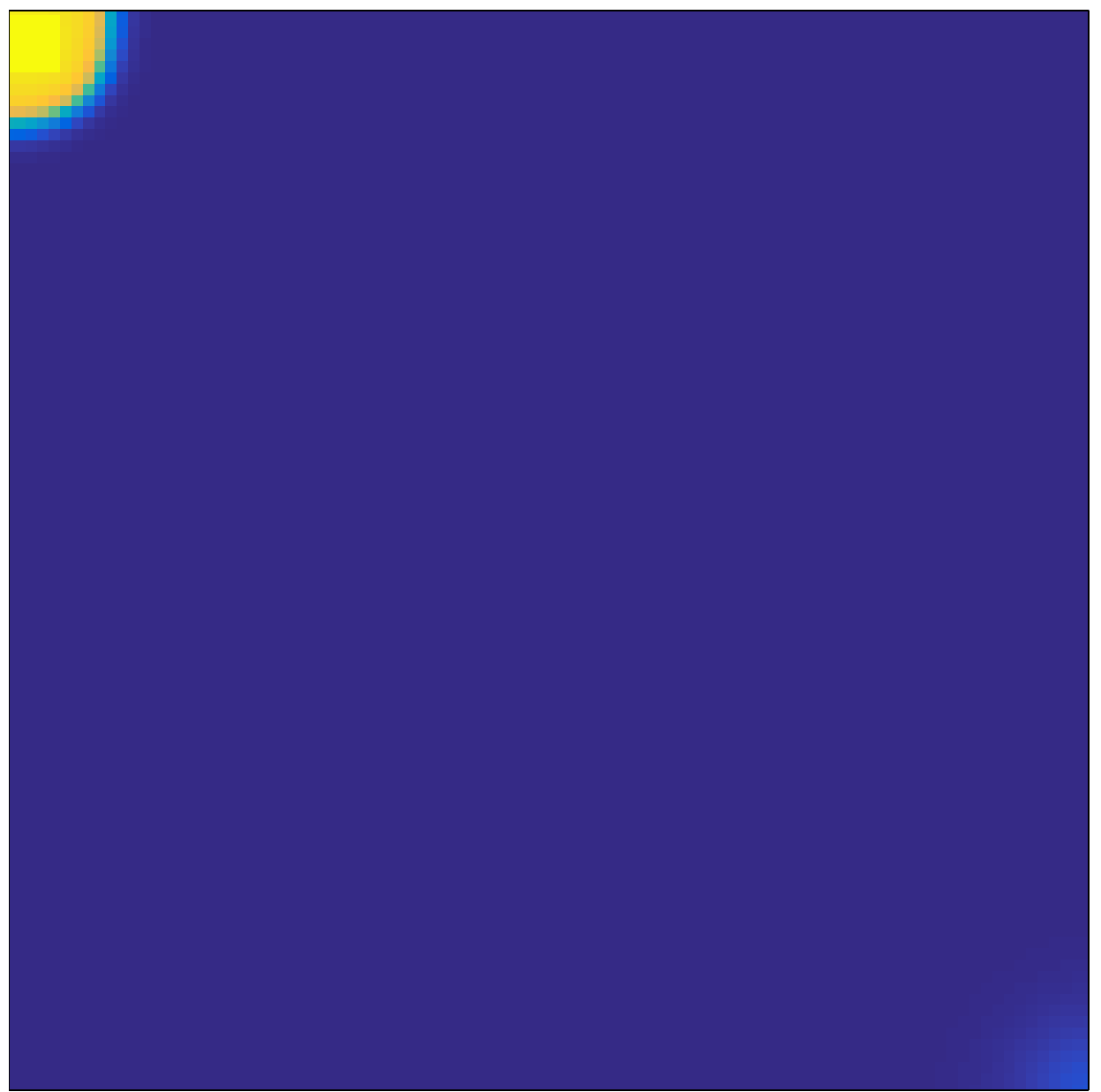}\vspace{1mm}\\
\includegraphics[width=1.1in]{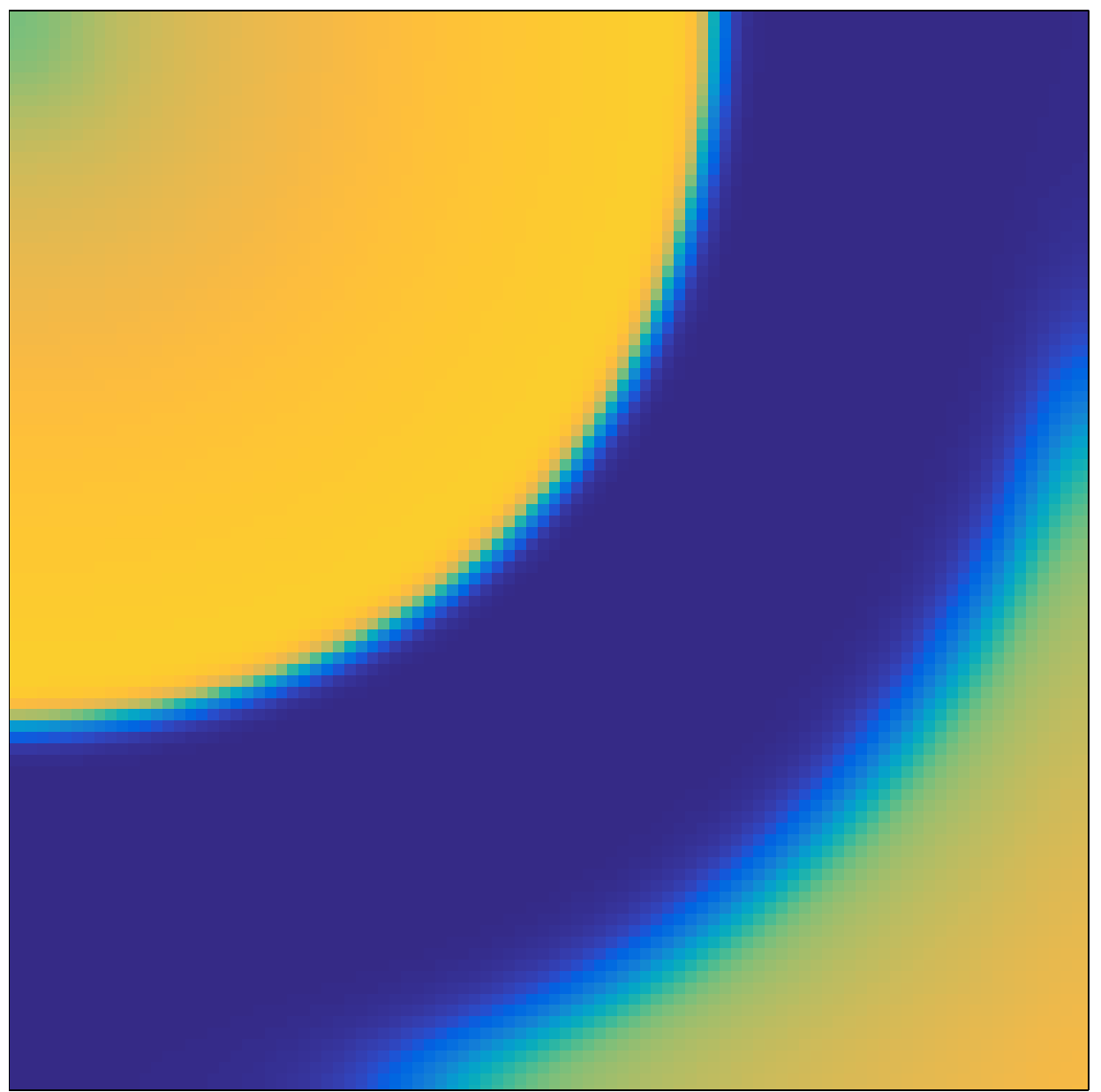}
\includegraphics[width=1.1in]{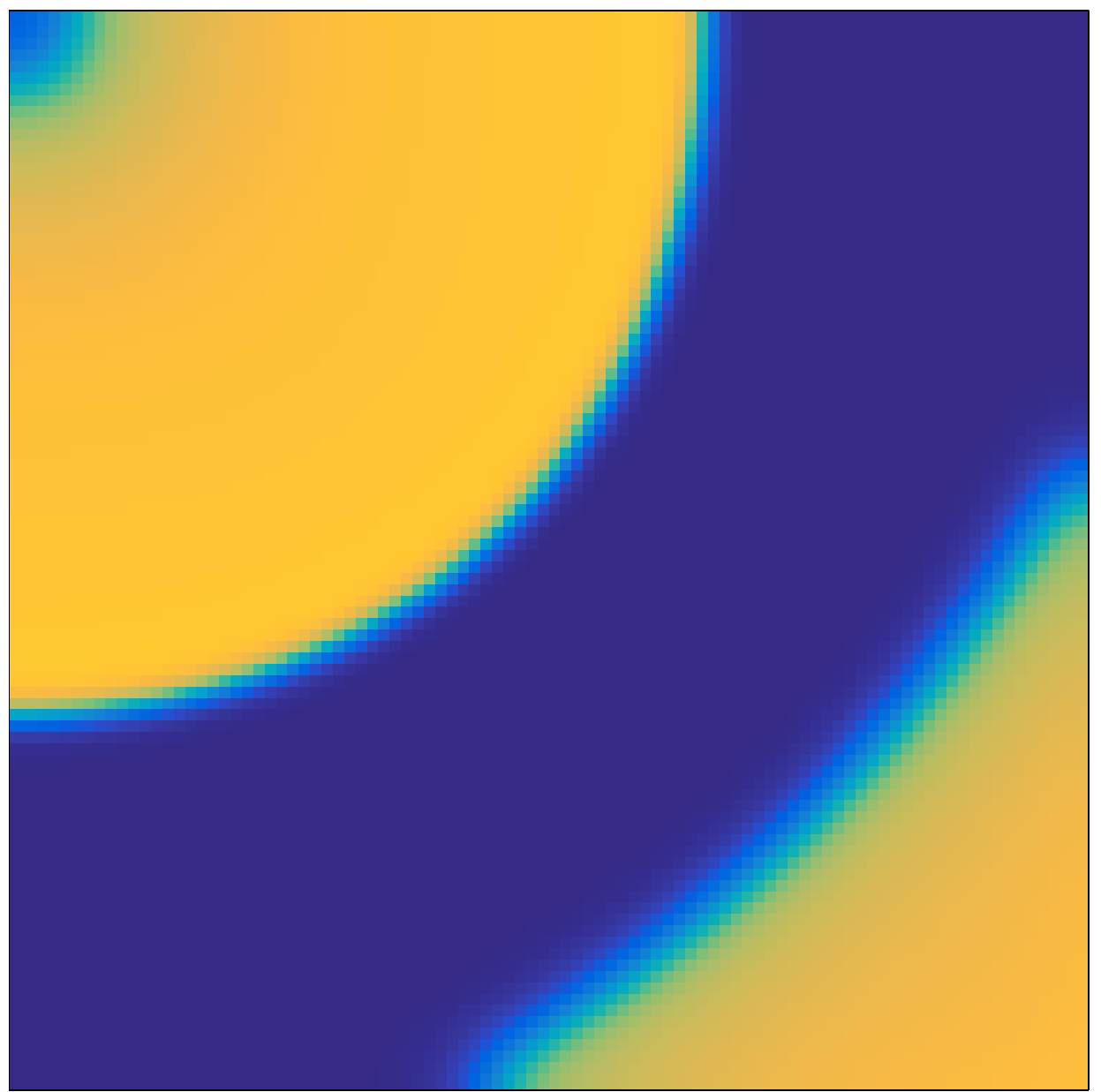}
\includegraphics[width=1.1in]{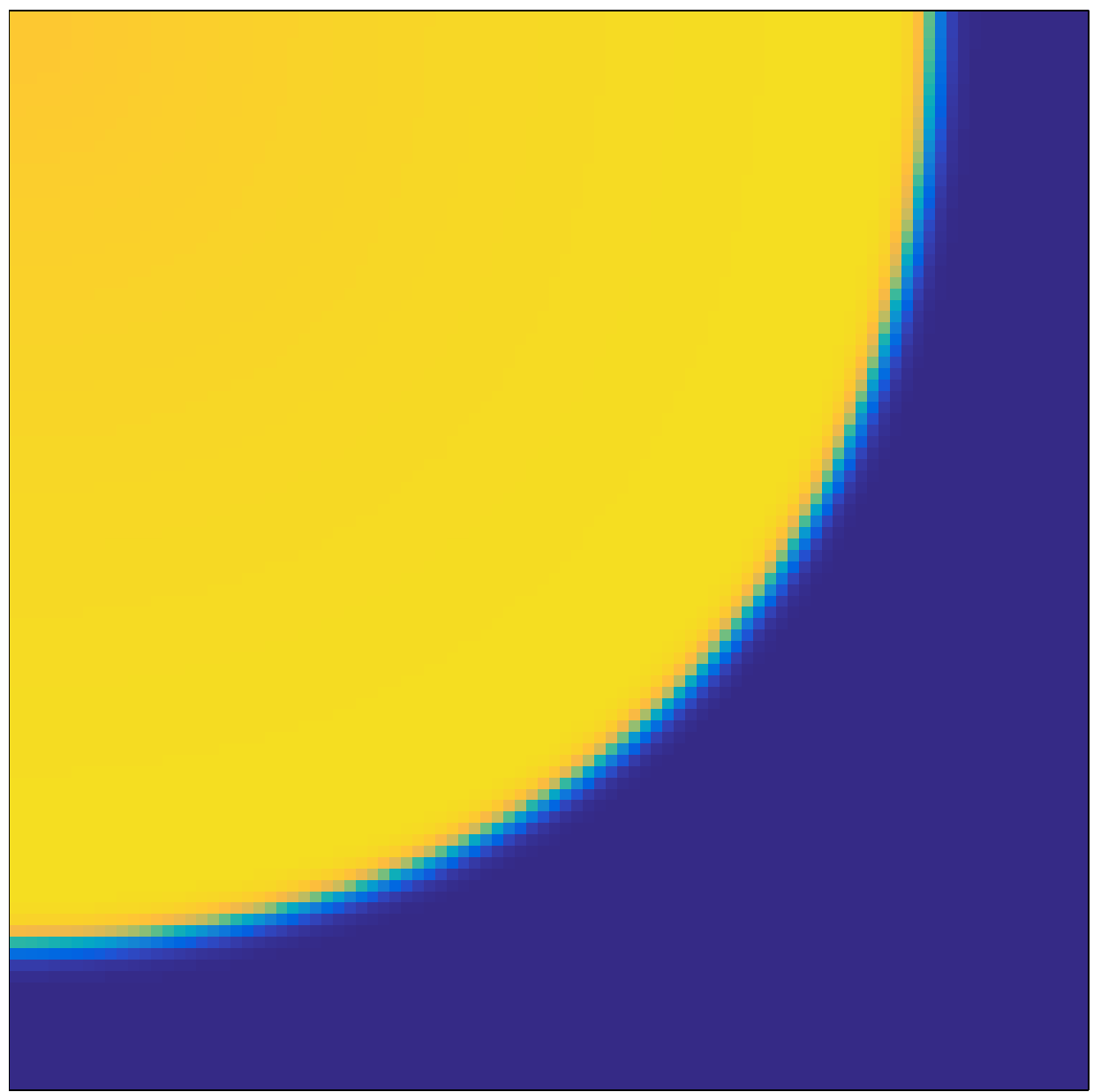}\vspace{1mm}\\
\includegraphics[width=1.1in]{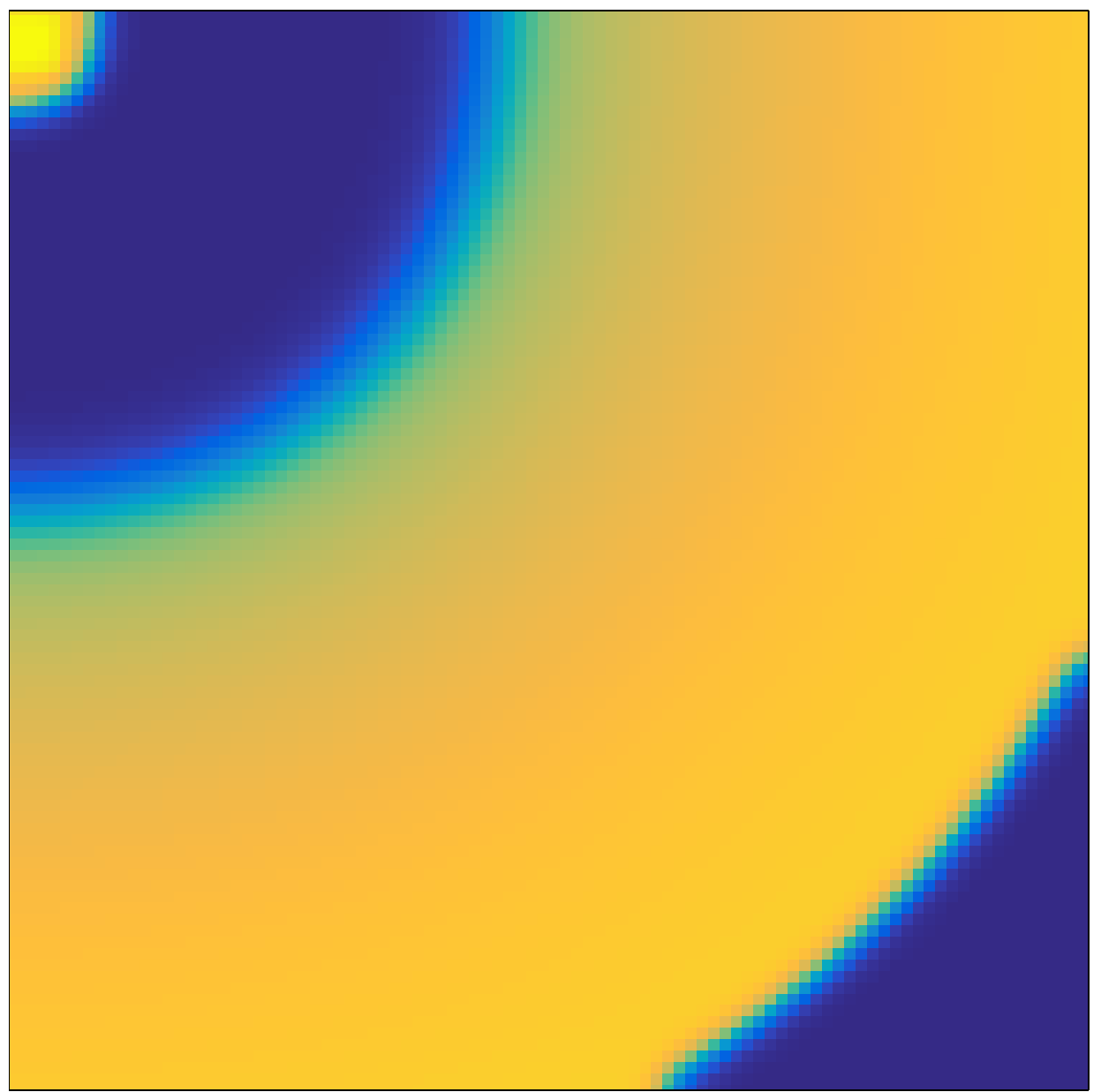}
\includegraphics[width=1.1in]{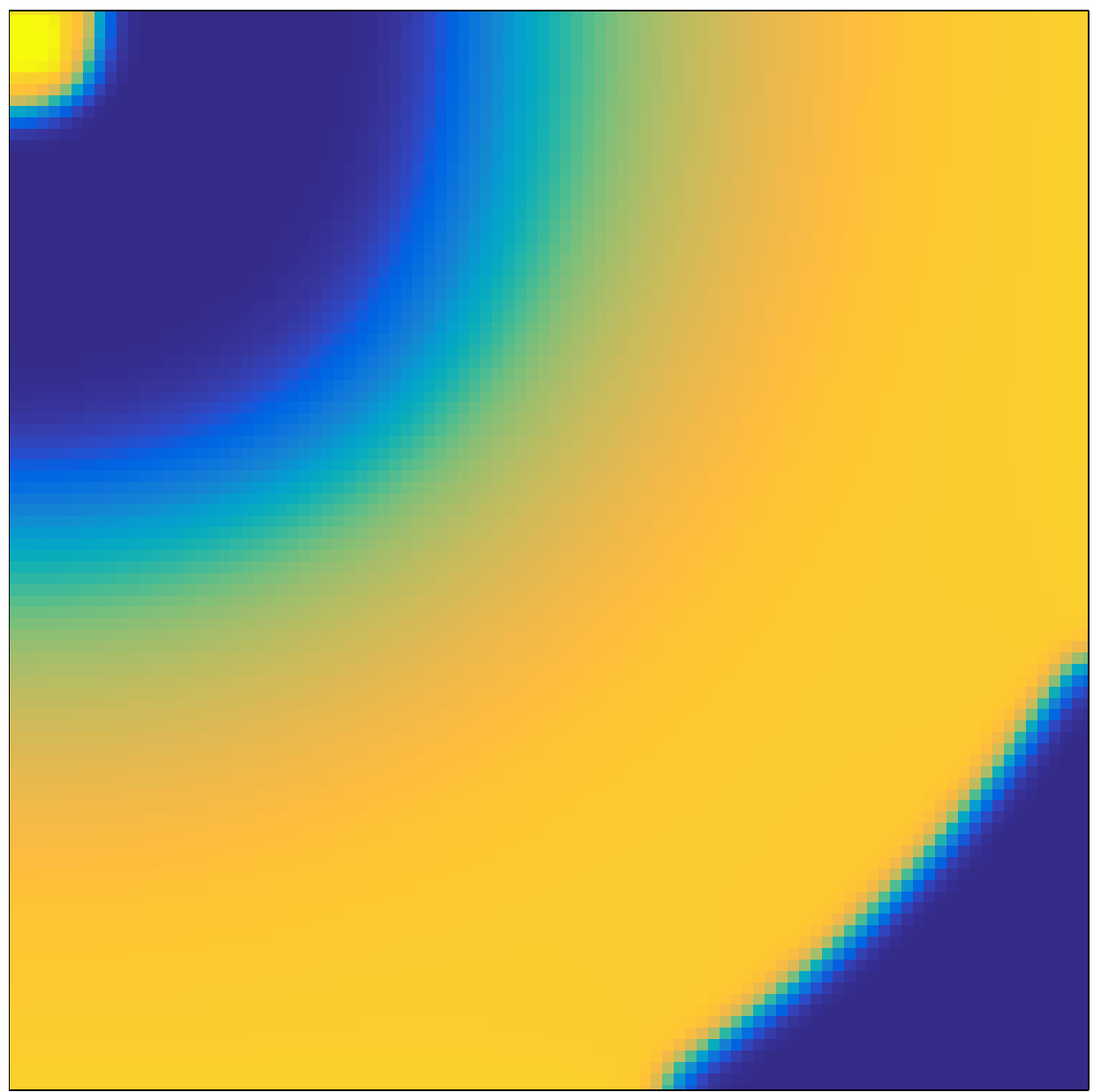}
\includegraphics[width=1.1in]{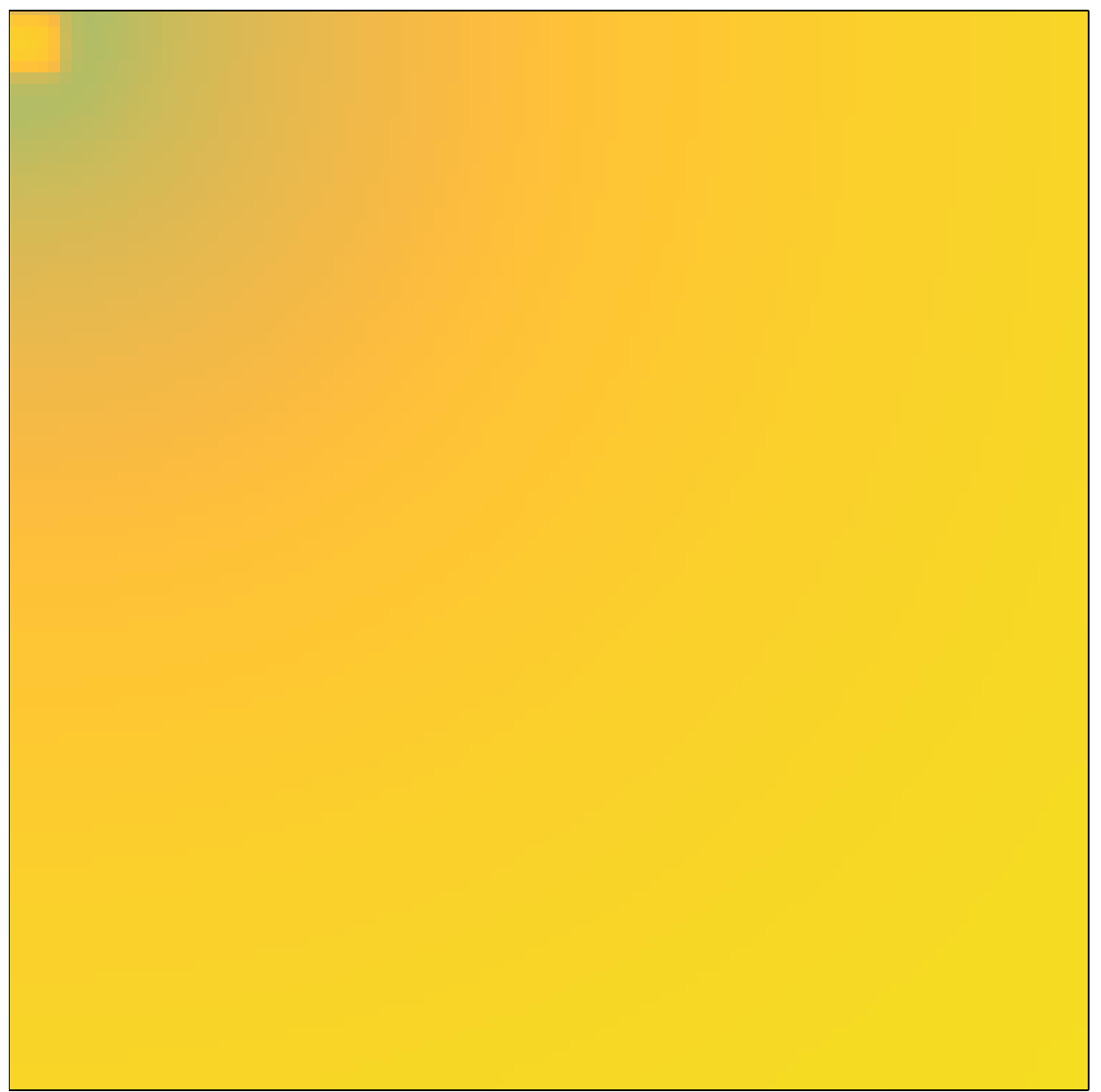}\vspace{1mm}\\
\includegraphics[width=1.1in]{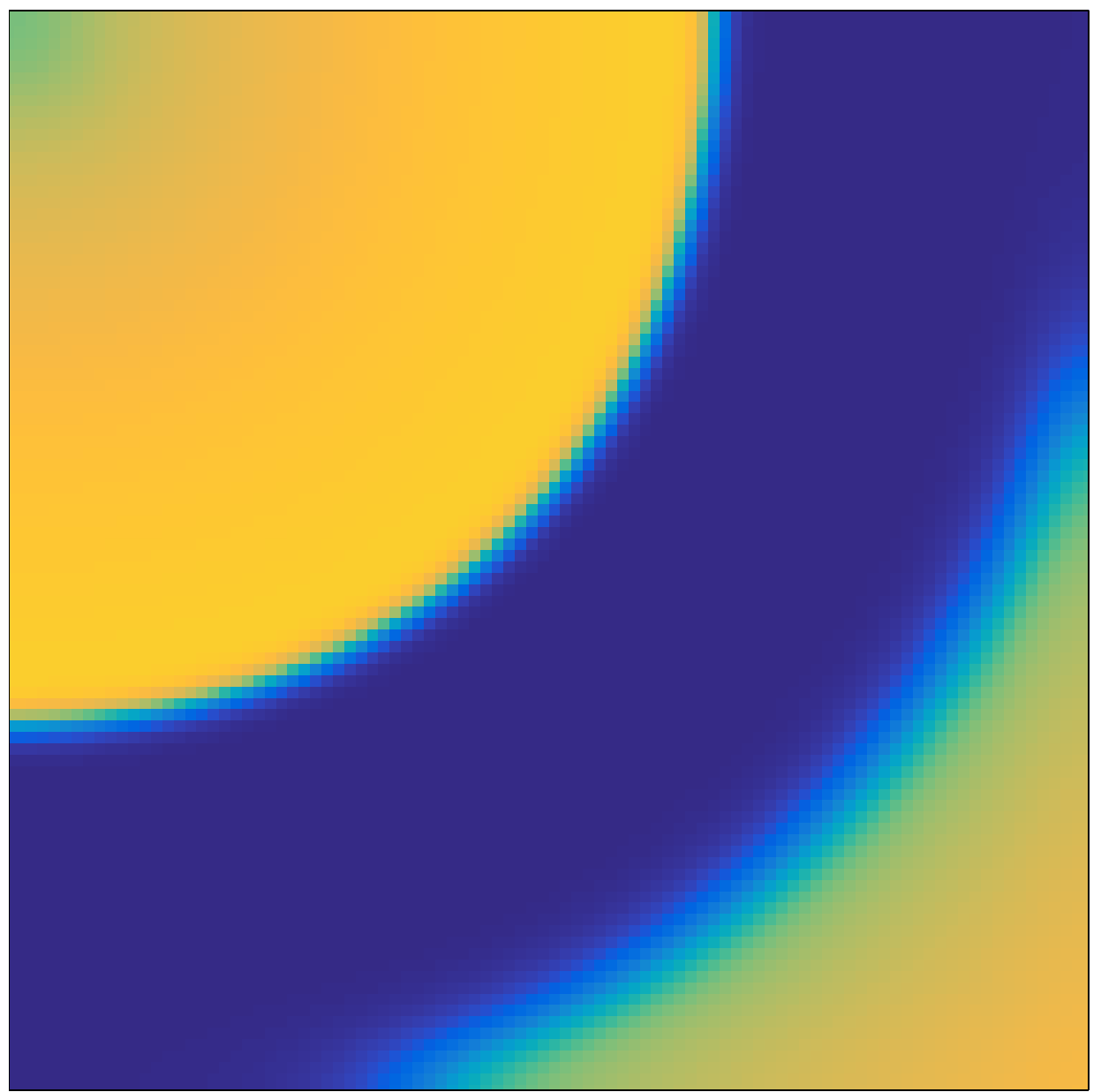}
\includegraphics[width=1.1in]{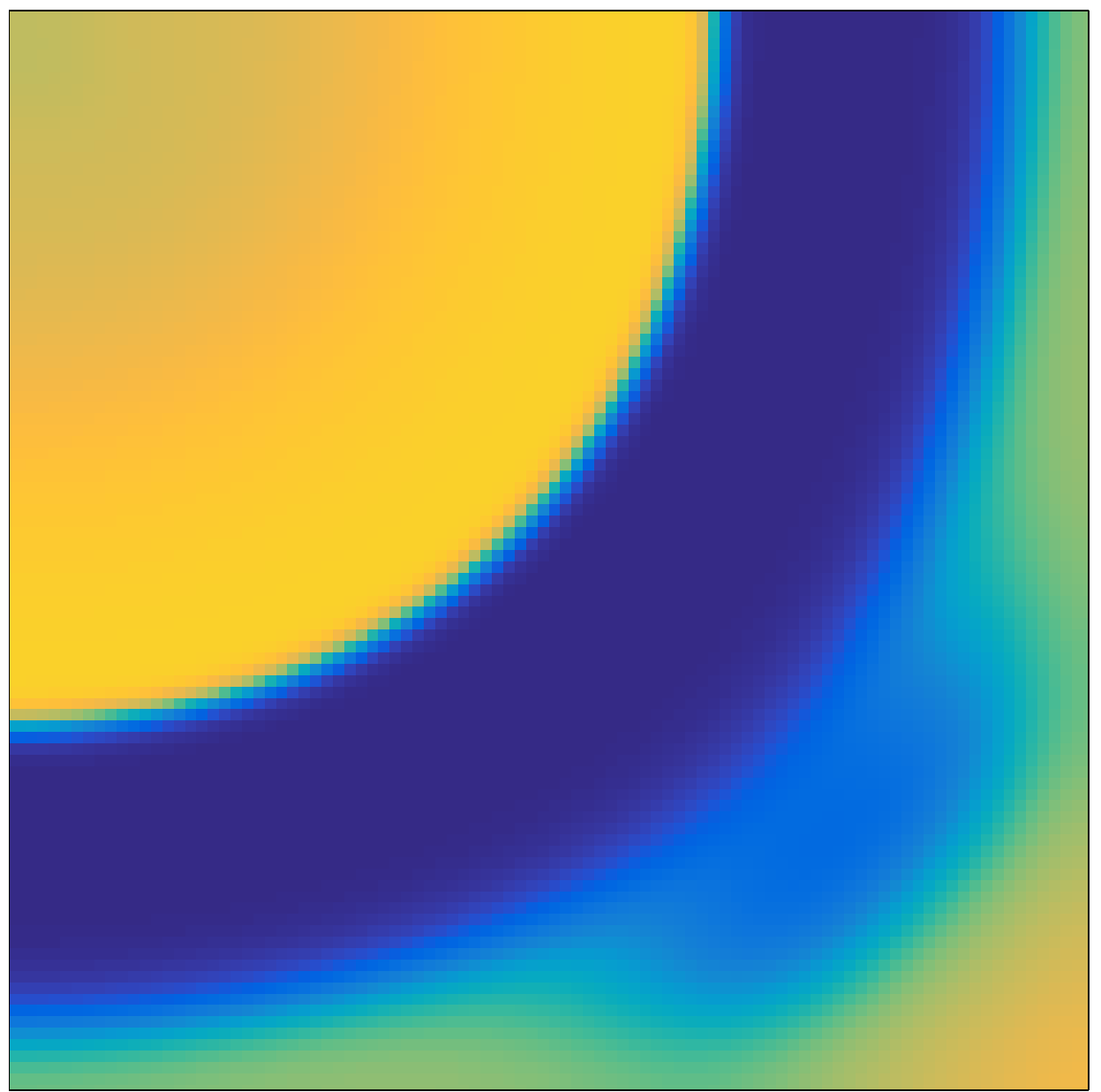}
\includegraphics[width=1.1in]{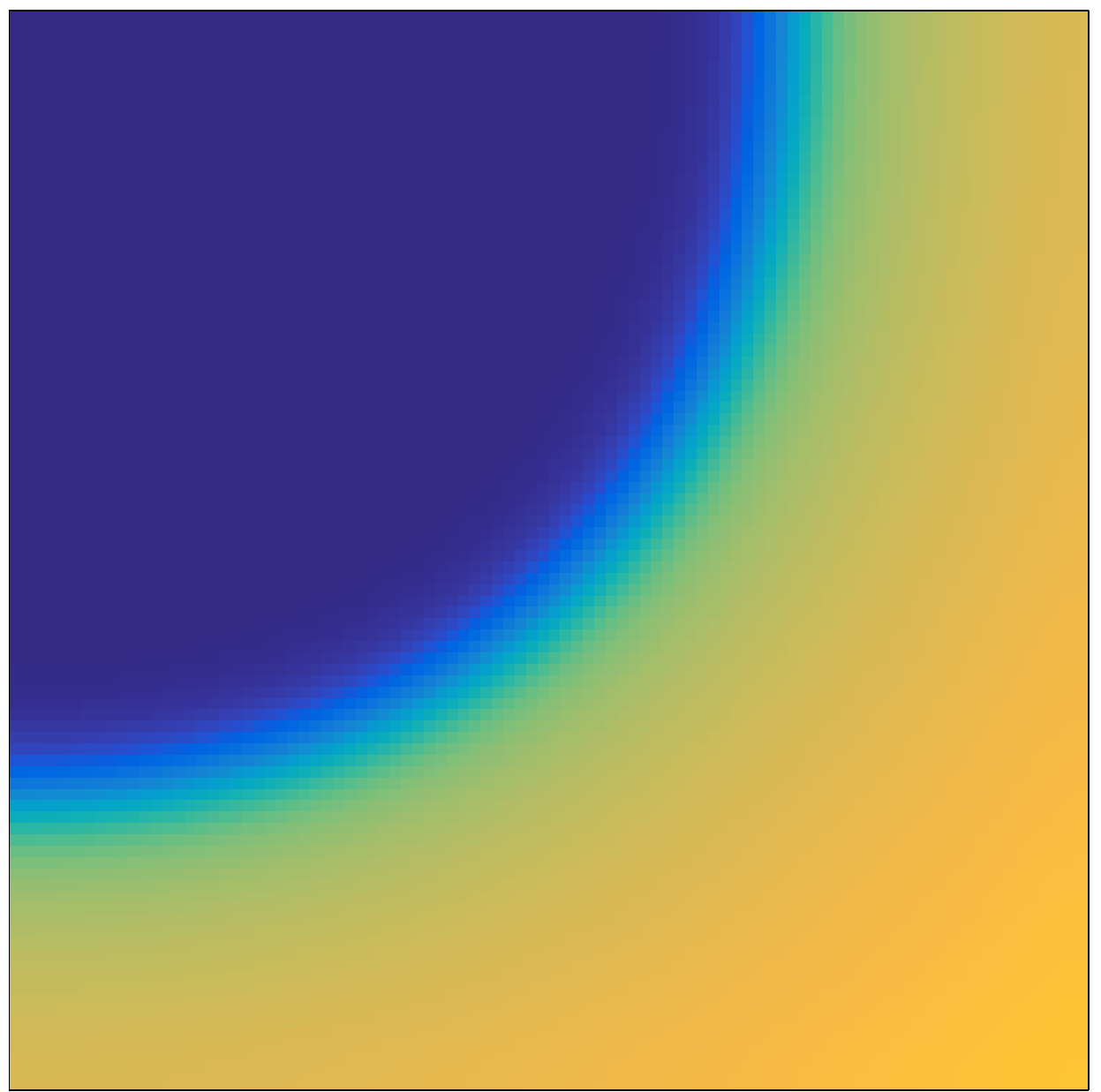}
\caption{\label{fig:rhythms}
Voltage component of the periodic solutions representing the 1:1 response (left column), 2:2 response (middle column), and 2:1 response (right column). These are shown at times $t=(n+0.05)T, \,(n+0.6)T,\, (n+1.05)T,$ and $(n+1.6)T$, where $n=1,3,5,...$ (from top to bottom row, respectively).}
\end{figure}

Let us start the analysis of the problem with a description of the various solutions to \eqref{eq:p_tu}.
To explicitly express the dependence of a solution ${\bf u}({\bf r},t)$ of (\ref{eq:p_tu}) on the initial condition ${\bf u}_0({\bf r})$, we define the time evolution operator
\begin{equation}\label{eq:Phi}
{\bf u}({\bf r},t) = \Phi^t[{\bf u}_0({\bf r})].
\end{equation}
For constant pacing, $T_{[1,\infty]}=T$, various time-periodic solutions can be defined as fixed points or $k$-cycles of the Poincar\'e map ${\bf u}^n\equiv{\bf u}({\bf r},nT)=\Phi^T[{\bf u}^{n-1}]$, which describes the discrete-time dynamics on the Poincar\'e section $t=nT$.
States ${\bf u}_k({\bf r},t)$ with different periods corresponds to solutions of the nonlinear equation
\begin{equation}
\label{eq:Pumu}
\Phi^{kT}[{\bf u}]={\bf u}
\end{equation}
with different $k$. In particular, the 1:1 response ${\bf u}_1({\bf r},t)$ corresponds to the fixed point of the Poincar\'e map ${\bf u}_{11}=\Phi^T[{\bf u}_{11}]={\bf u}_1({\bf r},0)$.
At faster pacing rates, two different solutions of \eqref{eq:Pumu} with $k=2$ appear. 
The 2-cycle $\{{\bf u}_{21},{\bf u}_{22}\}$, where ${\bf u}_{22}=\Phi^T[{\bf u}_{21}]={\bf u}_2({\bf r},0)$ and ${\bf u}_{21}=\Phi^T[{\bf u}_{22}]={\bf u}_2({\bf r},T)$, corresponds to the alternans solution ${\bf u}_2({\bf r},t)$ (2:2 response). 
A different 2-cycle $\{{\bf u}'_{21},{\bf u}'_{22}\}$, where ${\bf u}'_{22}=\Phi^T[{\bf u}'_{21}]={\bf u}'_2({\bf r},0)$ and ${\bf u}'_{21}=\Phi^T[{\bf u}'_{22}]={\bf u}'_2({\bf r},T)$, corresponds to the conduction block solution ${\bf u}'_2({\bf r},t)$ (2:1 response).
These three periodic solutions are compared in Fig. \ref{fig:rhythms}.

Stability of time-periodic solutions can be described in terms of the dynamics of deviations $\delta {\bf u}({\bf r},t)={\bf u}({\bf r},t)-{\bf u}_k({\bf r},t)$ from the reference solution ${\bf u}_k$. The dynamics of finite disturbances is described by
\begin{equation}
\label{eq:p_tdu}
\partial_t \delta{\bf u} = D\nabla^2 \delta{\bf u} + {\bf f}({\bf u}_k+\delta{\bf u})-{\bf f}({\bf u}_k).
\end{equation}
For infinitesimal disturbances, \eqref{eq:p_tdu} reduces to a linear PDE
\begin{equation}
\label{eq:p_tduL}
\partial_t \delta{\bf u} = \mathcal{L}\delta{\bf u},
\end{equation}
where $\mathcal{L}=D\nabla^2 + J$ and $J(t) = \partial_{\bf u}{\bf f}|_{{\bf u}_k(t)}$ is the time-dependent Jacobian of the nonlinear term describing cellular dynamics evaluated on the periodic orbit.

The linear time-evolution operator that advances the solution of (\ref{eq:p_tduL}) from $0$ to $t$ in the tangent space,
\begin{equation}
\label{eq:uUk}
\delta {\bf u}({\bf r},t) = U(t,0) \delta {\bf u}({\bf r},0)
\end{equation}
can be found by integrating \eqref{eq:p_tduL}, which yields a time-ordered exponential
\begin{equation}
\label{eq:Uk}
U(t,0)=\exp\left\{\int_0^t\mathcal{L}_k(t')dt'\right\}.
\end{equation}
Linear stability of ${\bf u}_k({\bf r},t)$ is determined by the eigenvalues (Floquet multipliers) $\lambda_{ki}$ of the operator $U(kT,0)$,
\begin{equation}
U(kT,0){\bf e}_{ki}({\bf r}) = \lambda_{ki} {\bf e}_{ki}({\bf r}),
\end{equation}
where ${\bf e}_{ki}({\bf r})$ are the corresponding eigenfunctions (Floquet modes).
In particular, if $|\lambda_{ki} | < 1$ for all $i$, any infinitesimal disturbance $\delta {\bf u}({\bf r},t)$ will eventually decay to zero, so that ${\bf u}({\bf r},t)\to{\bf u}_k({\bf r},t)$ for $t\to\infty$.
In the following discussion, we will index the eigenvalues in the order of decreasing absolute value, $|\lambda_{k1}| \ge |\lambda_{k2}| \ge |\lambda_{k3}| \ge \cdots$.

Asymptotic stability, however, does not imply that initial disturbances will decay monotonically. If the evolution operator $U(t,0)$ is nonnormal (i.e., $UU^\dagger\ne U^\dagger U$), some disturbances may grow transiently, before asymptotic decay takes over \cite{Trefethen1993}. As we have shown previously \cite{Garzon11,Garzon14} in the context of one-dimensional paced cardiac tissue (Purkinje fibers), such transient growth may arise, for example, due to the action of closed-loop feedback control, which makes the 1:1 solution linearly stable. In the present case, the Jacobian $J(t)$ is non-self-adjoint, so both the differential operator $\mathcal{L}(t)$ and the evolution operator $U(t,0)$ are nonnormal, hence, generalized linear stability theory \cite{Farrell1996,Farrell1996b} is required to describe transient response to perturbations associated with changes in the pacing rate.

The adjoint of the evolution operator
\begin{equation}
\label{eq:UkA}
U^\dagger(t,0)=\exp\left\{\int_0^t\mathcal{L}^\dagger(t-t')dt'\right\}
\end{equation}
 plays an important role in the generalized stability theory. It defines the evolution of disturbances $\delta{\bf v}$ in the adjoint of the tangent space,
\begin{equation}
\delta {\bf v}({\bf r},0) = U^\dagger(t,0)\delta {\bf v}({\bf r},t),
\end{equation}
backward in time. In practice it is more straightforward to compute the action of $U^\dagger(t,0)$ on a vector by solving the linear PDE
\begin{equation}
\label{eq:p_tduLa}
-\partial_t \delta{\bf v} = \mathcal{L}^\dagger \delta{\bf v}
\end{equation}
with ``initial'' condition defined at the final time.

Given that time-periodic states ${\bf u}_k({\bf r},t)$ could be stable or unstable, we computed them as solutions ${\bf u}_k$ of (\ref{eq:Pumu}) using a Newton-Krylov solver \cite{Garzon11,Marcotte2015}. 
The spectrum of the evolution operator $U(t,0)$ was computed using the Arnoldi method implemented by the Matlab function {\it eigs} \cite{lehoucq1996} with accuracy yielding eight significant digits.
In both instances matrix-free evaluation of the the matrix-vector product $U(t,0)\,\delta {\bf u}$ was accomplished via simultaneous numerical time-integration of \eqref{eq:p_tu} and \eqref{eq:p_tduL}. 
The matrix-free calculation of the product $U(t,0)^\dagger\delta {\bf v}$ involves backward integration and cannot be accomplished by integrating \eqref{eq:p_tu} and \eqref{eq:p_tduLa} simultaneously. 
Instead a precomputed and interpolated solution ${\bf u}_k({\bf r},t)$ was used to compute the solution to \eqref{eq:p_tduLa} using the procedure described in Ref. \onlinecite{Marcotte2016}.
The time integrators were implemented as OpenCL kernels \cite{Marcotte:2012} administered by host codes written in C and encapsulated in Matlab mex-files. This allowed substantially speeding up the calculations by executing them on general purpose Graphics Processing Units (GPUs), at the same time retaining convenient interface within Matlab scripts.

\begin{figure}
(a)\hspace{2mm}\includegraphics[width=2.5in]{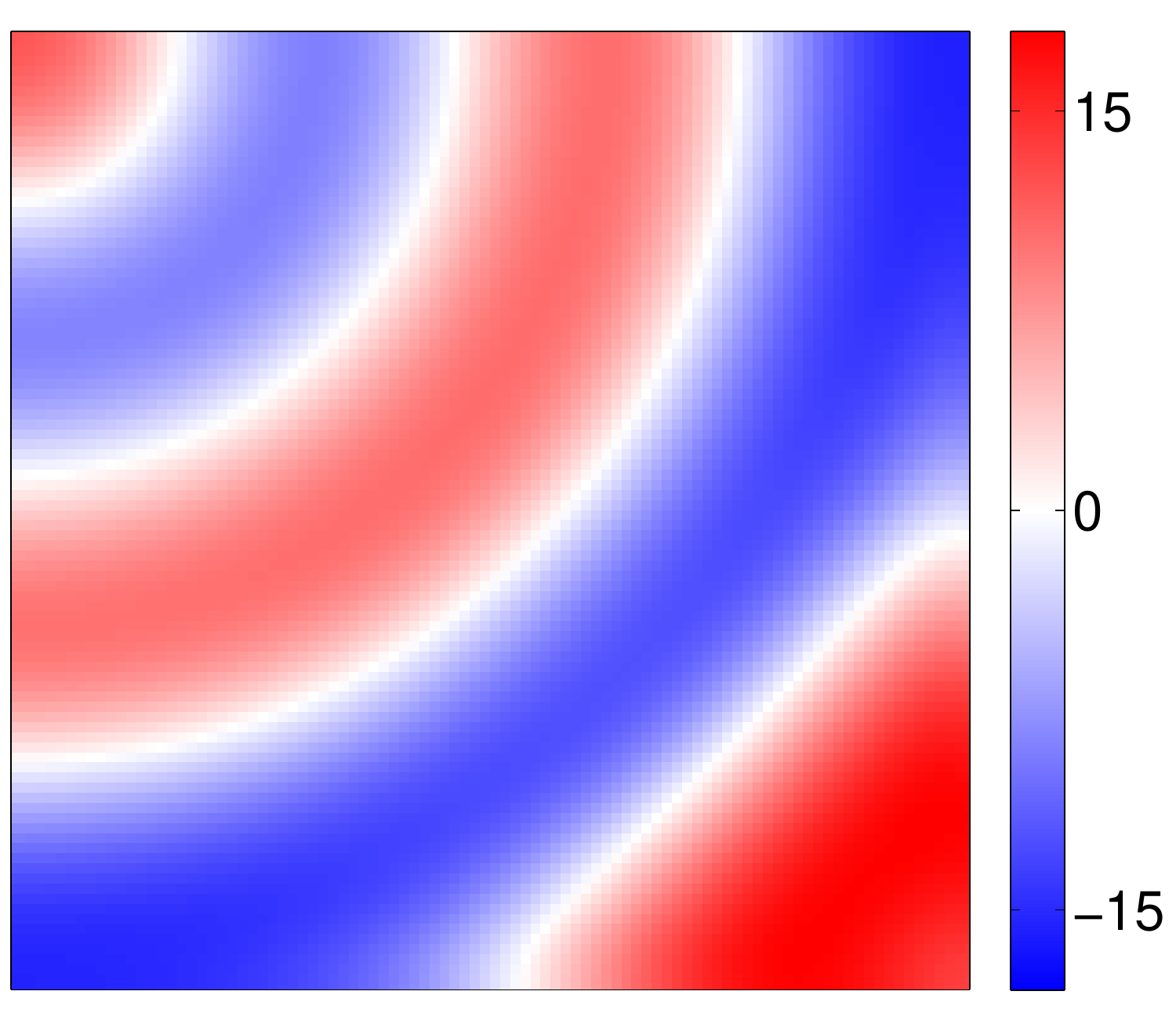}\vspace{2mm}
(b)\includegraphics[width=2.8in]{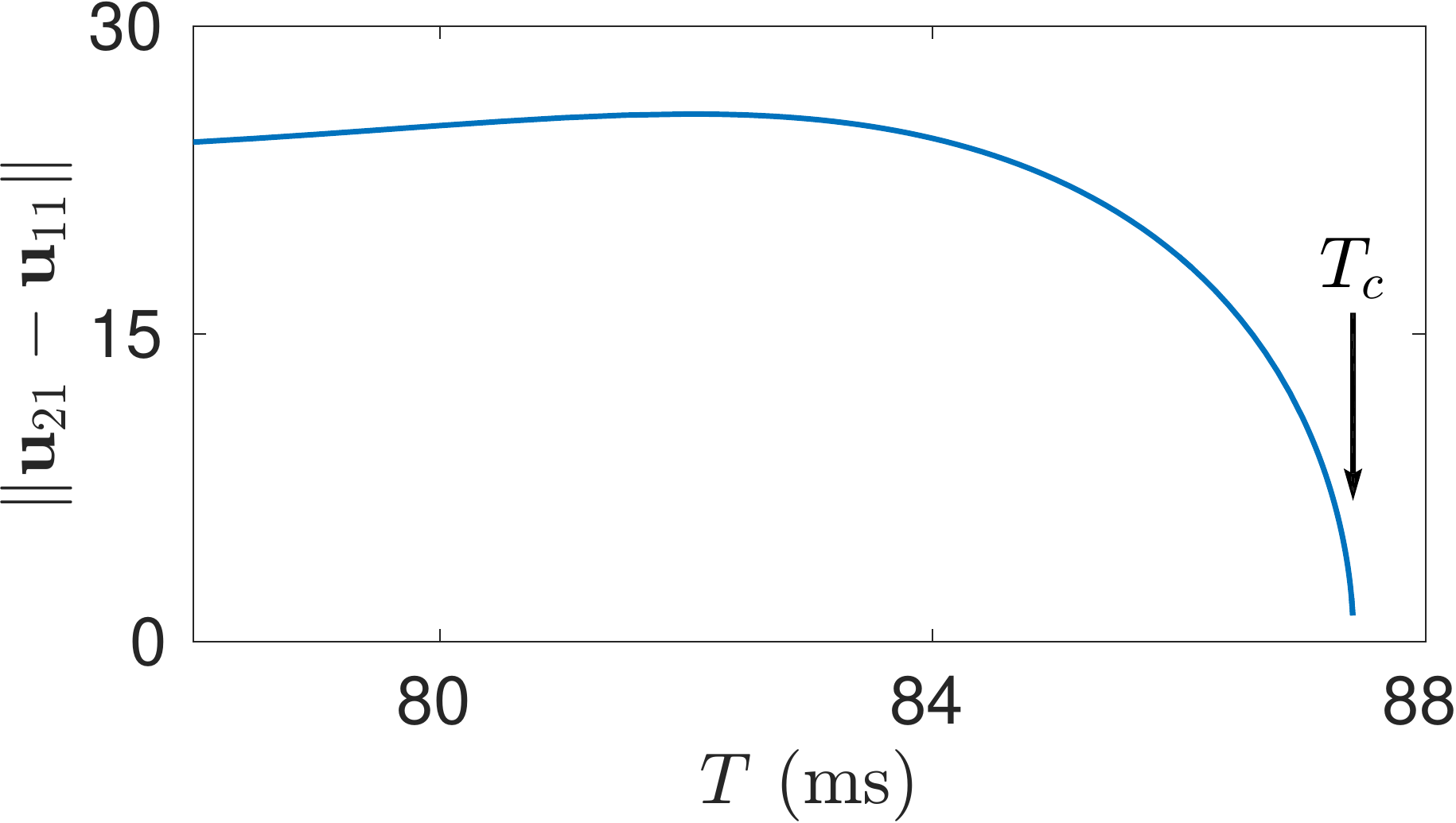}
\caption{\label{fig:alt_85} The 2:2 solution which corresponds to discordant alternans. (a) The difference (in ms) between two successive APDs at $T=80$ ms. (b) The distance between the 1:1 and 2:2 solutions in the state space. The arrow marks the critical period $T_c$.}
\end{figure}

\subsection{Stability spectra}

%
The 1:1 solution ${\bf u}_1({\bf r},t)$ exists at all $T$ considered in this study. 
It is stable for $T>T_c \approx 87.4$ ms and becomes unstable for $T<T_c$.
The leading eigenvalue crosses the unit circle along the negative real line, which corresponds to a period doubling bifurcation, at which point the alternans solution ${\bf u}_2({\bf r},t)$ with period $2T$ is created. It corresponds to discordant alternans with four stationary nodal lines, as Fig. \ref{fig:alt_85}(a) illustrates. 
The alternans is created discordant, since the domain is  sufficiently large, according to theoretical analysis \cite{Echebarria2002,echebarria2007}.
In experiments \cite{Pastore1999,walker2003} and simulations  \cite{Qu2000a,Watanabe2001,Gizzi2013} alternans is typically created concordant and becomes discordant as the pacing interval is decreased.
However, this subtle distinction has no bearing on the development of conduction block that plays the crucial role in transition to fibrillation. 

To further characterize this bifurcation, we computed the distance $\|{\bf u}_{21}-{\bf u}_{11}\|$ between the 1:1 and 2:2 solutions in the state space, defined using the Euclidean 2-norm
\begin{equation}\label{eq:2norm}
\|{\bf u}({\bf r})\| = \langle {\bf u}({\bf r}), {\bf u}({\bf r})\rangle^{1/2},
\end{equation}
where the scalar product is given by
\begin{equation}\label{eq:inprod}
\langle {\bf v}, {\bf u}\rangle = \int {\bf v}^*({\bf r}) \cdot {\bf u}({\bf r})\,dx\, dy,
\end{equation}
and the integral is over the entire computational domain $0< x,y < L$.
The bifurcation is supercritical (cf. Fig. \ref{fig:alt_85}(b)), also in agreement with the theoretical predictions\cite{Echebarria2002}, and the 2:2 solution exists for all $T<T_c$ (at least down to $T=50$ ms). 


\begin{figure}
\includegraphics[width=2.8in]{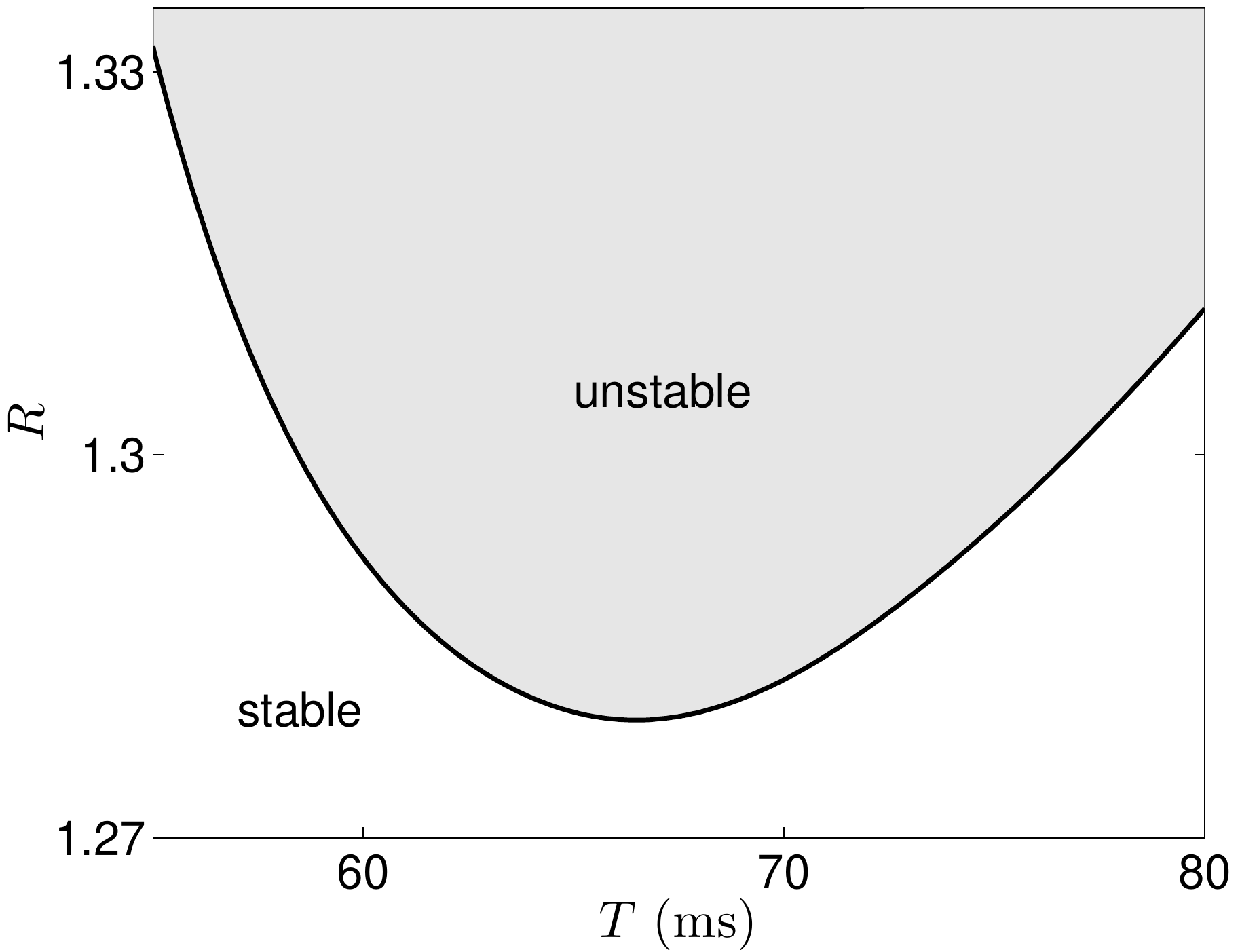}
\caption{\label{fig:T_R_dmax} Regions in the $(T,R)$ parameter plane showing where the 2:2 state is stable and where it is not. The solid line is the neutral stability curve which corresponds to $|\lambda_{21}|=1$.}
\end{figure}

\begin{figure}
\subfigure[]{\includegraphics[width=2.3in]{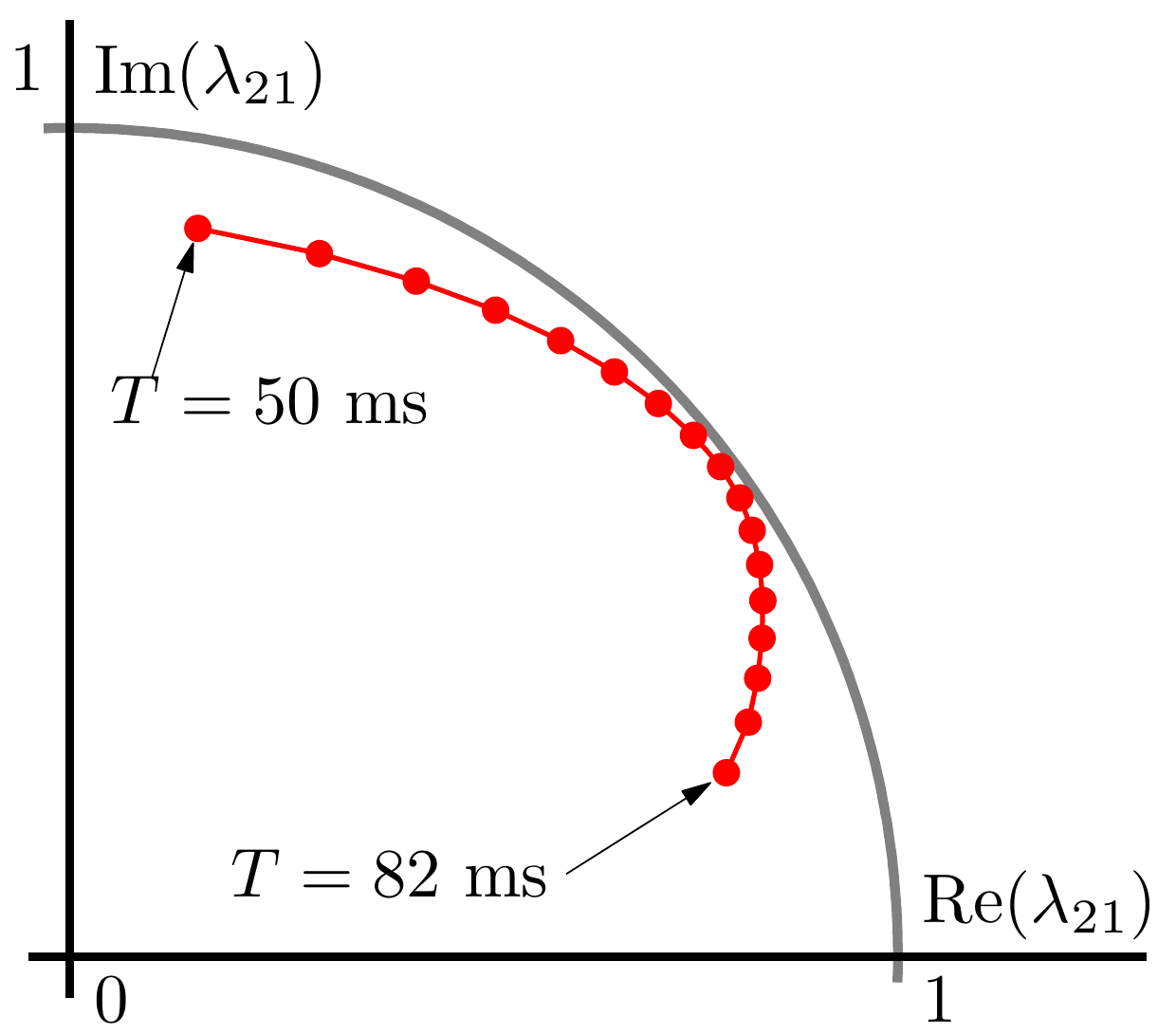}}\\
\subfigure[]{\includegraphics[height=1.3in]{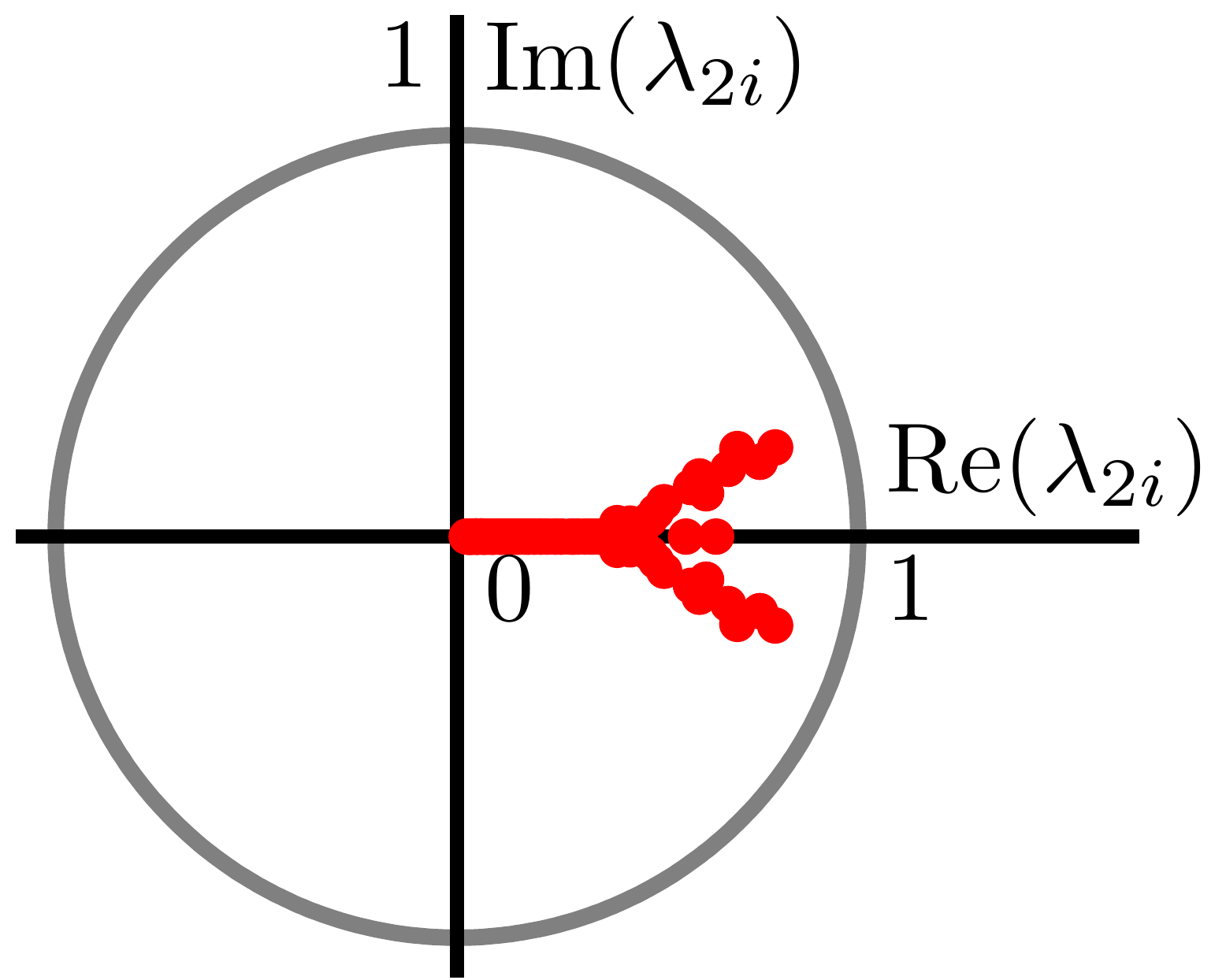}}
\subfigure[]{\includegraphics[height=1.3in]{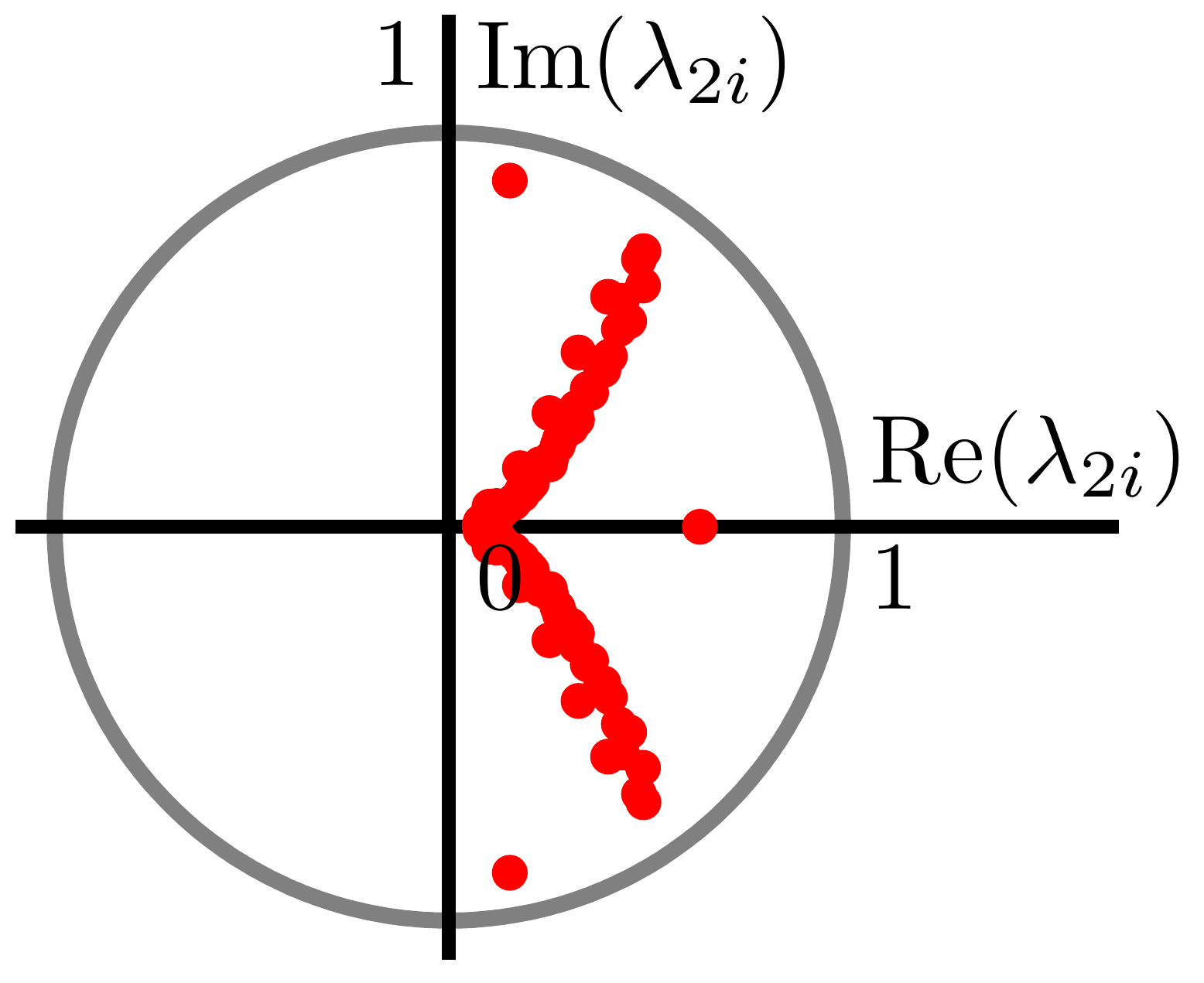}}
\caption{\label{fig:sp_2T} (a) Leading eigenvalue $\lambda_{21}$ shown for 50 ms $\le T \le$ 82 ms in steps of 2 ms. The spectrum $\{\lambda_{2i}\}, i=1,2,...,100$ for $T=82$ ms (b) and $T=50$ ms  (c).}
\end{figure}

The stability balloon for the 2:2 solution in the $(T,R)$ plane is shown in Fig. \ref{fig:T_R_dmax}. 
This solution undergoes a subcritical Hopf bifurcation when the neutral stability curve (shown as a solid line) is crossed.
The minimum of the neutral stability curve is at $R_c=1.279$ and $T=66.5$ ms. 
The value $R=1.273$ considered here is {\it below} $R_c$, so the 2:2 solution turns out to be linearly stable for all $T$. 
The corresponding Floquet spectra for different values of $T$ are shown in Fig. \ref{fig:sp_2T}. 
The change in the leading eigenvalue $\lambda_{21}$ associated with a decrease in $T$ is shown in Fig. \ref{fig:sp_2T}(a). 
Although this eigenvalue never leaves the unit circle, it approaches this circle for $T\approx 66.5$ ms.

Above the neutral stability curve infinitesimally small perturbations of the 2:2 state cause a direct transition to SWC. 
This chaotic state exists over a region in the plane that is bigger than the gray-shaded area in Fig. \ref{fig:T_R_dmax}, so our system indeed features both a subcritical instability and multistability -- the hallmarks of a bypass transition. 
Of particular relevance for this study, persistent SWC exists for $R=1.273$ and $T=80$ ms.
In addition, there is one more stable state that exists for the same parameter values, the one that describes a 2:1 response. 
For $R=1.273$, it is created in a saddle-node bifurcation at $T\approx 116.2$ ms and remains stable as $T$ decreases to at least 50 ms (that is, over the entire range of $T$ in Fig. \ref{fig:T_R_dmax}).

\begin{figure}
\includegraphics[trim=0 0 -49 0,clip,width=1.65in]{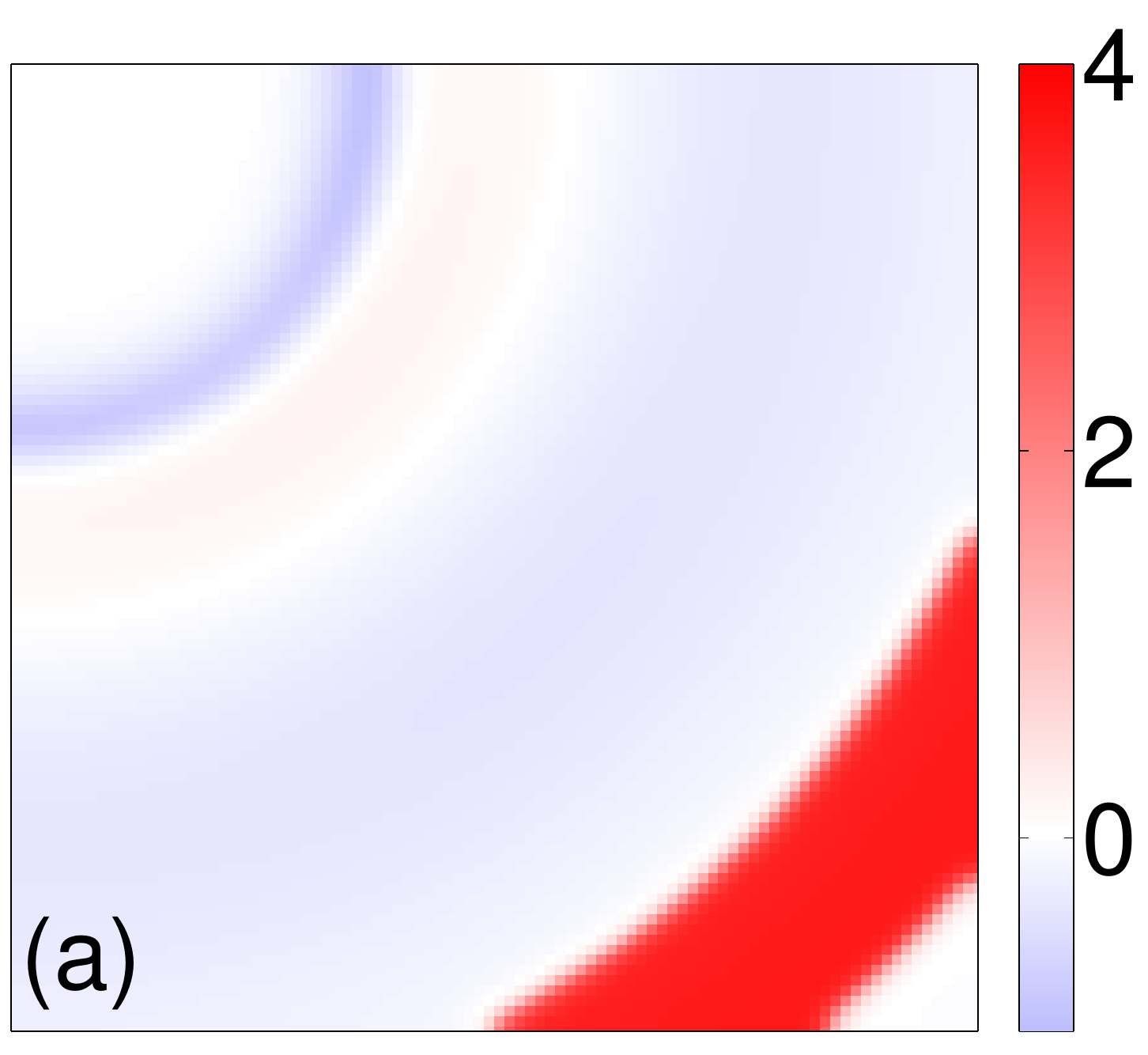} 
\includegraphics[width=1.65in]{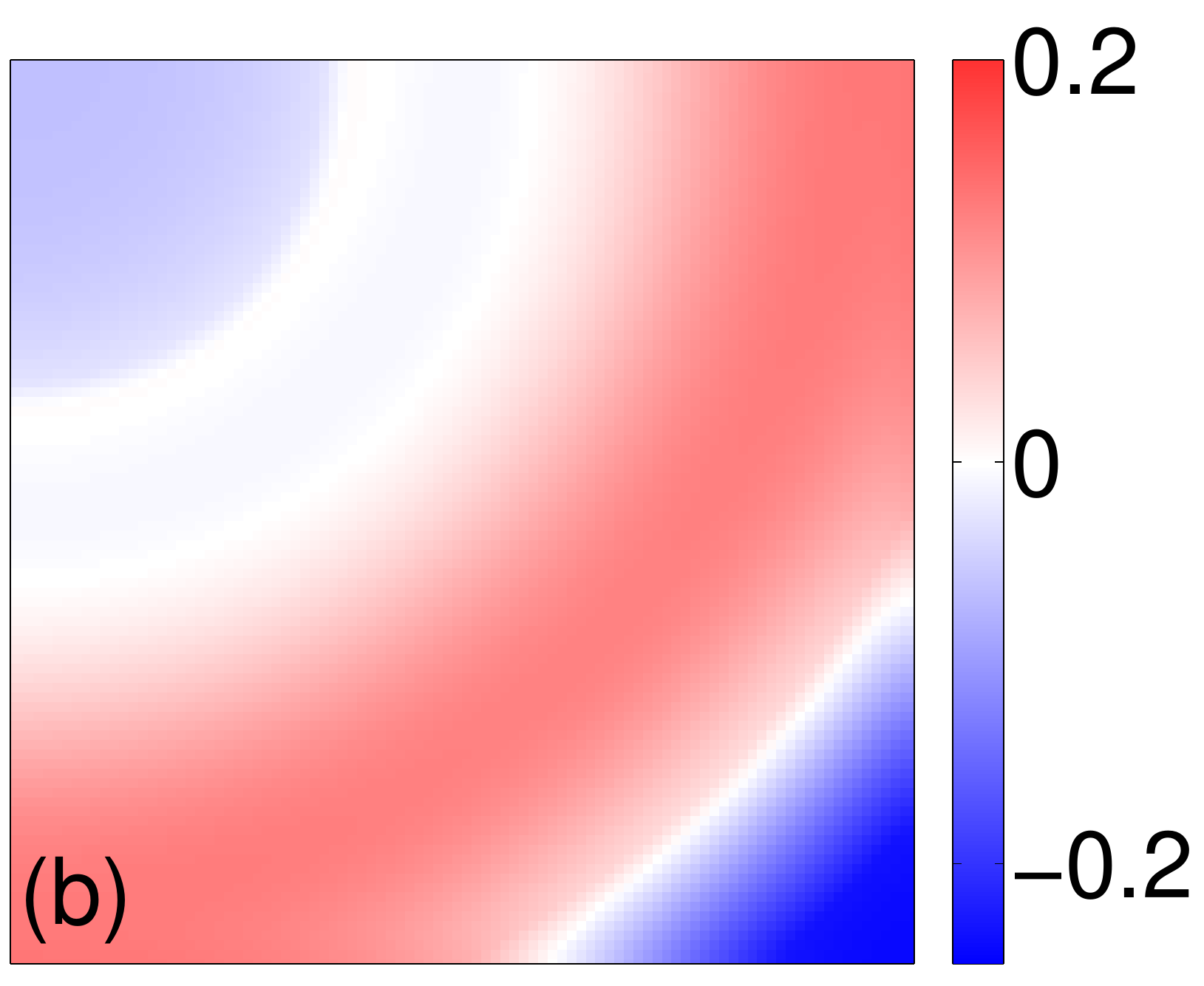} \\
\vspace{2mm}
\includegraphics[trim=0 0 -29 0,clip,width=1.65in]{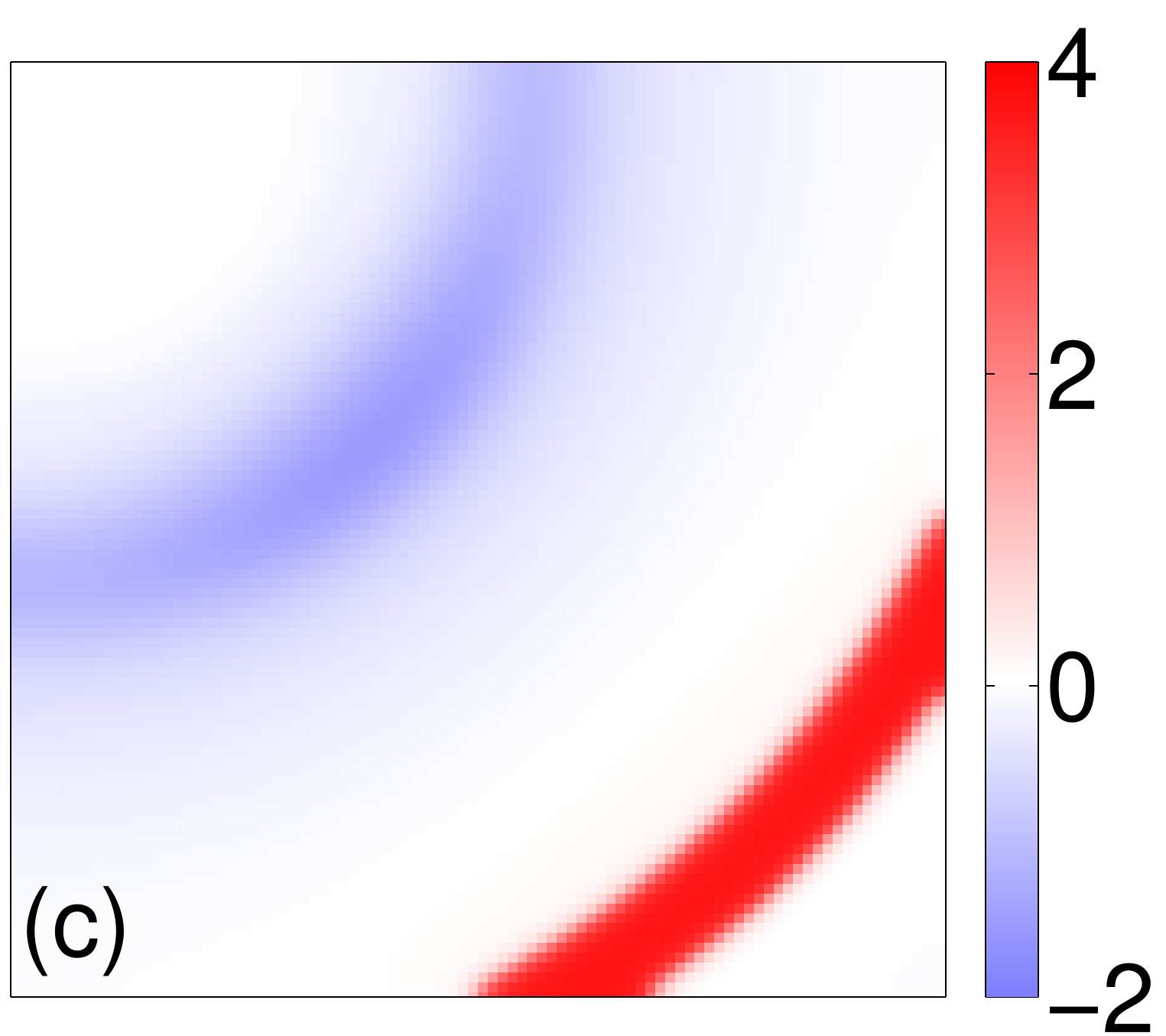} 
\includegraphics[trim=0 -5 0 0,clip,width=1.65in]{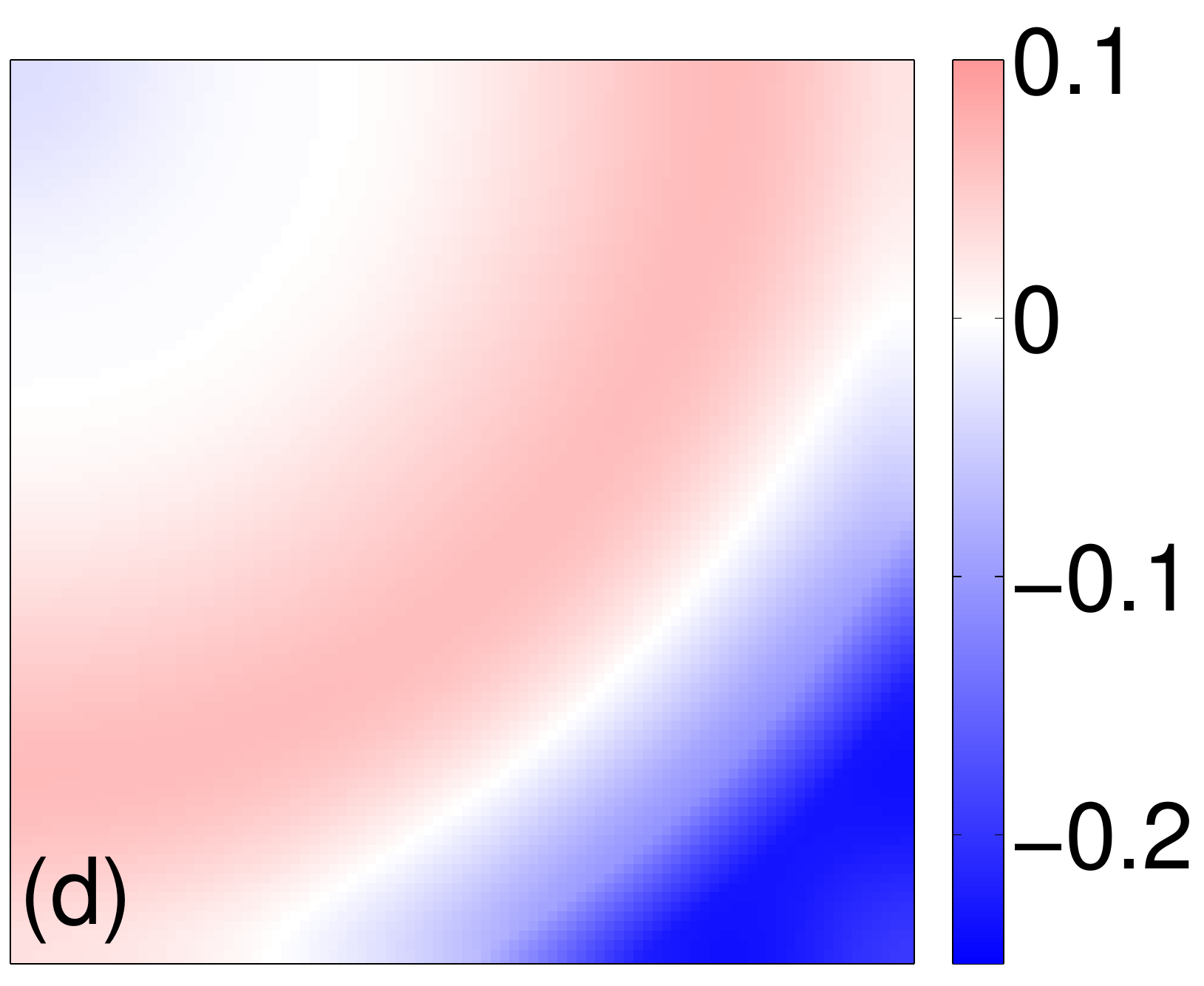}
\caption{\label{fig:protAB_IC}
Comparison of the initial perturbations $\delta{\bf u}_A^0({\bf r})={\bf u}_A^0-{\bf u}_{21}$ and $\delta{\bf u}_B^0({\bf r})={\bf u}_B^0-{\bf u}_{21}$, for the second stage of protocols A and B, respectively. (a) and (b) show $u$- and $v$-components of $\delta{\bf u}_A^0$. (c) and (d) show $u$- and $v$-components of $\delta{\bf u}_B^0$. 
}
\end{figure}

Given this information, how can we understand the difference in the outcomes of the pacing protocols A and B?
Both protocols can be viewed as being composed of two stages: the first stage (first 40/50 pacing intervals for protocol A/B) determines the initial condition for the second stage (pacing with constant $T=80$ ms that starts, respectively at $t_{40}$ or $t_{50}$). 
The perturbation $\delta{\bf u}_B^0={\bf u}_B({\bf r},t_{50})-{\bf u}_{21}$ (cf. Fig. \ref{fig:protAB_IC}(c-d)) in protocol B eventually decays, in agreement with the prediction of linear stability analysis, while the perturbation $\delta{\bf u}_A^0={\bf u}_A({\bf r},t_{40})-{\bf u}_{21}$ (cf. Fig. \ref{fig:protAB_IC}(a-b)) in protocol A grows, producing conduction block and wave breakup after 15 pacing intervals and, eventually, transition to persistent SWC.
(Note that in the 2:1 state conduction block is global, annihilating an entire excitation wave, while in protocol A conduction block is local and leads to a breakup, but not a total annihilation, of an excitation wave.)

Since three different stable solutions (2:2, 2:1, and SWC) coexist at $T=80$ ms, the asymptotic regime is determined by the initial condition. In particular, there is a neighborhood $\Omega_{2:2}=\Omega_{21}\cup\Omega_{22}$ of the 2-cycle 
$\{{\bf u}_{21},{\bf u}_{22}\}$, such that for all initial conditions ${\bf u}^0\in\Omega_{2:2}$, the orbit ${\bf u}^n$ approaches this 2-cycle as $n \to \infty$. 
$\Omega_{2:2}$ is known as the basin of attraction of the 2-cycle and has a finite size. 
Similarly, the 2-cycle $\{{\bf u}'_{21},{\bf u}'_{22}\}$ has a basin of attraction $\Omega_{2:1}=\Omega'_{21}\cup\Omega'_{22}$.
In particular, the initial condition ${\bf u}_B^0\equiv{\bf u}_B({\bf r},t_{50})$ in protocol B lies inside $\Omega_{2:2}$, so the asymptotic state corresponds to the 2:2 solution.
On the other hand, the initial condition ${\bf u}_A^0\equiv{\bf u}_A({\bf r},t_{40})$ in protocol A lies outside of both $\Omega_{2:2}$ and $\Omega_{2:1}$, so the asymptotic state corresponds to SWC.

\begin{figure}
\includegraphics[width=\columnwidth]{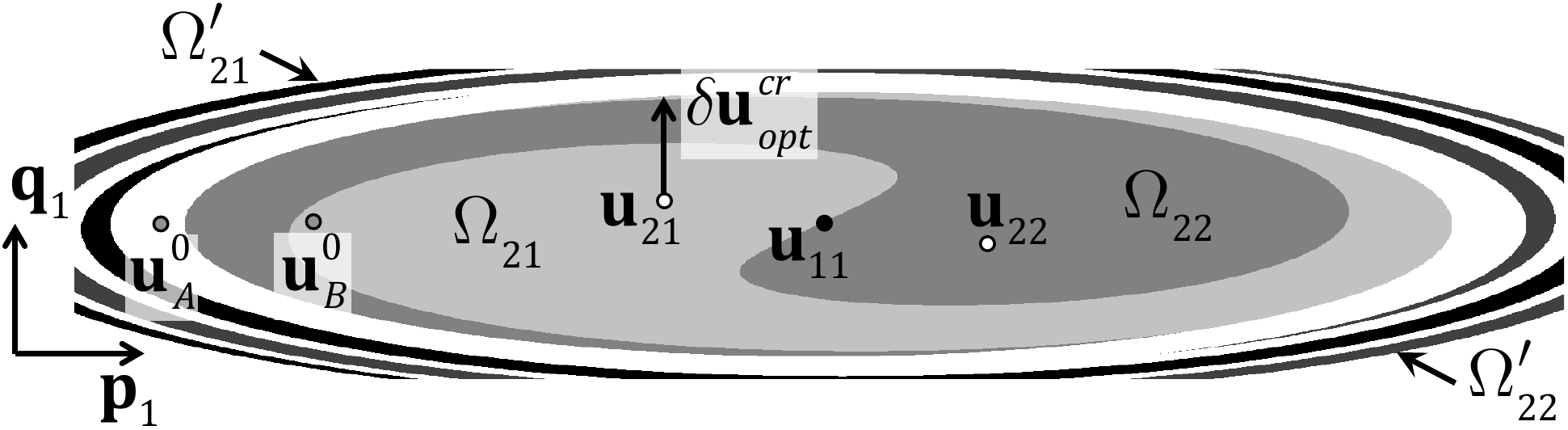}
\caption{\label{fig:basin} Schematic representation of the basins of attraction of the stable 2T-periodic solutions in a two-dimensional projection of the state space. Gray dots represent the initial conditions for the second stage of protocols A and B. The vector $\delta{\bf u}_{opt}^{cr}$ shows the critical linearly optimal perturbation. The white region corresponds to the basin of attraction $\Omega_{swc}$ of the SWC state.}
\end{figure}

Since the evolution operator describing the linearized dynamics is nonnormal, the basin of attraction $\Omega_{2:2}$ becomes quite thin in some directions. Fig. \ref{fig:basin} shows a two-dimensional cartoon of this; the actual dimensionality of $\Omega_{2:2}$ is the same as that of the state space of the discretized PDE, 18432 in the present case. 
We are primarily interested in the direction in the state space (which corresponds to perturbation shape in the physical space) for which the distance $\|\delta{\bf u}\|$ from ${\bf u}_{21}$ to the boundary $\partial\Omega_{2:2}$ of the basin of attraction is the smallest (we will refer to it as the optimal direction), since the dynamics are the most sensitive to perturbations corresponding to this direction. 
The point on $\partial\Omega_{2:2}$ that is the closest to the 2-cycle defines the critical optimal perturbation $\delta{\bf u}_{opt}^{cr}$. 
Although the shape of $\Omega_{2:2}$, and hence the perturbation $\delta{\bf u}_{opt}^{cr}$, is defined by the nonlinear evolution operator \eqref{eq:Phi}, as we illustrate below, both the direction of $\delta{\bf u}_{opt}^{cr}$ and the strong transient growth associated with perturbations in this direction can be understood reasonably well within the linear approximation.

\subsection{Generalized linear stability analysis}

Let us compute how strongly an initial disturbance $\delta{\bf u}({\bf r},0)$ is amplified as a result of transient growth. With the help of \eqref{eq:uUk} and \eqref{eq:2norm} we find
\begin{align}
\|\delta{\bf u}({\bf r},t)\|^2 &=
   \langle U(t,0)\delta{\bf u}({\bf r},0),
           U(t,0)\delta{\bf u}({\bf r},0)
   \rangle , \nonumber \\
  \label{eq:UdU}
  &= \langle U^\dagger(t,0)\,U(t,0)\delta{\bf u}({\bf r},0),
     \delta{\bf u}({\bf r},0)
     \rangle .
\end{align}
Since $U^\dagger(t,0)U(t,0)$ is self-adjoint, it has real eigenvalues $\sigma_i^2(t)\ge 0$ and orthogonal eigenvectors ${\bf q}_i({\bf r},t)$,
 \begin{equation}
\label{eq:UUq}
U^\dagger(t,0) U(t,0) {\bf q}_i({\bf r},t) = \sigma^2_i(t) {\bf q}_i({\bf r},t),
\end{equation}
with eigenvalues sorted from largest to smallest. Expanding the initial condition in the eigenvector basis yields
\begin{equation}
\label{eq:uxi}
\delta{\bf u}({\bf r},0)= \sum_i \eta_i(t) {\bf q}_i({\bf r},t),
\end{equation}
where $\eta_i(t) = \langle {\bf q}_i({\bf r},t), \delta {\bf u}({\bf r},0) \rangle$, such that 
\begin{equation}
\label{eq:sigxi}
\|\delta {\bf u}({\bf r},t) \|^2 =\sum_i \sigma^2_i(t)\eta_i^2(t).
\end{equation}
The maximum of $\|\delta {\bf u}({\bf r},t)\|$ over all initial disturbances with a fixed norm $\|\delta {\bf u}({\bf r},0)\|$ is reached when all $\eta_i(t)=0$, except for $i=1$, i.e., $\delta{\bf u}_{opt}({\bf r},0)\propto{\bf q}_1({\bf r},t)$.

\begin{figure}
\includegraphics[width=3in]{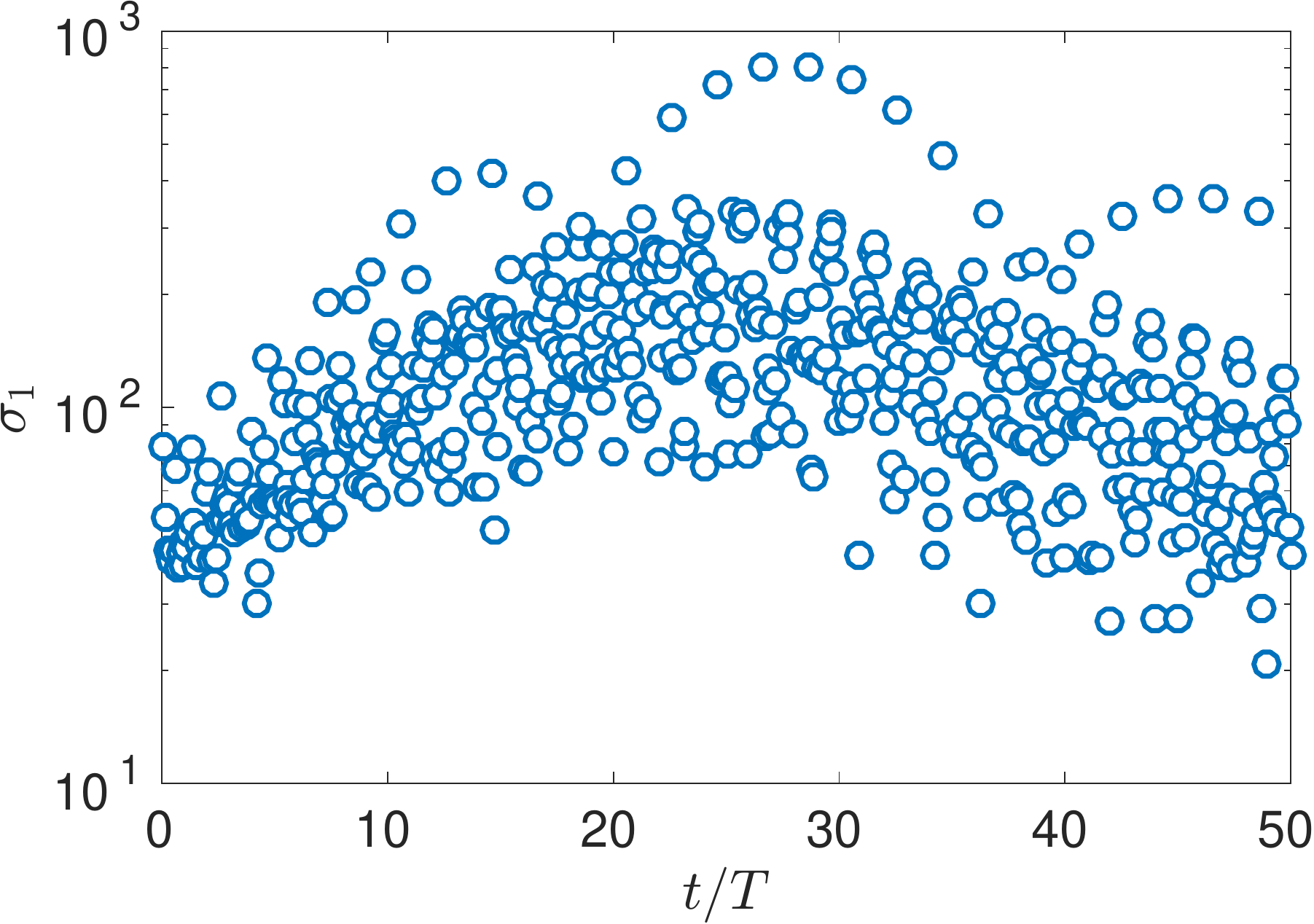}\\
\caption{\label{fig:sig_80} Transient amplification magnitude for $T=80$ ms. The singular value $\sigma_1(t)$ is shown over the discrete set of times with step $\Delta t=T/10=8$ ms.}
\end{figure}

Let $\|{\bf q}_i({\bf r},t)\|=1$ and define ${\bf p}_i({\bf r},t)  = U(t,0) {\bf q}_i({\bf r},t)$. The action of the evolution operator $U(t,0)$ can now be expressed in a simple manner. Multiplying both sides of (\ref{eq:uxi}) by $U(t,0)$ and recognizing that $\delta{\bf u}({\bf r},0)$ is arbitrary, we obtain
\begin{equation}
\label{eq:svd}
U(t,0) = \sum_i{\bf p}_i({\bf r},t)\, \sigma_i(t) \,\langle {\bf q}_i({\bf r},t), \cdot \rangle,
\end{equation}
which is the singular value decomposition (SVD) of $U(t,0)$ with $\sigma_i(t)$, ${\bf p}_i({\bf r},t)$, and ${\bf q}_i({\bf r},t)$ being, respectively, the singular values, left singular vectors, and right singular vectors of $U(t,0)$. The leading singular vectors and singular value have a useful interpretation: as we have already found, the right singular vector ${\bf q}_1({\bf r},t)$ determines the shape of the optimal initial perturbation $\delta{\bf u}_{opt}({\bf r},0)$ that is amplified the most (in the sense of the 2-norm \eqref{eq:UdU})  at time $t$. The left singular vector ${\bf p}_1({\bf r},t)$ determines the shape this optimal perturbation acquires at time $t$. Finally, the singular value determines the magnitude of transient amplification $\sigma_i(t)=\|\delta {\bf u}_{opt}({\bf r},t)\|/\|\delta{\bf u}_{opt}({\bf r},0)\|$ which corresponds to this particular initial condition.

For small initial disturbances, the time at which the maximal transient amplification $\sigma_1(t)$ is achieved is determined by the balance between transient growth and asymptotic decay. The weaker the asymptotic decay, the larger the amplification that can be achieved. Computation of $\max_t\sigma_1(t)$ is numerically rather expensive, since SVD requires alternate direction integration of the linear PDEs \eqref{eq:p_tduL} and \eqref{eq:p_tduLa} on the time interval $[0,t]$ and there is no simple relation that allows one to compute the singular values at time $t+dt$ based on their values at time $t$. We therefore performed this calculation over many pacing periods on a coarse grid with $\Delta t=T/10$. 

\begin{figure}
\includegraphics[width=3in]{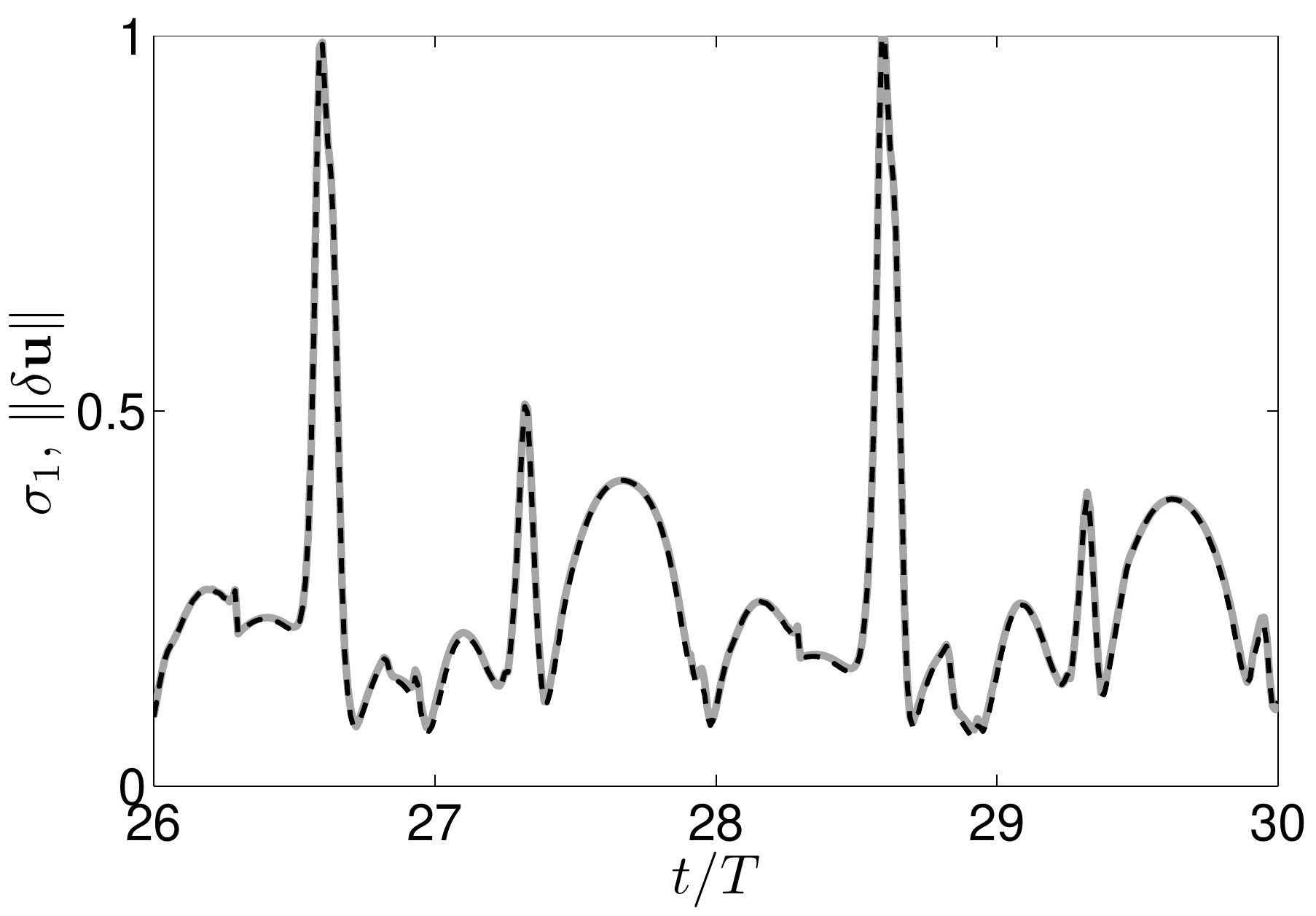}\\
\caption{\label{fig:sig_norm_np26_4T} Time-resolved transient amplification magnitude over the interval $t \in [26T,30T]$. The singular value $\sigma_1(t)$ (solid gray line) and the norm of the perturbation
$\delta{\bf u}({\bf r},t)=U(t,0){\bf q}_1({\bf r},t_{\rm max})$ (dashed black line) for $T=80$ ms. Both quantities were normalized by their maxima.}
\end{figure}

The leading singular value $\sigma_1(t)$ is shown in Fig. \ref{fig:sig_80}. We see that transient growth is a relatively slow process. The maximal transient amplification takes place only after almost 30 pacing stimuli. 
At this level of resolution, $\sigma_1(t)$ appears to be a discontinuous function of time. In fact, this is not the case, as the time-resolved calculation over the interval $[26T, 30T]$ illustrates (cf.  Fig. \ref{fig:sig_norm_np26_4T}). 
We find that $\sigma_1(t)$ has a rather fine structure on a time scale smaller than one pacing period. 
The {\it maxima} of transient amplification have a pattern that is $2T$-periodic; they are located at $t\approx(2n+0.6)T$, where $n$ is an integer. The peaks' values, on the other hand, are modulated with a period $2\pi/{\rm arg}(\lambda_{21})\approx18.9T$ determined by the leading Floquet multiplier.
The largest transient amplification ($\sigma_1=804$) is achieved at $t_{\rm max}=28.6T$ and the second highest value is achieved at $t=26.6T$.
Recall that for protocol A it takes a time interval of roughly $15T$ until the dynamics transitions from the neighborhood of the 2:2 state to SWC. 
These times are all of the same order of magnitude, suggesting that memory effect extends over times scales of order $20T=1600$~ms, much longer than the naive estimate $\epsilon^{-1}\tau_u=250$~ms would suggest.

\begin{figure}
\includegraphics[width=1.65in]{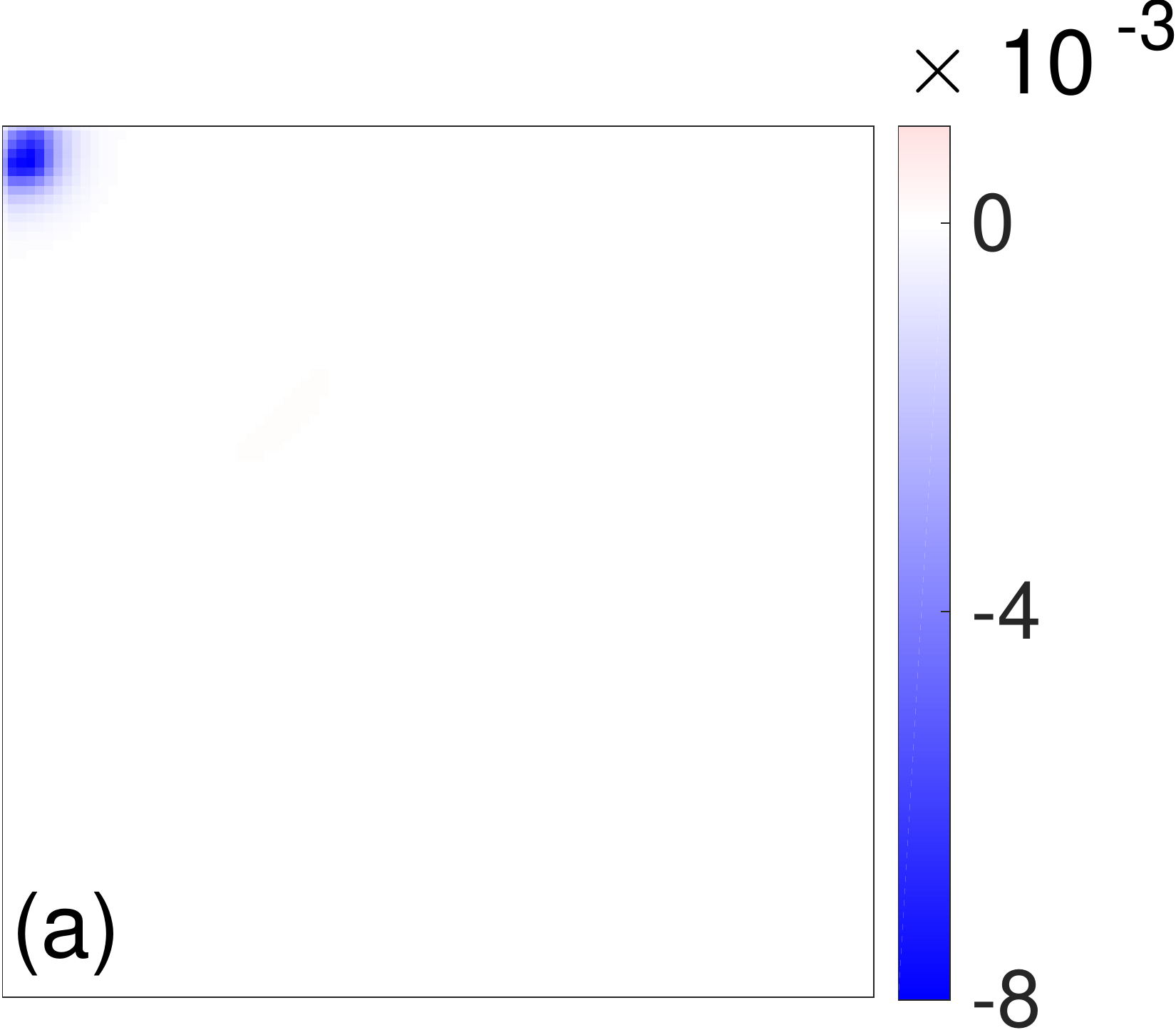}
\vspace{3mm}
\includegraphics[width=1.65in]{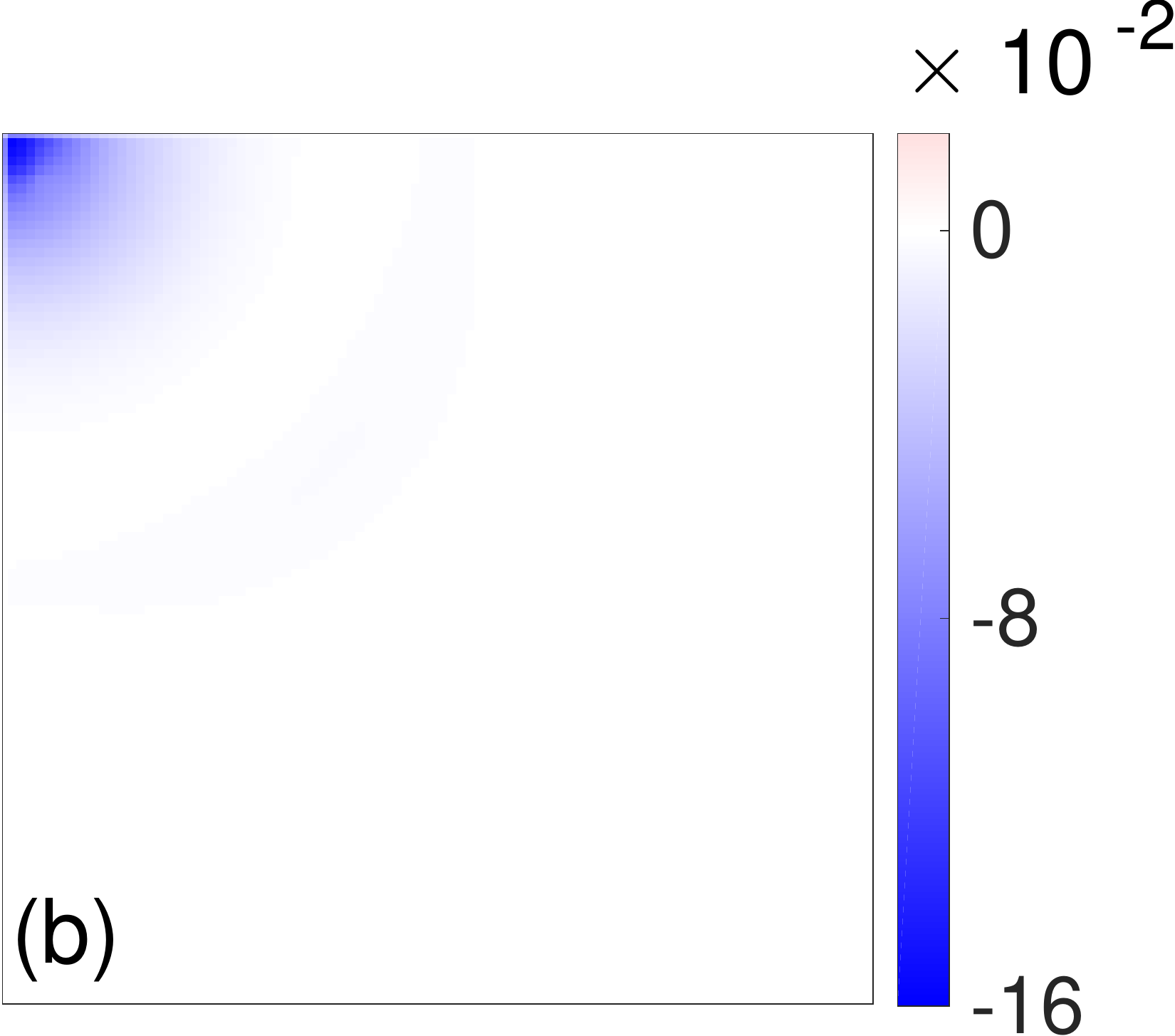}\\
\includegraphics[width=1.65in]{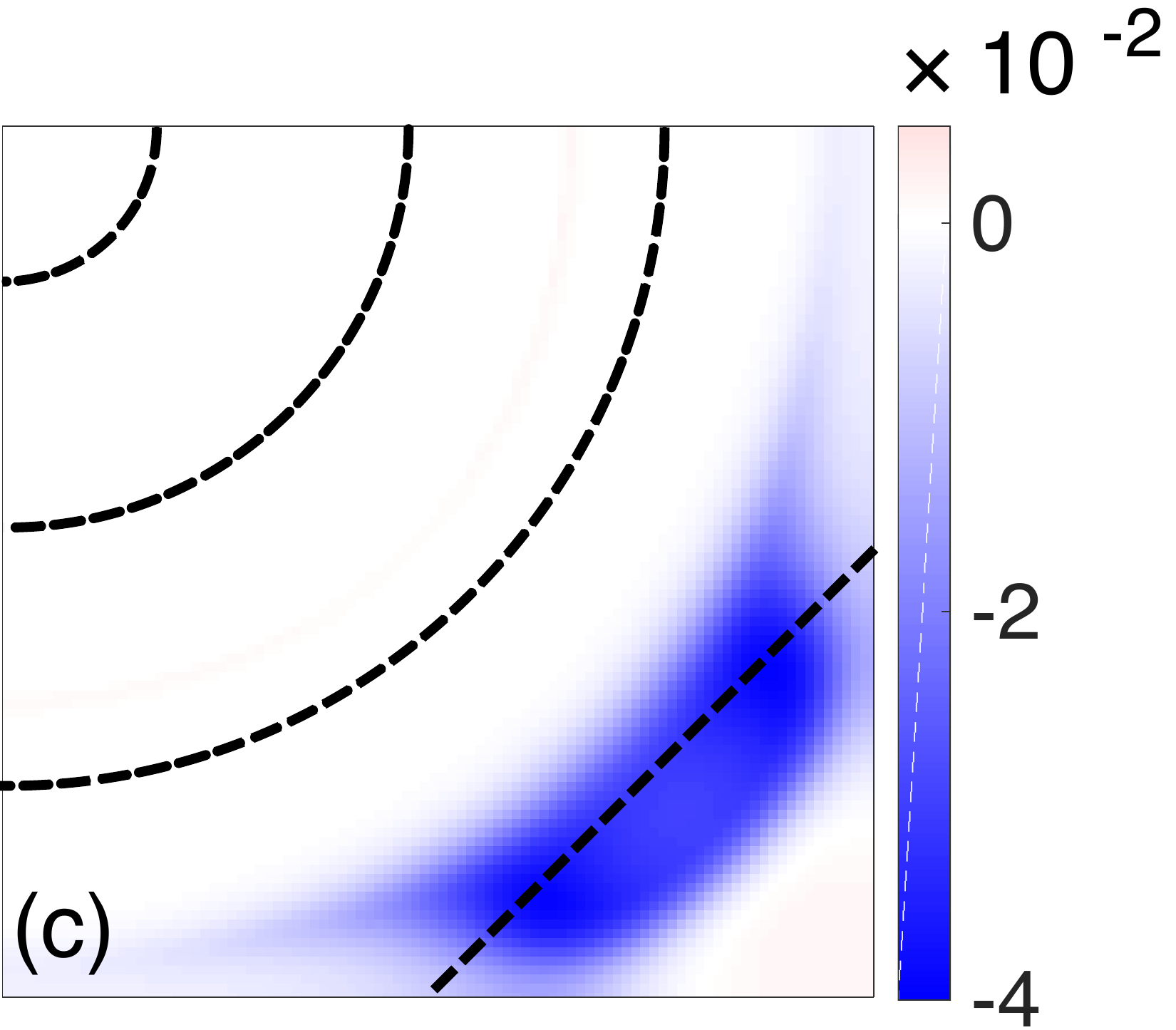}
\includegraphics[trim=0 -10 0 0,clip,width=1.65in]{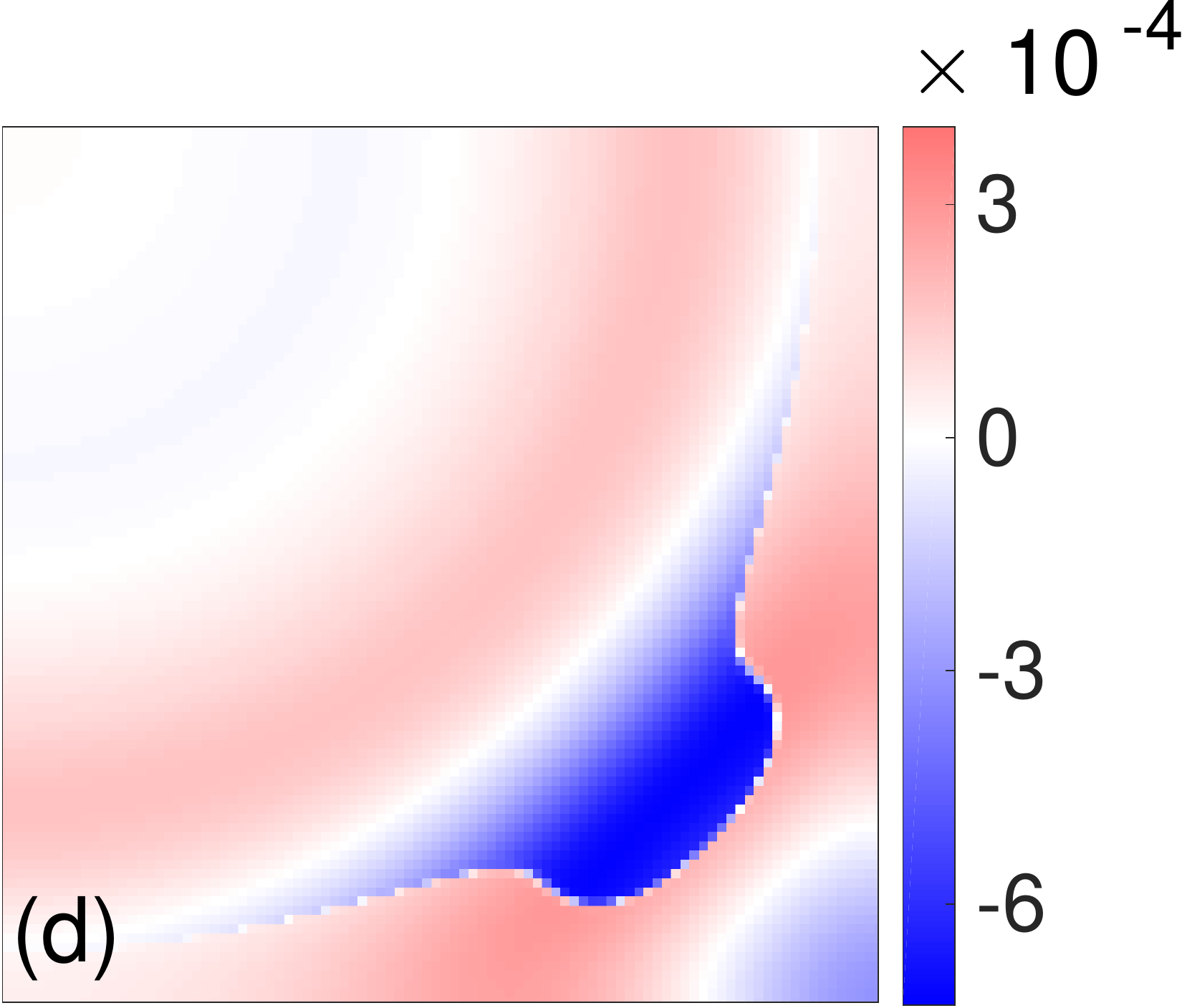}
\caption{\label{fig:singvec}
Leading singular vectors. (a) and (b) show $u$- and $v$-component of the right singular vector ${\bf q}_1({\bf r},t_{\rm max})$. (c) and (d) show $u$- and $v$-component of the left singular vector ${\bf p}_1({\bf r},t_{\rm max})$.
Dashed black lines indicate the positions of the nodal lines of the 2:2 solution.}
\end{figure}

The right singular vectors (cf. Fig. \ref{fig:singvec}(a-b)) associated with the time instances when $\sigma_1(t)$ is near the peak value $\sigma_1(t_{\rm max})$ are indistinguishable, implying that their spatial structure, which determines the optimal initial perturbation $\delta{\bf u}_{opt}({\bf r},0)$ is a robust feature of the dynamics. 
Indeed, the time-dependence of $\sigma_1(t)$ for $t\gtrsim 20T$ can be reproduced with extremely high accuracy by the norm of the disturbance which corresponds to the optimal one, e.g., $\delta{\bf u}_{opt}({\bf r},0)=s{\bf q}_1({\bf r},t_{\rm max})$, where $s$ is the amplitude of the initial disturbance. 
As Fig. \ref{fig:sig_norm_np26_4T} illustrates, the norm $\|\delta{\bf u}_{opt}({\bf r},t)\|$ is indistinguishable from $\sigma_1(t)$ once both have been normalized by their maximal values. Effectively, this means that the initial disturbance $s{\bf q}_1({\bf r},t_{\rm max})$ is in fact optimal for {\it all} times $t\gtrsim 20T$. 
This optimal disturbance is strongly localized around the pacing site, which means that the dynamics are very sensitive to the perturbation in the timing of the pacing stimuli.

The corresponding left singular vector ${\bf p}_1({\bf r},t_{\rm max})$ is shown in Fig. \ref{fig:singvec}(c-d)). It is worth noting that, unlike the right singular vector which is strongly localized, the left singular vector is not. 
This spatial delocalization is to a large extent responsible for the apparent large transient amplification with respect to the 2-norm \eqref{eq:2norm}, which involves integration over the entire spatial domain. 
Furthermore, the right singular vector is dominated by the $v$ component, so that the dynamics are more sensitive to the perturbation of the $v$ variable than the $u$ variable. 
For the left singular vector the opposite is true, hence the perturbation affects the $u$ variable to a much larger degree than the $v$ variable.
A similar asymmetry was found for the left and right marginal eigenvectors (response functions and Goldstone modes) of spiral wave solutions, which similarly describe the cause-and-effect relationship for small perturbations \cite{Marcotte2016}.

\begin{figure}
\includegraphics[width=3in]{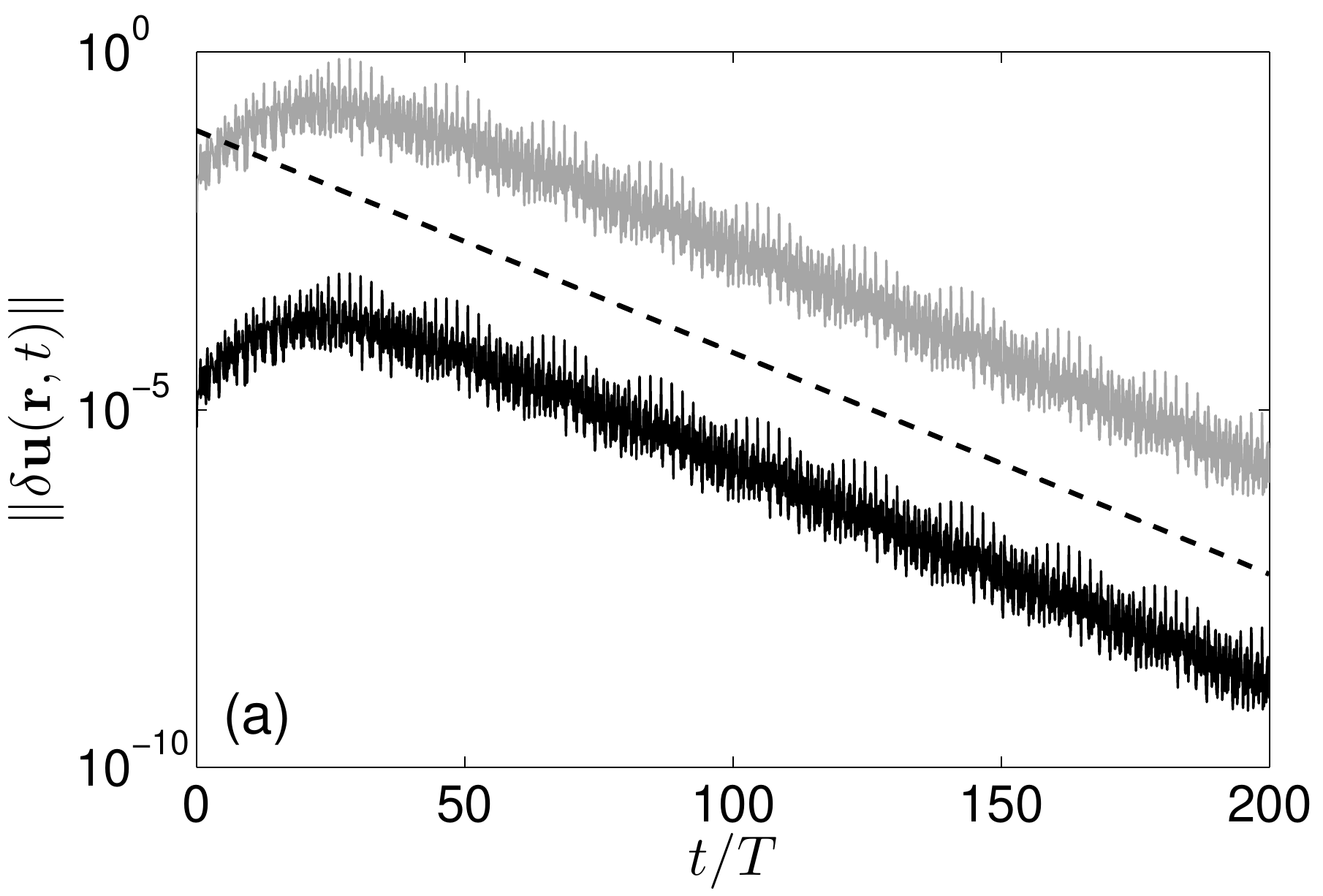}\\
\vspace{3mm}
\includegraphics[trim=-12 0 0 0,clip,width=3in]{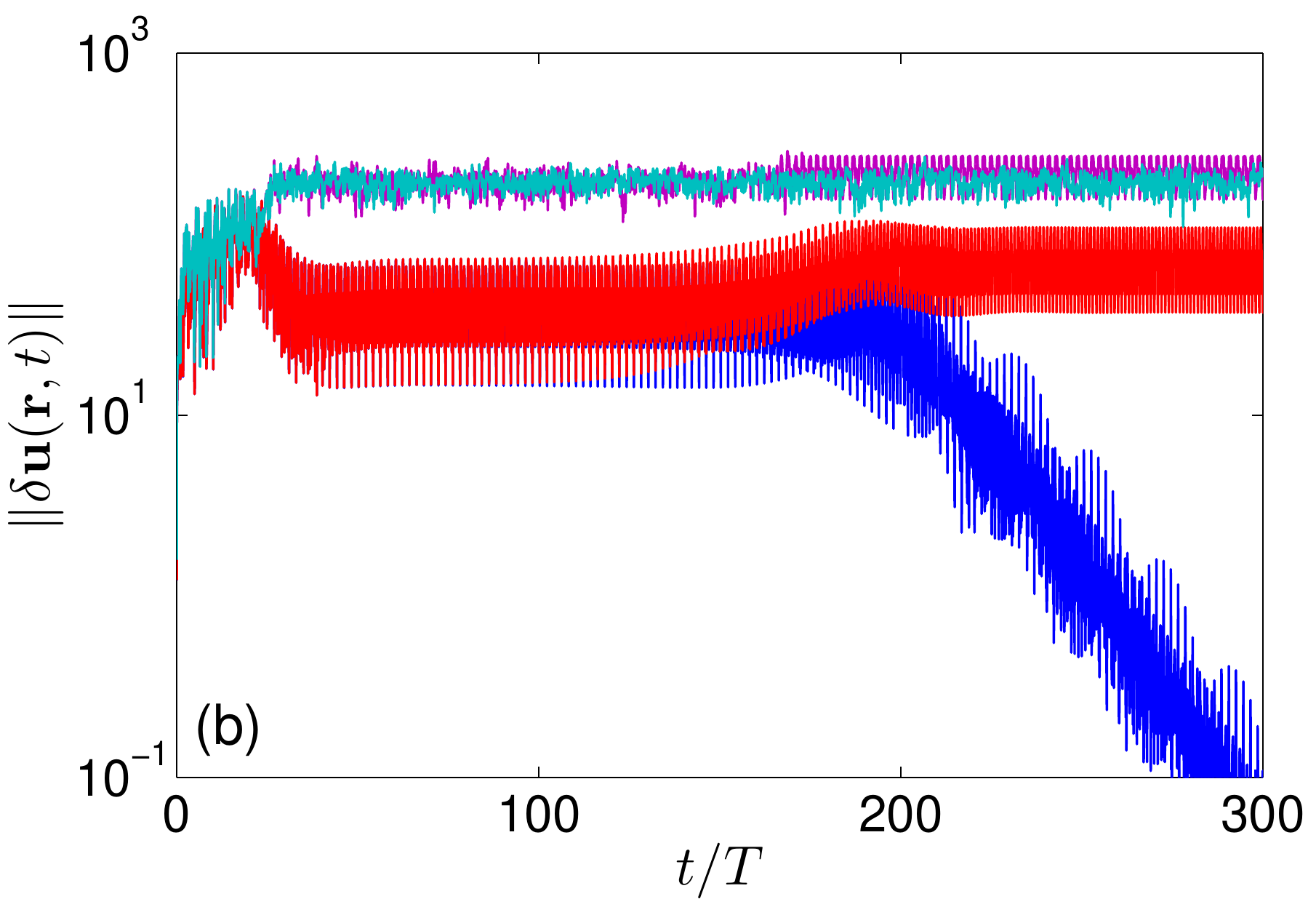}
\caption{\label{fig:norm_dyt_epsi} Transient growth of the optimal initial perturbation $\delta{\bf u}_{opt}({\bf r},0)=s\,{\bf q}_1({\bf r},t_{\rm max})$ for 
different perturbation strengths $s$. (a) $s=10^{-6}$ (black line) and $s=10^{-3}$ (solid gray line). The dashed line 
depicts the predicted asymptotic decay $\|\delta {\bf u}\|e^{\kappa t}$. 
(b) $s=1.227$ (blue line), $s=1.228$ (red line),  $s=1.594916$ (cyan line), and $s=1.59492$ (magenta line).}
\end{figure}

As expected, generalized linear stability analysis correctly predicts the dynamics produced by optimal initial disturbances of sufficiently small amplitude $s$. 
Evolving the optimal initial perturbation with magnitude $s=10^{-6}$ using the nonlinear equation \eqref{eq:p_tdu}, we find that $\delta{\bf u}_{opt}({\bf r},t )$ grows transiently, achieving the predicted amplification $\sigma_1(t_{\rm max})$ at time $t_{\rm max}$ (cf. Fig. \ref{fig:norm_dyt_epsi}(a)). 
For $t\gtrsim 30T$ the asymptotic exponential decay $\|\delta{\bf u}_{opt}({\bf r},t )\|\propto e^{\kappa t}$ with the predicted rate $\kappa=\ln|\lambda_{21}|/(2T)$ sets in, where $|\lambda_{21}|=0.8669$ 
is the leading Floquet multiplier. A stronger perturbation with $s=10^{-3}$ leads to the same result, with the entire curve simply shifted up, indicating that nonlinear effects are still negligible.

These results are qualitatively similar to transient amplification found for traveling wave solutions describing thin fluid films spreading on a solid substrate \cite{bertozzi1997linear}. In that case as well, transient growth of small localized perturbations was traced to the nonnormality of the evolution operator \cite{Grigoriev2005}. Strong transient amplification of disturbances also characterizes traveling excitation wave in models of paced Purkinje fibers with feedback, although in that case the base state corresponds to the 1:1 response \cite{Garzon11,Garzon14}. In fact, increasing transient amplification was identified as the reason for breakdown of feedback control for long fibers controlled at the pacing site.

\subsection{Finite amplitude disturbances}

Since the 2:2 state is linearly stable, transition to SWC is subcritical and requires a finite amplitude perturbation. 
The critical magnitude of the perturbation away from ${\bf u}_{21}$ that causes a transition to a different solution (${\bf u}_{22}$, ${\bf u}'_{21}$, ${\bf u}'_{22}$, or SWC) is given by the distance from ${\bf u}_{21}$ to the nearest boundary of the corresponding basin of attraction (cf. Fig. \ref{fig:basin}).
For linearly optimal initial perturbations, $\delta{\bf u}_{opt}({\bf r},0)=s{\bf q}_1({\bf r},t_{\rm max})$, we find that the asymptotic state is ${\bf u}_{21}$ for $s\le 1.227$, as Fig. \ref{fig:norm_dyt_epsi}(b) shows. 
For $1.228\le s\le 1.5949155$ the asymptotic state is ${\bf u}_{22}$, which is also a 2:2 state, but with a different phase.
For $s=1.5949156$ the asymptotic state is again ${\bf u}_{21}$, but for $s=1.5949157$ the asymptotic state is SWC, which persists for at least $400T$.
Increasing $s$ further we find either persistent SWC or transient SWC that eventually gives way to one of the 2:1 states (${\bf u}'_{21}$ or ${\bf u}'_{22}$).


The extreme sensitivity of the outcome to the choice of initial conditions is a characteristic feature of chaotic attractors that have fractal basin boundaries \cite{Mcdonald1985}.
Since ${\bf u}_{21}$ and ${\bf u}_{22}$ are parts of the same 2-cycle, both $\Omega_{21}$ and $\Omega_{22}$ share a part of their boundary with the basin of attraction $\Omega_{swc}$ of the SWC state. 
Like a quilt, basin boundaries are composed of a patchwork of pieces, each one corresponding to the stable manifold of an edge state -- a saddle solution embedded into the boundary, which has a one-dimensional unstable manifold. 
The relation between optimal perturbations, transient growth, edge states, and basin boundaries has originally been investigated in the context of simple nonlinear PDEs \cite{Handel2006}. 
More recently, edge states have been studied extensively as gateways mediating subcritical transition to turbulence in shear fluid flows \cite{schneider2007,GHCW07}.

\begin{figure}
\includegraphics[width=3in]{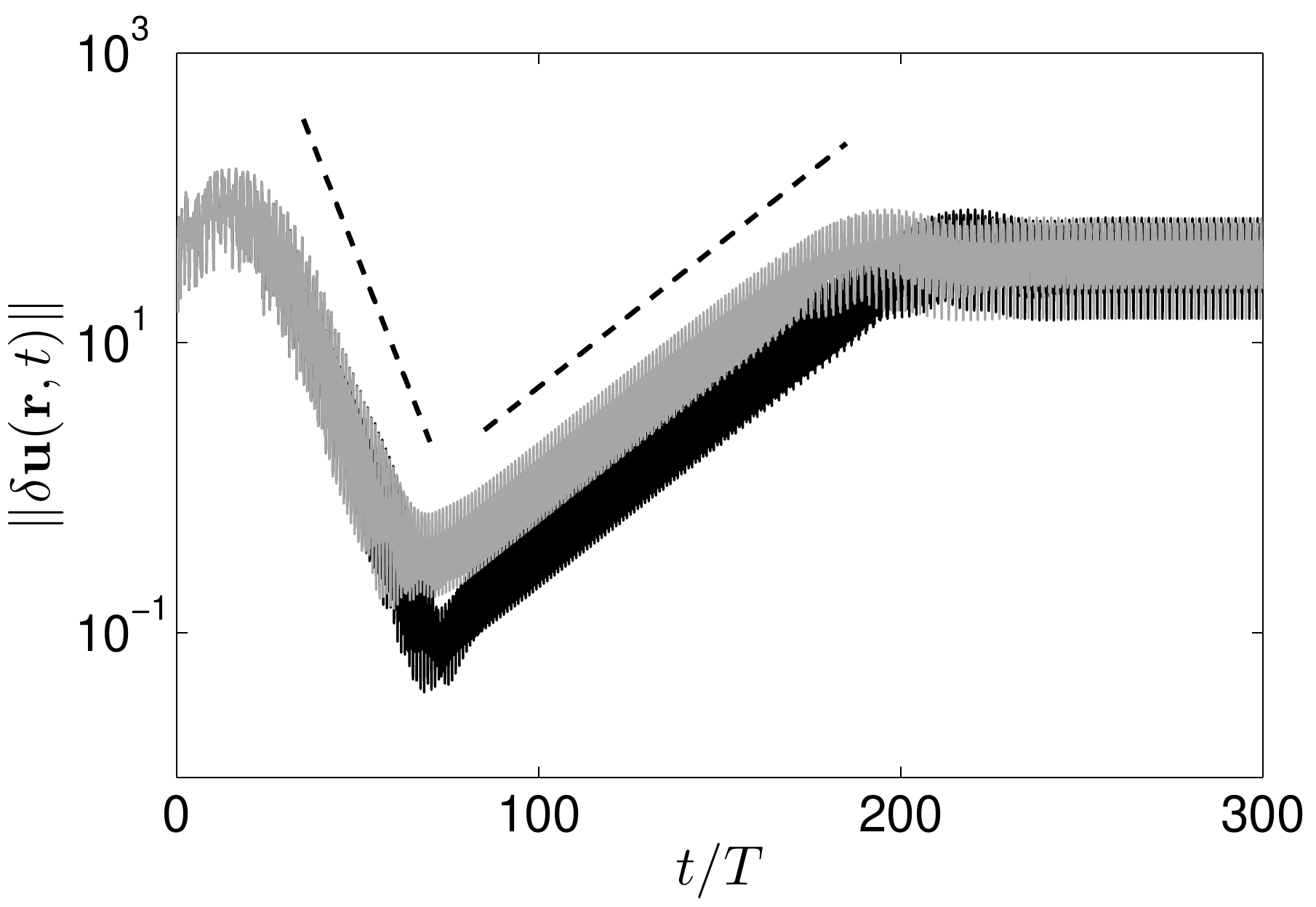}
\caption{\label{fig:edge1}
The norm of the deviation $\delta{\bf u}({\bf r},t) = {\bf u}({\bf r},t)-{\bf u}_1({\bf r},t)$ from the 1:1 solution for initial conditions ${\bf u}_{21}({\bf r}) + s {\bf q}_1({\bf r};t_{\rm max})$ with $s=1.227$ (black line) and $s=1.228$ (gray line). The dashed lines describe the exponential decay $\|\delta{\bf u}\|\propto e^{\kappa_{12} t}$ and exponential growth $\|\delta{\bf u}\|\propto e^{\kappa_{11}t}$ with the rates $\kappa_{1i} = \ln|\lambda_{1i}|/T$ predicted by the linearization around ${\bf u}_1({\bf r},t)$.} 
\end{figure}

The 1:1 state ${\bf u}_{11}$ is an example of such an edge state. 
It has one unstable direction and its stable manifold forms the boundary between $\Omega_{21}$ and $\Omega_{22}$. 
As Fig. \ref{fig:norm_dyt_epsi}(b) illustrates, the solutions corresponding to perturbations with $s=1.227$ and $s=1.228$ approach and closely follow the same periodic solution for $40T\lesssim t\lesssim 160T$, before eventually separating and approaching, respectively, ${\bf u}_{21}$ and ${\bf u}_{22}$. 
This periodic solution is indeed ${\bf u}_{11}$, as shown in Fig. \ref{fig:edge1}. 
As expected, the two trajectories approach ${\bf u}_1({\bf r},t)$ along the least stable direction and separate along the unstable direction, with the rates predicted by the linearization.

The basin boundary $\partial\Omega_{swc}$ of the chaotic attractor lies further away from ${\bf u}_{21}$ than the basin boundary $\partial\Omega_{22}$ of the 2:2 state ${\bf u}_{22}$. 
An upper bound for the distance to $\partial\Omega_{swc}$ is given by the norm $s_{cr}=\|\delta{\bf u}_{opt}^{cr}({\bf r},0)\|=1.5949157$ of the critical linearly optimal initial disturbance. 
It is worth noting that the norm of this disturbance is more than two orders of magnitude less than the norm of the state $\|{\bf u}_{21}\|=O(200)$ itself.
The ratio of the norms is not as large as the maximal transient amplification $\sigma_1(t_{\rm max})=804$, but it is of the same order of magnitude, which suggests that linear theory cold be used to estimate the critical disturbance magnitude. 
However, a fully nonlinear calculation is required to obtain a quantitatively accurate prediction for both the magnitude and the shape of the smallest initial disturbance required to trigger a transition from the 2:2 state to SWC.

\begin{figure}
\subfigure[]{\includegraphics[width=0.32\columnwidth]{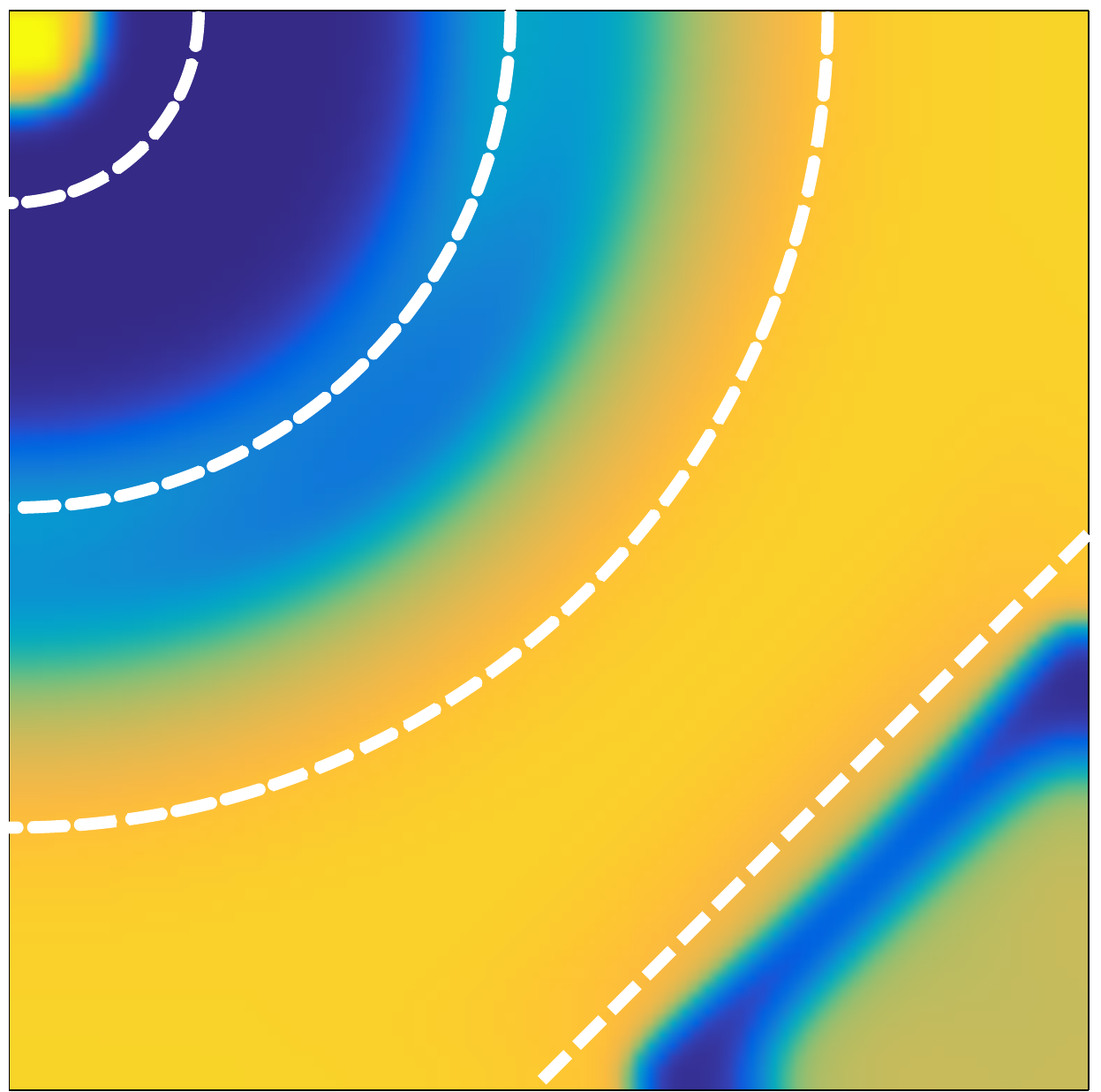}}
\subfigure[]{\includegraphics[width=0.32\columnwidth]{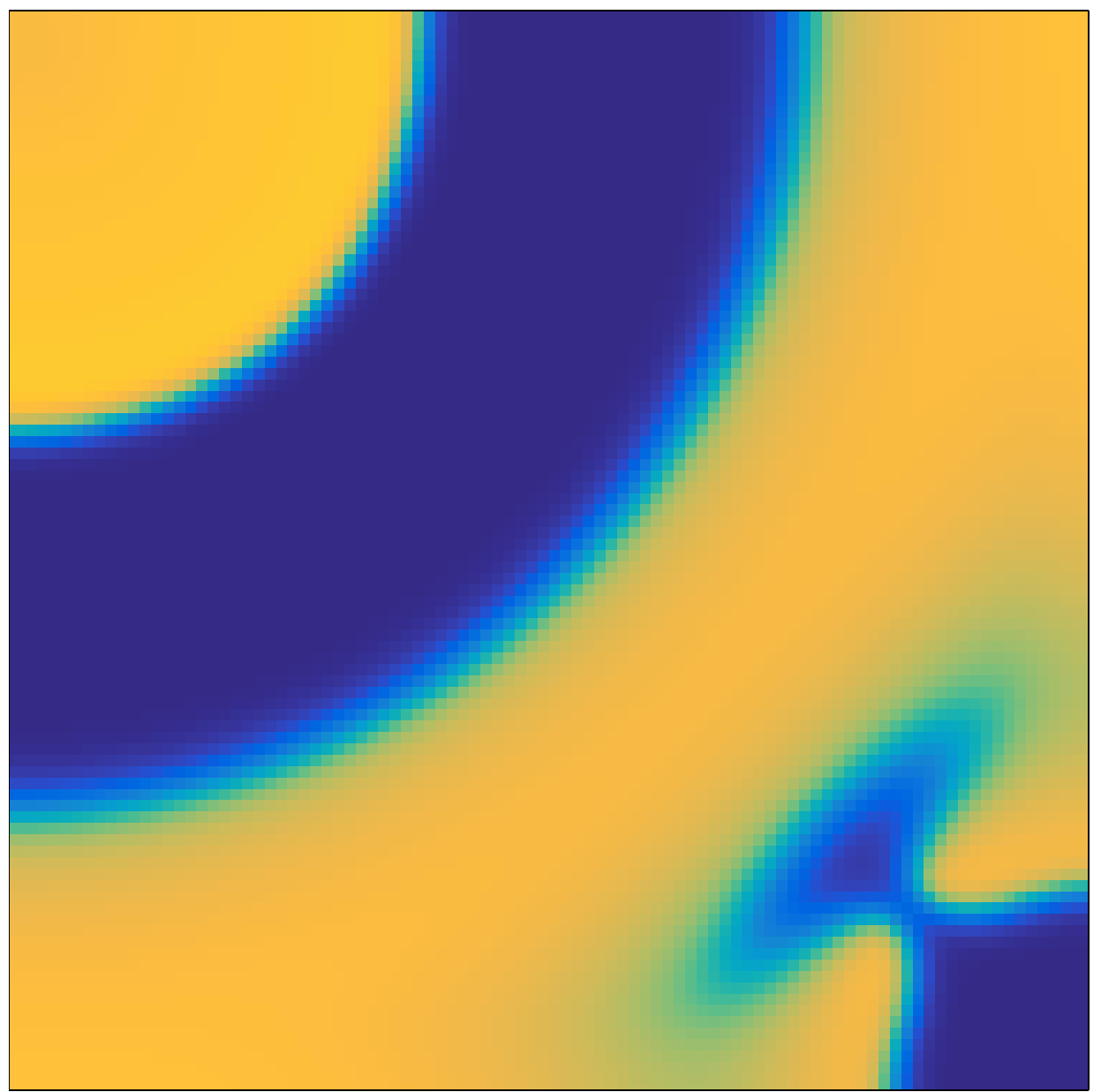}}
\subfigure[]{\includegraphics[width=0.32\columnwidth]{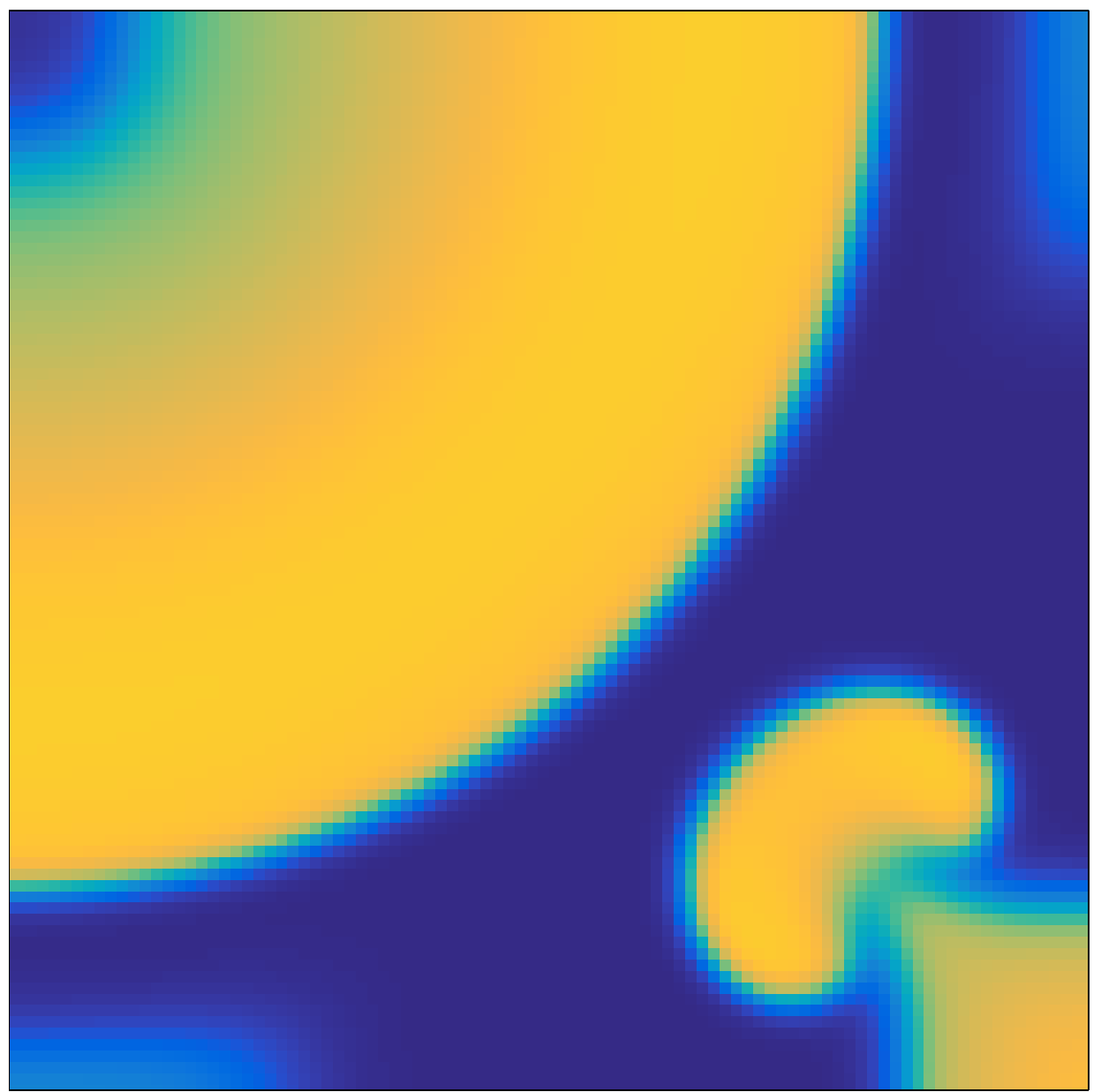}}\\
\subfigure[]{\includegraphics[width=0.32\columnwidth]{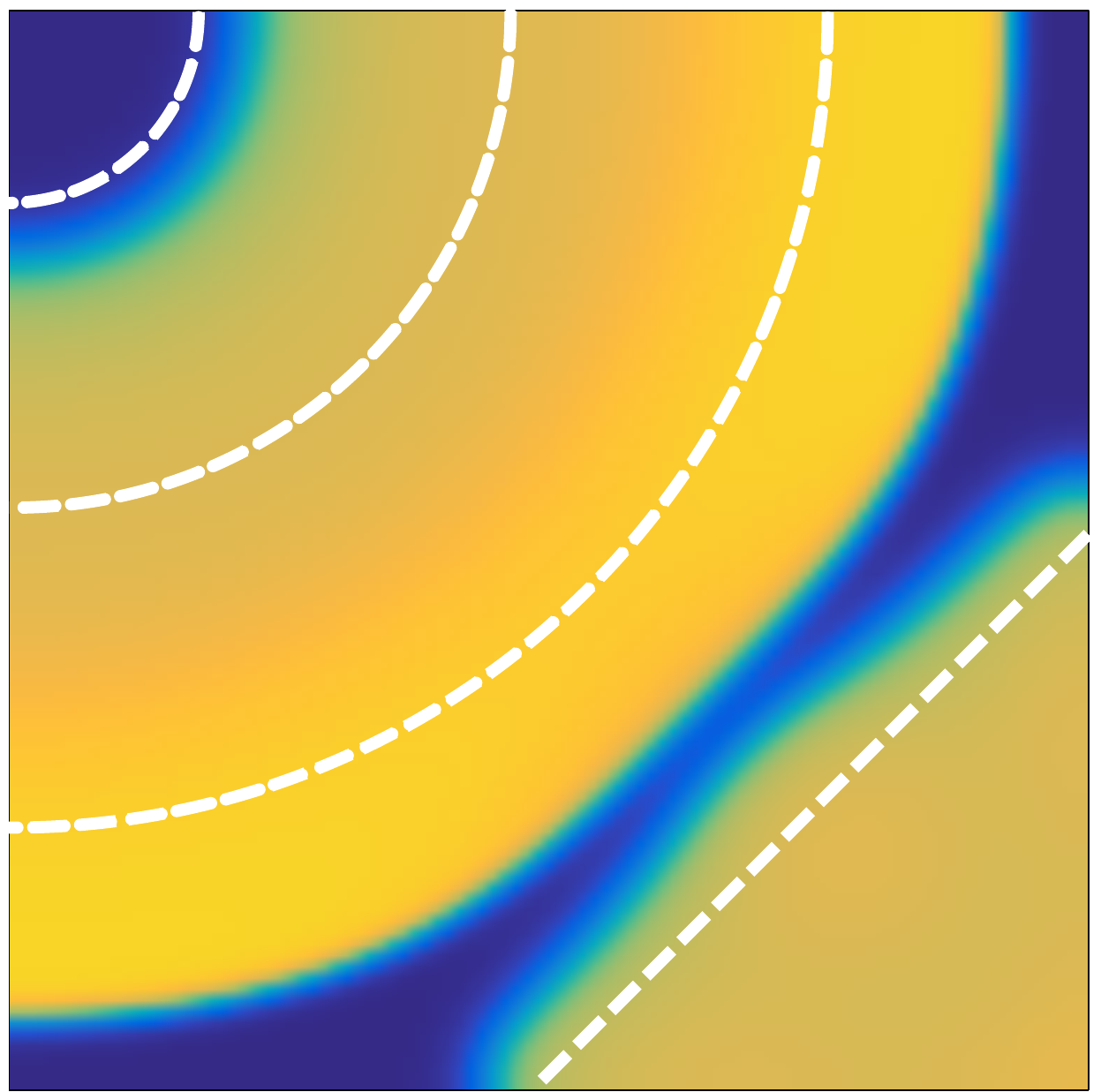}}
\subfigure[]{\includegraphics[width=0.32\columnwidth]{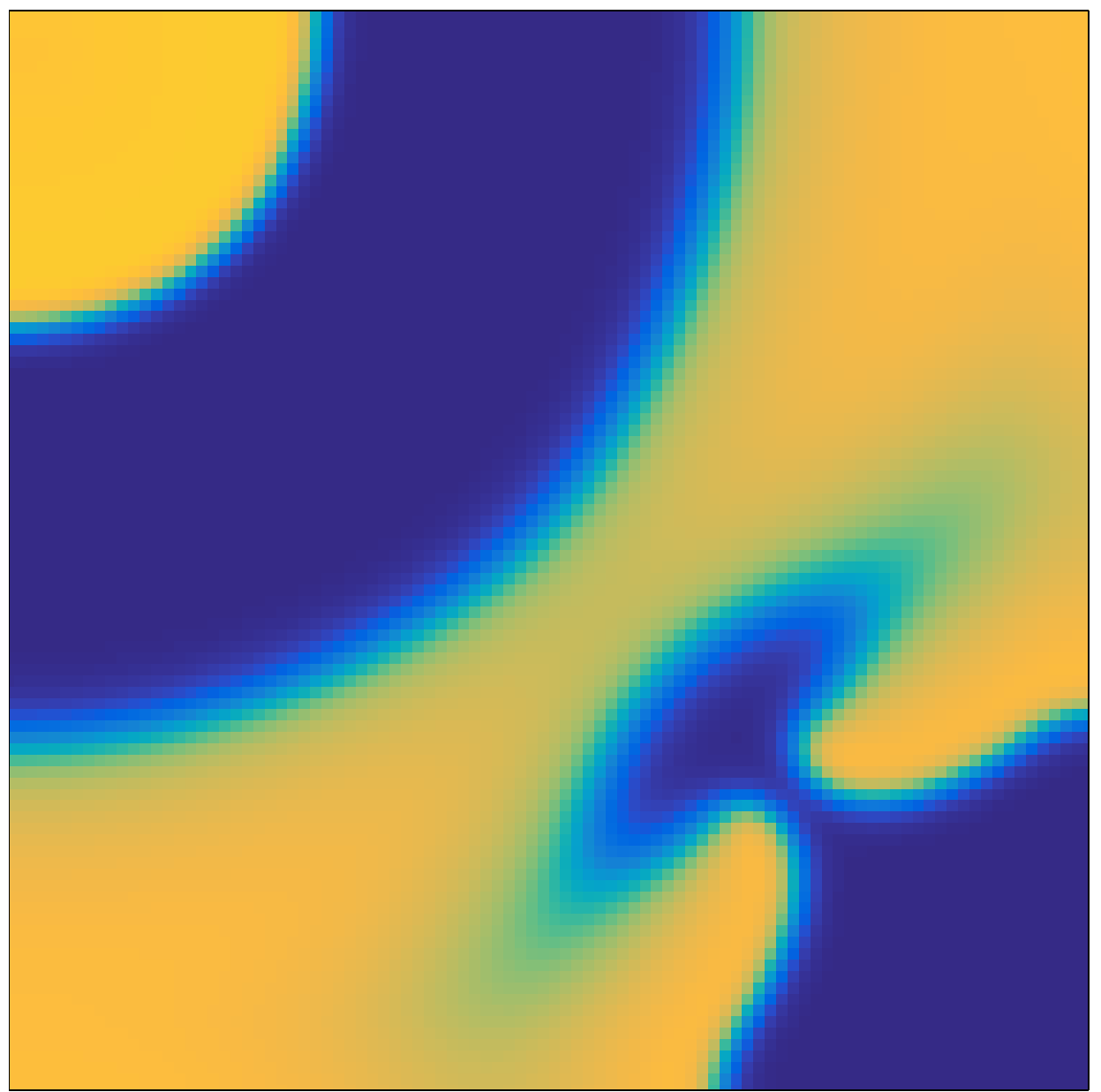}}
\subfigure[]{\includegraphics[width=0.32\columnwidth]{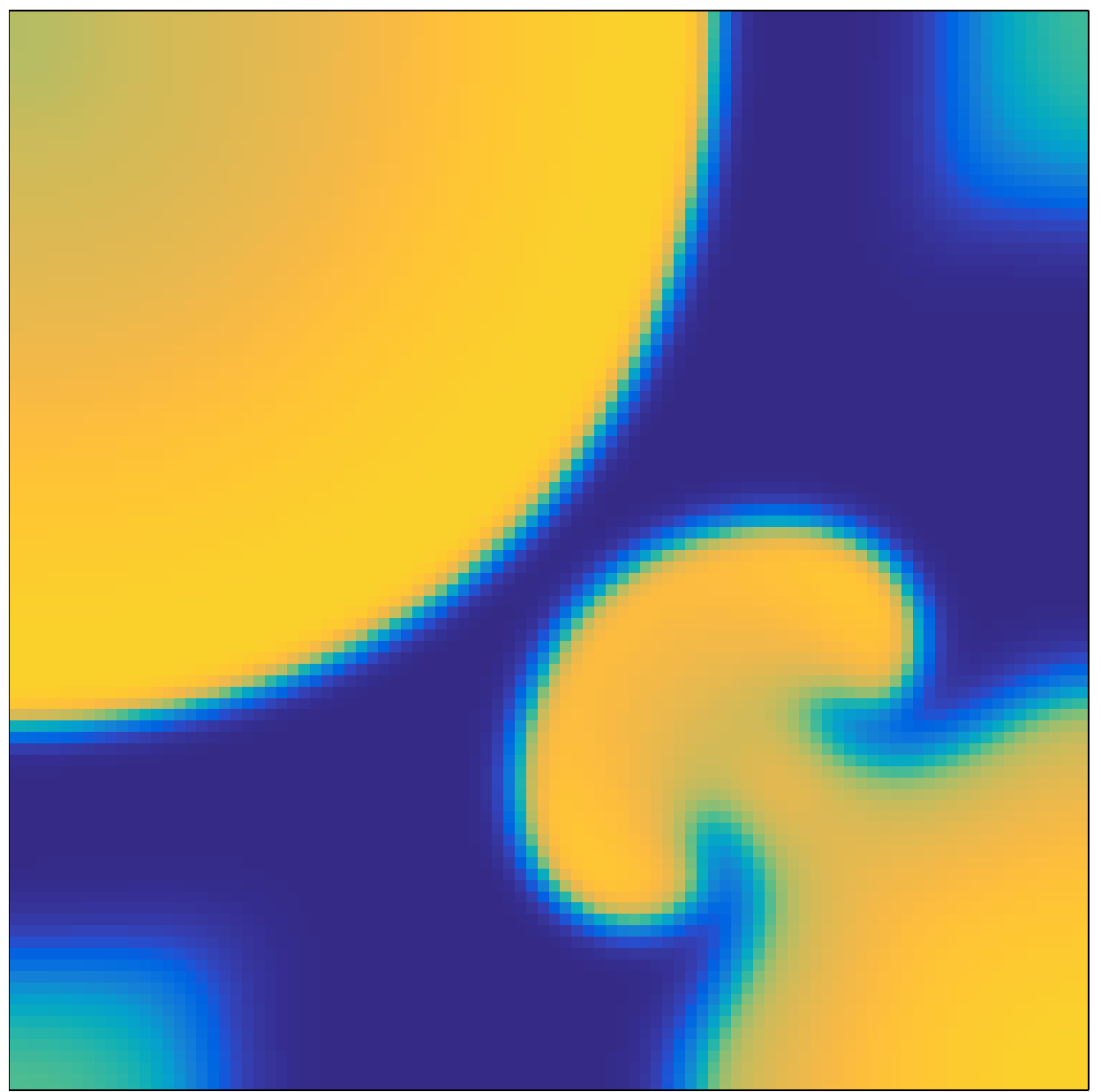}}\\
\subfigure[]{\includegraphics[width=0.32\columnwidth]{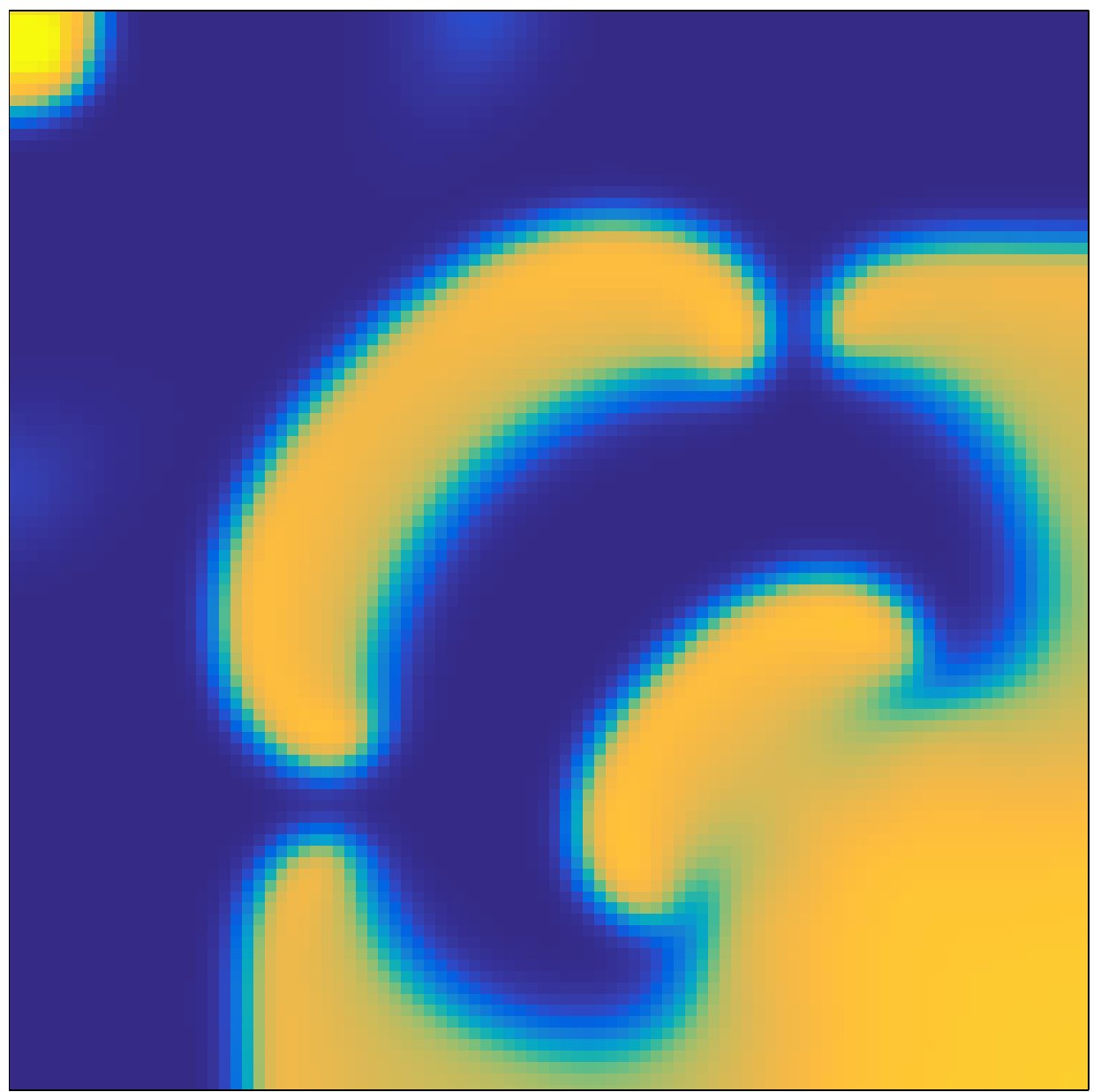}}
\subfigure[]{\includegraphics[width=0.32\columnwidth]{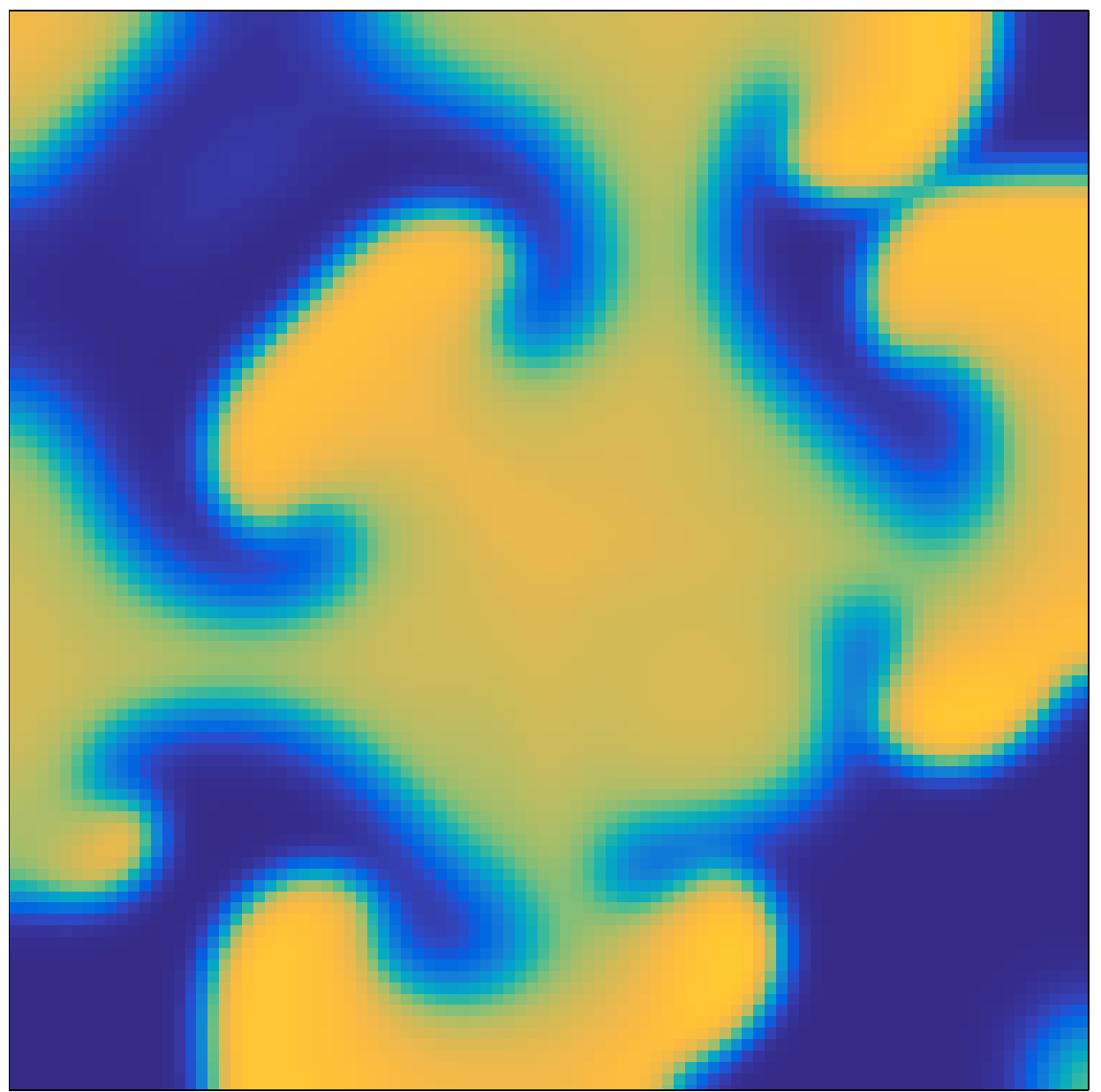}}
\subfigure[]{\includegraphics[width=0.32\columnwidth]{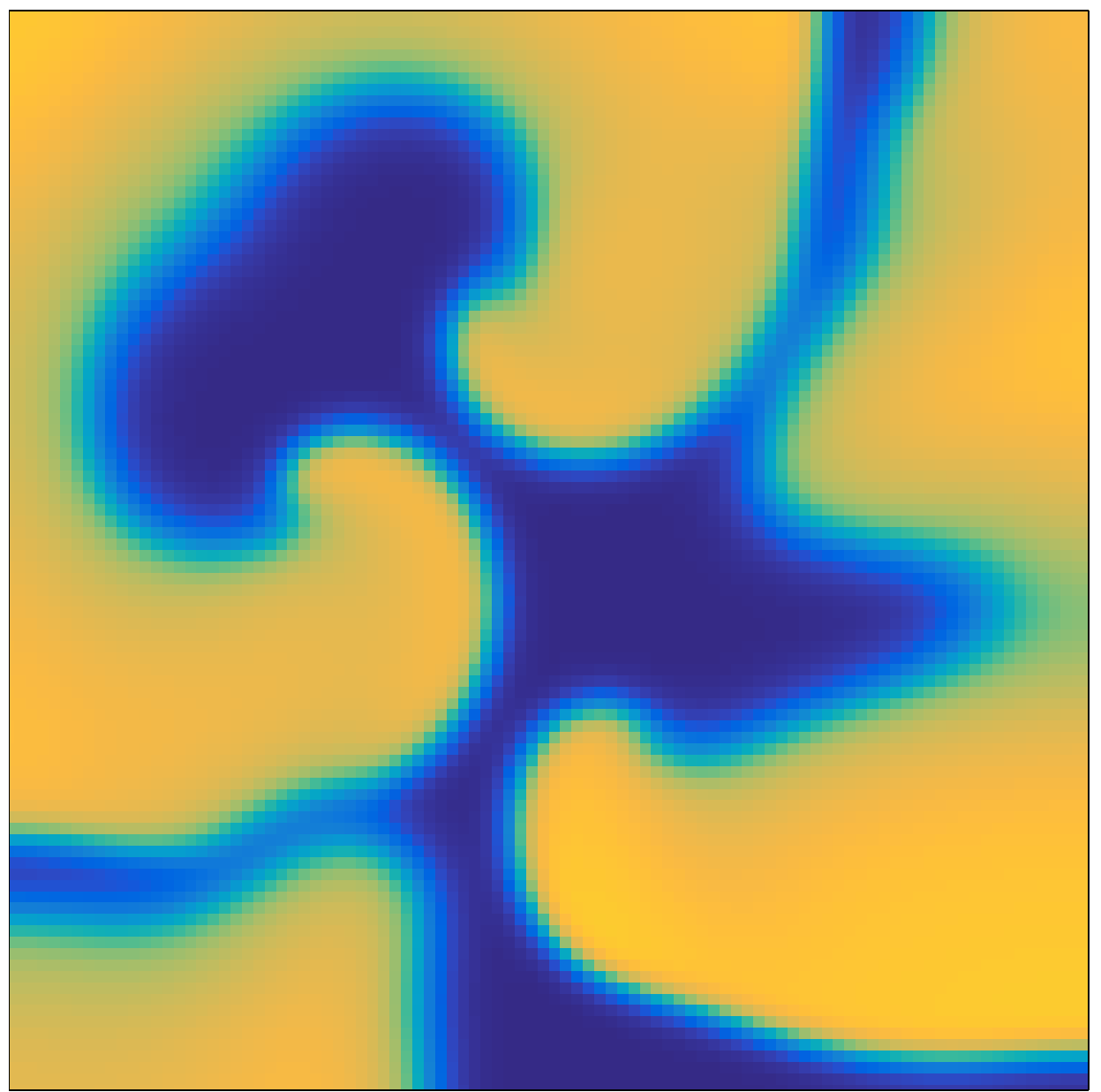}}
\caption{\label{fig:crit_dist} Wave breakup produced by the linearly optimal disturbance $\delta{\bf u}_{opt}({\bf r},0)=s{\bf q}_1({\bf r},t_{\rm max})$ with, $s=1.59492$.
The voltage component of the solution ${\bf u}({\bf r},t)$ is shown at (a) $t=19.05T$, (b) $t=19.35T$, (c) $t=19.75T$, (d) $t=22.85T$, (e) $t=23.25T$, (f) $t=23.6T$, (g) $t=27.05T$, (h) $t=30.2T$, and (i) $t=34T$. 
Dashed white lines indicate the positions of the nodal lines of the 2:2 solution.} 
\end{figure}

\begin{figure}
\includegraphics[height=1.4in]{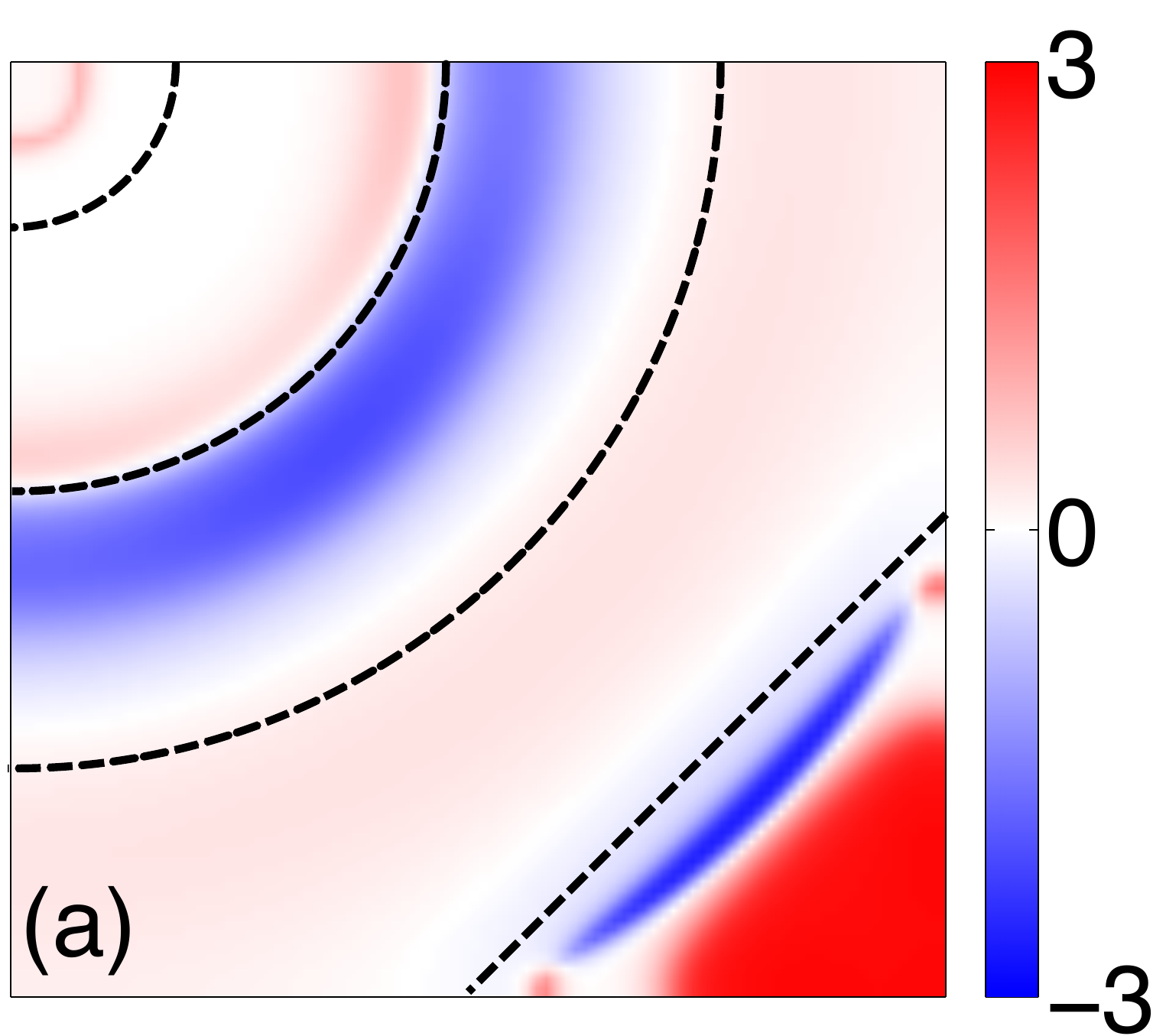}
\includegraphics[height=1.4in]{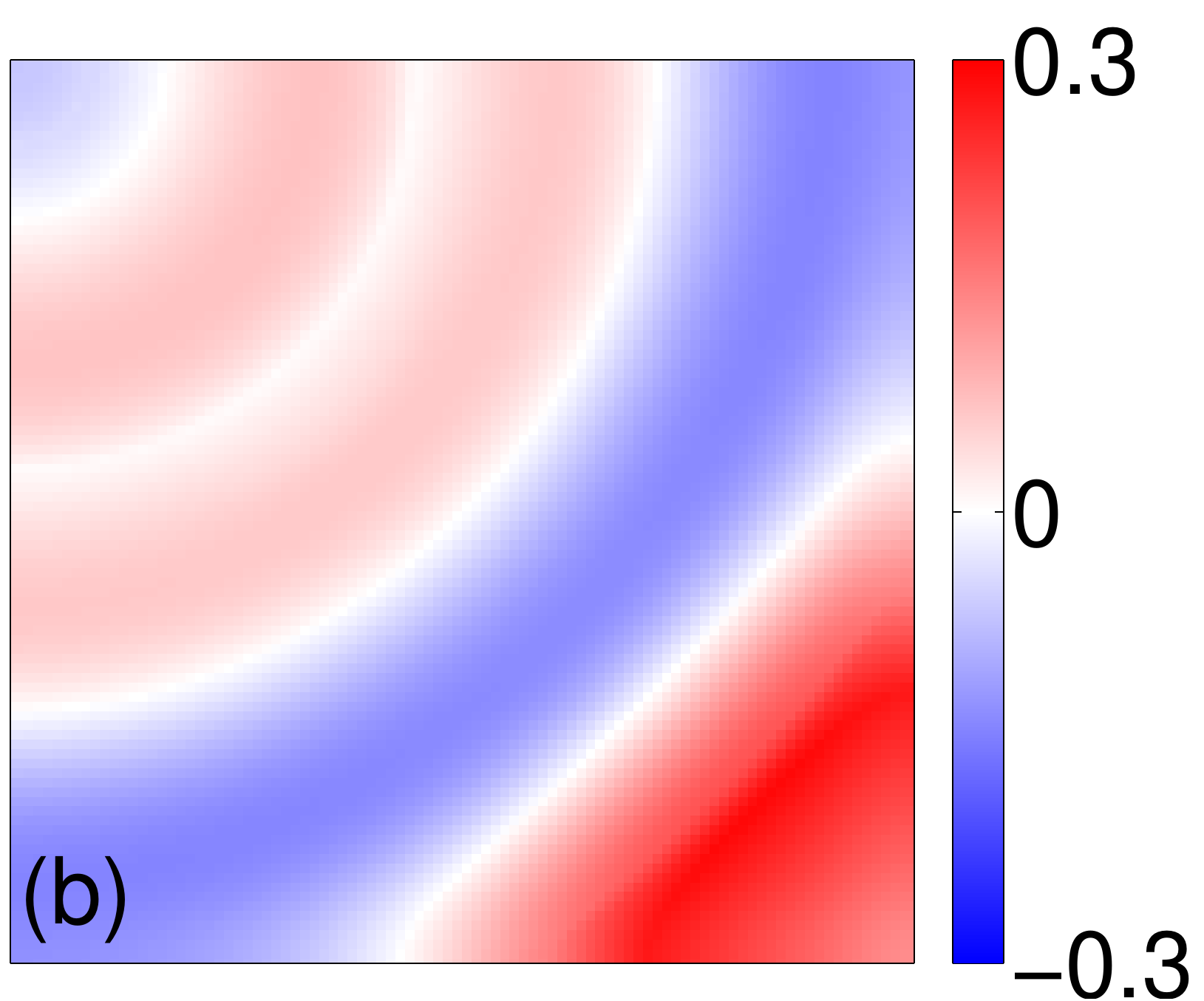}
\caption{\label{fig:condblk}
The perturbation $\delta{\bf u}_{opt}({\bf r},t_r)$ describing conduction block at $t=t_r\approx 19.05T$, which corresponds to the initial condition $\delta{\bf u}_{opt}({\bf r},0)=s{\bf q}_1({\bf r},t_{\rm max})$ with $s=1.59492$. (a) and (b) show the $u$- and $v$-component, respectively.
Dashed black lines indicate the positions of the nodal lines of the 2:2 solution.}
\end{figure}

Figure \ref{fig:crit_dist} shows the evolution of the initial condition ${\bf u}({\bf r},0)={\bf u}_{21}+\delta{\bf u}_{opt}({\bf r},0)$, which corresponds to the linearly optimal initial disturbance with a magnitude $s=1.59492$ that places it just outside of $\Omega_{2:2}$.
Even though the 2:2 state is linearly stable and the perturbation is quite small, transient amplification of the initial disturbance leads to pronounced alternans which causes conduction block in the lower right corner of the domain at $t=t_r\approx 19.05T$. 
Recall that, according to generalized linear stability theory, the shape of the perturbation at the moment when it reaches the largest amplification is given by the left singular vector of $U(t_{\rm max},0)$: $\delta{\bf u}_{opt}({\bf r},t_{\rm max})=\sigma_1(t_{\rm max})s{\bf p}_1({\bf r},t_{\rm max})$.
The location of conduction block is indeed found to be consistent with the shape of the left singular vector ${\bf p}_1({\bf r},t_{\rm max})$, which has maximal amplitude in the same corner of the domain, opposite the pacing site (cf. Fig. \ref{fig:singvec}(c-d)).
However, the overall shape of the actual deviation $\delta{\bf u}_{opt}^{cr}({\bf r},t_r)={\bf u}({\bf r},t)-{\bf u}_2({\bf r},t)$ (cf. Fig. \ref{fig:condblk}) is noticeably different from ${\bf p}_1({\bf r},t_{\rm max})$ (cf. Fig. \ref{fig:singvec}(c,d)).

While the first instance of conduction block generates a pair of spiral waves at $t\approx 19.35T$, this does not cause an immediate transition to SWC. 
The spiral waves have too little room to sustain themselves: they collide with the next wavefront emitted by the pacing site and are annihilated at $t\approx 20T$.
Another instance of conduction block occurs at $t\approx 22.85T$, again generating two spiral waves at $t\approx 23.25T$. 
Now conduction block occurs sooner in the cycle and further away from the lower right corner of the domain, and the spiral waves survive the collision with the next wavefront emitted by the pacing site.
(Note, however, that in both instances conduction block occurs near the nodal line that is the furtherest from the pacing site.)
Since the period of the spiral waves is shorter than the pacing period, they gradually invade the domain, generating further breakups and initiating sustained SWC.

\begin{figure}
\includegraphics[width=3in]{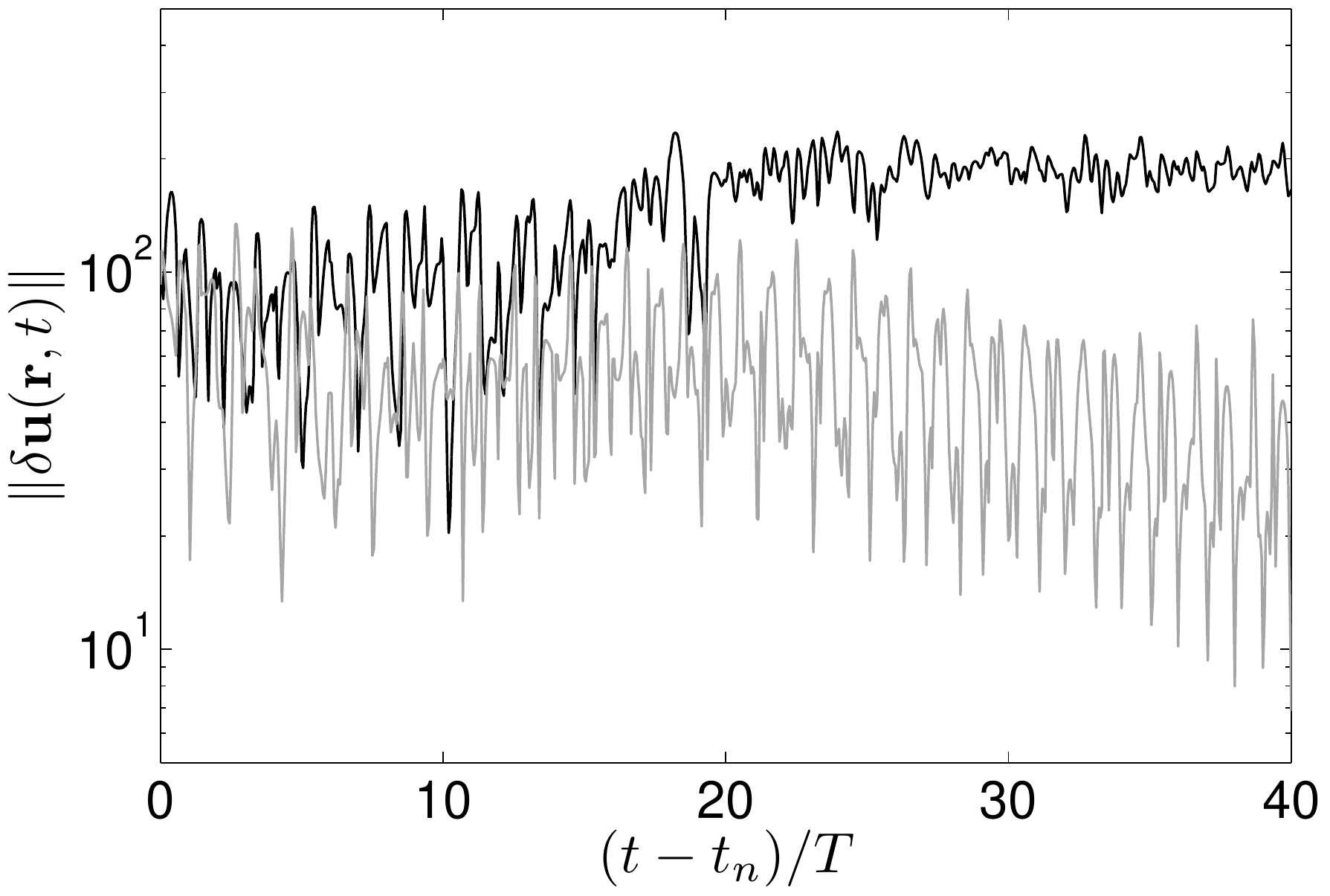}
\caption{\label{fig:normAB} The norm of the perturbations away from the 2:2 solution describing the protocols A (black line, $n=40$) and B (gray line, $n=50$).}
\end{figure}

As we discussed previously, from $t=t_{40}$ on, protocol A corresponds to constant pacing with $T=80$ ms. 
The initial condition ${\bf u}_A^0={\bf u}({\bf r},t_{40})$ for the last stage of protocol A corresponds to a fairly significant  perturbation away from the 2:2 state ${\bf u}_{21}$. 
The norm $\|\delta{\bf u}_A^0\|=\|{\bf u}_A^0-{\bf u}_{21}\|=104.4$ of this perturbation is two orders of magnitude larger than $s_{cr}=\|\delta{\bf u}_{opt}^{cr}\|$ (and is close to the norm of ${\bf u}_{21}$ itself).
On the other hand, the spatial structure of this perturbation (cf. Fig. \ref{fig:protAB_IC}(a-b)) is quite different from that of the optimal perturbation (cf. Fig. \ref{fig:singvec}(a-b)) and therefore it does not experience transient amplification, as Fig. \ref{fig:normAB} illustrates.
This difference can be quantified using the scalar product $\langle\delta\hat{\bf u}_A^0,\hat{\bf q}_1\rangle=0.0110$, where the hat denotes normalization (e.g., $\delta\hat{\bf u}=\delta\hat{\bf u}/\|\delta\hat{\bf u}\|$), which shows that $\delta{\bf u}_A^0$ has essentially no component along ${\bf q}_1$. 
At the same time $\langle\delta\hat{\bf u}_A^0,\hat{\bf p}_1\rangle=-0.4001$, so that $\delta{\bf u}_A^0$ has a significant component in the direction of ${\bf p}_1$ (cf. Fig. \ref{fig:basin}). 
   
Despite the dramatic difference in the amplitude and spatial structure of the perturbations $\delta{\bf u}_A^0$ and $\delta{\bf u}_{opt}^{cr}$, the sequence of states generated by protocol A (cf. Fig. \ref{fig:circ_wave}(b-d))
looks very similar to that produced by the near-critical optimal perturbation (cf. Fig. \ref{fig:crit_dist}).
One possible explanation is that, before approaching the chaotic attractor, both trajectories closely follow the boundary between $\Omega_{22}$ and $\Omega_{swc}$ for an extended period of time (around $20T$ for both $\delta{\bf u}_A^0$ and $\delta{\bf u}_{opt}^{cr}$) and finally they start to separate from it along the unstable manifold of the same edge state that lies on this boundary.

Although itself unstable, the basin boundary can be followed using the bisection algorithm \cite{schneider2007} that has been successfully used in the context of fluid turbulence.
For the system considered here we found that the critical optimal perturbation $\delta{\bf u}_{opt}^{cr}$ approaches a chaotic edge state that lies on the boundary of $\Omega_{2:2}$.  
It may in principle be possible to find nonchaotic edge states embedded into $\partial\Omega_{2:2}$ as well.
For instance, the unstable quasiperiodic solution born an the subcritical Hopf bifurcation of the 2:2 state lies on this boundary and, if it is has one unstable eigenvalue, it would be an example of an edge state.
Even simpler (e.g., time-periodic) edge states may be found by applying the bisection algorithm to initial conditions other than ${\bf u}_{21}+\delta{\bf u}_{opt}^{cr}$, however this is far outside the scope of this study.

The perturbation that corresponds to protocol B is also much stronger than the critical optimal one, but since its spatial structure (cf. Fig. \ref{fig:protAB_IC}(c-d)) is also quite different from $\delta{\bf u}_{opt}$, it does not experience transient amplification either, as Fig. \ref{fig:normAB} illustrates.
Since $\langle\delta\hat{\bf u}_B^0,\hat{\bf q}_1\rangle=0.0034$ and  $\langle\delta\hat{\bf u}_B^0,\hat{\bf p}_1\rangle=-0.5132$, $\delta{\bf u}_B^0$ has essentially no component along ${\bf q}_1$, but a significant component along ${\bf p}_1$. 
The norm of the perturbation, $\|\delta{\bf u}_B^0\|=\|{\bf u}_B^0-{\bf u}_{21}\|=92.8$, is small enough for the initial condition ${\bf u}_B^0$ to fall within the basin of attraction of the 2:2 state (cf. Fig. \ref{fig:basin}).
Hence the perturbation eventually decays, and the dynamics return to concordant alternans instead of transitioning to SWC.

\section{Discussion}\label{sec:summary}

The results presented in this paper help improve our understanding of the transition to fibrillation in paced atrial tissue and the role of alternans.
Specifically, these results confirm that the first step in the transition caused by an increase in the pacing rate is the development of discordant alternans \cite{Pastore1999,Weiss2006,Gizzi2013}.
Theoretical analysis of alternans in one dimension predicts \cite{Echebarria2002,fox2002conduction} that conduction block should occur at the location that is the furthest from the pacing site. 
On the other hand, experimental and numerical evidence suggests that reentry is promoted by steep repolarization gradients \cite{konta1990significance,tachibana1998,Weiss2006} and therefore conduction block should happen near a nodal line.
Indeed, we find that conduction block that initiates wave breakup and reentry happens to lie close to the nodal line of the stable discordant alternans that is the furtherest from the pacing site.
It should be pointed out, however, that this correspondence is not precise, since nodal lines can (and do) move during the transient evolution from stable discordant alternans to conduction block.

However, in a marked departure from the prevailing paradigm, the transition from discordant alternans to conduction block (followed by wave breakup, reentry, and SWC) is not associated with a bifurcation leading to either an instability or disappearance of the alternans state.
Instead the transition is triggered by strong transient amplification of disturbances away from stable alternans.
At the pacing interval of 80 ms, certain infinitesimal disturbances were found to be amplified by almost three orders of magnitude.
Transient amplification increases as the pacing interval is decreased, reaching a maximum of four orders of magnitude for a pacing interval of 66 ms.
This is very similar to the trend found for linearly stable fluid flows \cite{meseguer2003linearized}, where transient amplification increases with the Reynolds number, which plays the role analogous to the pacing rate. 

Furthermore, we found that the spatial profile of the optimal perturbations that experience the strongest transient amplification corresponds to a tiny change in the pacing interval.
Hence even small changes in the pacing interval will generate a strong perturbation away from the 2:2 response.  
For a protocol with a decreasing pacing interval, the strength of a perturbation is controlled by the rate at which the pacing interval changes.
Since transient amplification increases with decreasing pacing interval, transition from alternans to SWC will happen sooner (for a longer pacing interval) if the pacing interval is decreased faster.

Although generalized linear stability analysis is capable of describing transient dynamics qualitatively, it cannot produce a quantitatively accurate description of the transition, which involves intrinsically nonlinear effects, such as conduction block.
Using numerical simulations, we were able to predict that, for a pacing interval of 80 ms, transition to SWC is triggered by a linearly optimal perturbation the norm of which is two orders of magnitude smaller than the norm of the alternans state itself. 
Even smaller, nonlinearly optimal perturbations about the alternans state could trigger a transition to SWC.
Computing their magnitude and spatial structure, however, would require a fully nonlinear analysis, following, for example, an adjoint-based minimization approach used to characterize transition to turbulence in shear fluid flows \cite{Pringle2010}.


Our results also have interesting implications for the problem of low-energy defibrillation.
Although some of the modern approaches \cite{Winkle1989,Fenton2009,Luther2011} have demonstrated that the energy required for defibrillation could be reduced substantially compared with the standard defibrillation protocols, further dramatic reduction may be possible by exploiting the geometric structure of the state space. 
Understanding the dynamics in the region of state space surrounding an edge state that lies on the basin boundary between SWC and time-periodic solutions can help design protocols that can terminate fibrillation with, technically, infinitely small perturbations.
In practice, the perturbation strength will depend on how close the chaotic dynamics approach the basin boundary, but still there is a potential for orders-of-magnitude improvement. 

The identification of edge states should also help predict the opposite process -- when a seemingly tiny perturbation to a stable time-periodic rhythm leads to a transition into SWC.
More immediately, our results also support an approach designed to prevent fibrillation, rather than terminate fibrillation after it started.
In that approach feedback control is used to stabilize the normal rhythm, thereby preventing the development of alternans in the first place \cite{Karma2002,dubljevic2008,Garzon11,Garzon14}. 
This approach should work regardless of whether the alternans state is stable or not (e.g., for $R>1.279$ in the model considered here).

The analysis presented here was based on a greatly simplified model of cardiac tissue, so while its results may have a profound impact on how we view the dynamical mechanisms that describe the onset of fibrillation, they need to be verified using more detailed electrophysiological models of cardiac tissue in two and three dimensions.
In particular, it is important to check whether our main results can be reproduced in bi-domain models that are required to correctly describe intra- and extra-cellular potentials.
Finally, it would be interesting to see how the difference between atria and ventricles (both in terms of the electrophysiology and in terms of dimensionality) affect the dynamical description of transition from discordant alternans to fibrillation.

\begin{acknowledgments}
This material is based upon work supported by the National Science Foundation under Grant No. CMMI-1028133.  
The Tesla K20 GPUs used for this research were donated by the ``NVIDIA Corporation'' through the academic hardware donation program.
\end{acknowledgments}

\section*{References}
%

\end{document}